\documentclass[onecolumn]{article}
\usepackage{arxiv}
\usepackage[backend=biber,
style=numeric,
sorting=none,
maxcitenames=2,
maxbibnames=100, 
language=english,
isbn=false,
url=false,
firstinits=true,
uniquename=false
]{biblatex} 
\DeclareNameAlias{default}{last-first}
\DeclareLanguageMapping{english}{english-apa} 

\bibliography{references.bib} 

\usepackage{subcaption}
\usepackage{placeins}
\usepackage{mathtools}
\usepackage{amssymb}
\usepackage{amsmath}
\usepackage{amsfonts}   
\usepackage{fancyhdr}
\usepackage{times}
\usepackage{amsthm}
\usepackage{siunitx}
\usepackage{nicefrac}

\usepackage{algorithm}
\usepackage{algorithmic}

\newcommand{\given}{\,|\,}

\newcommand{\model}{\mathcal{M}}

\usepackage{float}

\usepackage{accents}

\usepackage{enumitem}

\usepackage{xcolor}
\definecolor{darkblue}{RGB}{0,0,128}
\definecolor{darkgreen}{RGB}{0, 128, 0}
\definecolor{darkred}{RGB}{128, 0, 0}
\definecolor{black}{RGB}{0, 0, 0}
\definecolor{errorcolor}{HTML}{481567}
\definecolor{viridisgreen}{HTML}{55C667}

\usepackage{hyperref}
\usepackage{url}
\usepackage{changes}
\hypersetup{
	colorlinks=true,
	linkcolor=darkblue,
	filecolor=darkblue,
	urlcolor=darkblue,
	citecolor=darkblue,
	pdfauthor={L. Schumacher, M. Schnürch, A. Voss, S. T. Radev},
	pdftitle={Validation and Comparison of Non-Stationary Models}
}

\title{Validation and Comparison of Non-Stationary Cognitive Models: A Diffusion Model Application}
\author{
    Lukas Schumacher\thanks{For correspondence, please contact Lukas Schumacher (\href{mailto:schuma.luk@gmail.com}{schuma.luk@gmail.com})}\\
    Institute of Psychology\\
    Heidelberg University\\
    \And
    Martin Schnuerch\\
    Institute of Psychology\\
    University of Mannheim\\
    \And
    Andreas Voss\\
    Institute of Psychology\\
    Heidelberg University\\
    \And
    Stefan T.~Radev\\
    Department of Cognitive Science\\
    Rensselaer Polytechnic Institute\\
}

\begin{document}

\maketitle
\begin{abstract}
    Cognitive processes undergo various fluctuations and transient states across different temporal scales.
    Superstatistics are emerging as a flexible framework for incorporating such \textit{non-stationary dynamics} into existing cognitive model classes.
    In this work, we provide the first experimental validation of superstatistics and formal comparison of four non-stationary diffusion decision models in a specifically designed perceptual decision-making task.
    Task difficulty and speed-accuracy trade-off were systematically manipulated to induce expected changes in model parameters.
    To validate our models, we assess whether the inferred parameter trajectories align with the patterns and sequences of the experimental manipulations.
    To address computational challenges, we present novel deep learning techniques for amortized Bayesian estimation and comparison of models with time-varying parameters.
    Our findings indicate that transition models incorporating both gradual and abrupt parameter shifts provide the best fit to the empirical data.
    Moreover, we find that the inferred parameter trajectories closely mirror the sequence of experimental manipulations.
    Posterior re-simulations further underscore the ability of the models to faithfully reproduce critical data patterns. 
    Accordingly, our results suggest that the inferred non-stationary dynamics may reflect actual changes in the targeted psychological constructs.
    We argue that our initial experimental validation paves the way for the widespread application of superstatistics in cognitive modeling and beyond.
\end{abstract}

\section*{Introduction}
\label{sec:introduction}

The human brain operates in a perpetual state of activity, whether it is focused on a particular task or wandering in the inner world of thoughts.
This activity reflects the non-stationary nature of neuronal dynamics, which are characterized by a complex interplay between transient, evoked states, and ongoing spontaneous fluctuations \autocite{galadi2021, melanson2017}.
The complex cognitive processes that emerge from this neuronal activity also tend to exhibit non-stationary dynamics \autocite{vanorden2003, wagenmakers2004, castro-alvarez2023, craigmile2010}.
In other words, proverbial cognitive processes, such as attention, memory, and decision-making, are not constant over time, but instead undergo fluctuations, shifts, and alterations in their functions \autocite{schurr2024}.

Lapses of attention are a canonical cause of such non-stationary dynamics.
Even when actively engaged in a task, our focus can drift or momentarily falter \autocite{weissman2006}.
Moreover, our capacity to sustain attention and concentrate may vary, influenced by factors such as fatigue, motivation, and external distractions \autocite{esterman2019, ratcliff2011, walsh2017}.
These fluctuations can have a significant impact on our cognitive functioning, but they are often overlooked or simplified in traditional models of cognition.
And while these often assume cognitive processes to be stable and time-invariant, there has been a growing recognition that traditional models do not fully capture the complexity and variability of real-world cognition \autocite{schumacher2023, beer2023, evans2017, li2023, kucharsky2021, gunawan2022, cochrane2023}.
Common approaches to address variability in the components of cognitive models can be broadly classified into four categories: \textit{stationary variability}, \textit{trial binning}, \textit{regression approach}, and \textit{frontend-backend} models.

\begin{figure*}[t]
\centering
\includegraphics[width=0.99\textwidth]{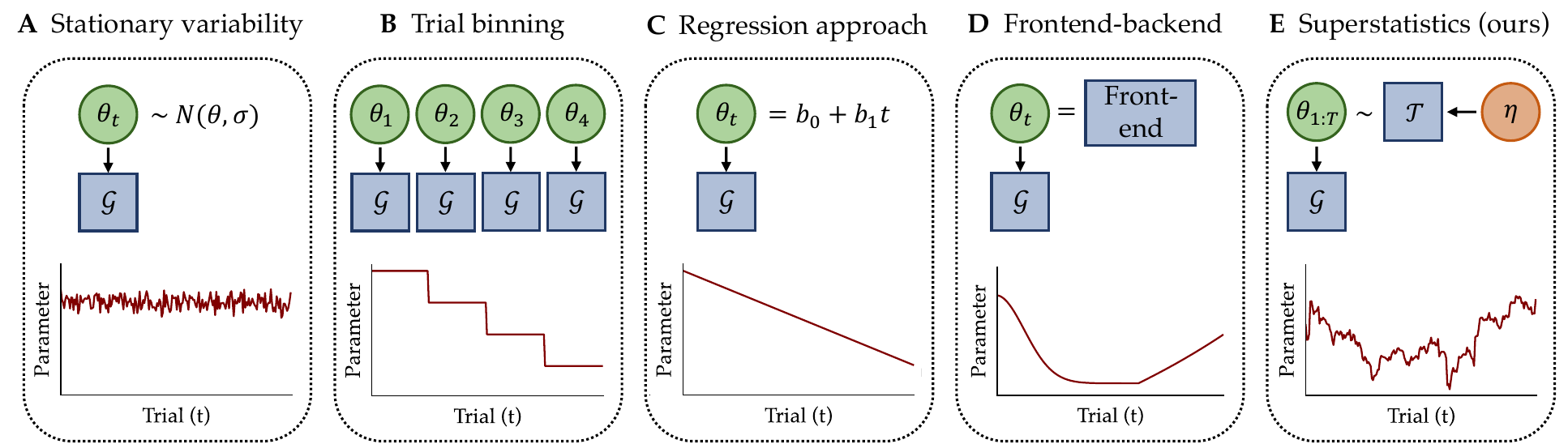}
\caption{
    A conceptual illustration of the five main approaches to model temporal variation in the parameters $\theta$ of a cognitive model $\mathcal{G}$.
    \textbf{a} \textit{Stationary variability}, also known as inter-trial variability, assumes that parameter values fluctuate around a stable mean.
    \textbf{b} \textit{Trial binning} involves organizing the data into distinct bins and fitting a cognitive model $\mathcal{G}$ to each bin individually.
    \textbf{c} \textit{Regression approach} employs time (and sometimes additional contextual variables) as predictors for the parameters $\theta$.
    \textbf{d} \textit{Frontend-backend} models employ a mechanistic model, referred to as the frontend, to elucidate the dynamics of the parameter of the cognitive model (i.e., the backend).
    \textbf{e} \textit{Superstatistics} involve a superposition of multiple stochastic processes operating on different temporal scales. They comprise a low-level observation model $\mathcal{G}$ and a high-level transition model $\mathcal{T}$ that specifies how the parameters $\theta_t$ evolve stochastically.}
\label{fig:modeling_categories}
\end{figure*}

The first approach assumes random fluctuations around a stable mean, referred to as stationary variability (see Fig~\ref{fig:modeling_categories}\textbf{a}). A prominent example of this approach is the ``full'' diffusion decision model (DDM), which allows for inter-trial variability of its core parameters \autocite{ratcliff1998, ratcliff2002}.
However, stationary inter-trial variability mainly improves in-sample model fit and cannot identify systematic changes or sudden shifts in model parameters.
Moreover, the resulting model family still treats behavioral data as independent and identically distributed (IID) responses, making it unsuitable for investigating systematic changes in cognitive constructs.

Another approach for detecting systematic changes in cognitive model components is trial binning \autocite{evans2017, evans2019, kahana2018}.
This method involves organizing data into discrete bins and then applying a stationary model to each of these data subsets separately (see Fig~\ref{fig:modeling_categories}\textbf{b}).
One can then examine variations in parameter estimates across these bins.
The challenge in employing this approach is selecting the number of time steps within each bin, which introduces an unwelcome trade-off between temporal resolution and estimation quality.
For instance, if only a few time steps are chosen, the analysis can yield relatively fine-grained, but very uncertain estimates due to the low number of data points.
A further shortcoming of trial binning is that estimates within a specific bin are not informed by data from neighboring bins.  However, the appeal of dynamic modeling lies in the distinctive capability to utilize both past and future data to constrain the estimated parameter trajectories.

The third approach involves a generalized linear model (GLM) with time (and possibly other contextual factors) as a predictor of model parameters \autocite{cochrane2023, evans2018}. 
The GLM approach is more appealing than trial binning, as it can detect linear or non-linear changes in model parameters without loss of resolution (see Fig~\ref{fig:modeling_categories}\textbf{c}). 
However, the underlying regression function makes strong assumptions about the nature of the relationship between model parameters and time.
Thus, even though a modeler will typically fit and compare a few plausible specifications (e.g., linear vs. exponential), it is often difficult to determine all plausible specifications \textit{a priori}, and the overall flexibility of the GLM model as a process characterization remains severely limited \autocite{gunawan2022}.

Differently, the frontend-backend approach aims to account for changes in model parameters, while providing a mechanistic explanation for the dynamic nature of the target system (see Fig~\ref{fig:modeling_categories}\textbf{d}).
Here, the backend model pertains to the cognitive model which formalizes how the behavioral data are generated (e.g., a DDM).
The frontend constitutes a mechanistic model, elucidating how the parameters of the backend model adapt over time, in different contexts, and in response to additional factors \autocite{fontanesi2019, osth2018, schumacher2023a, brown2008a}.
This approach has several advantages, as it not only accommodates the dynamic nature of the parameters, but also provides a mechanistic description for their temporal variation through a set of static parameters and deterministic functions. For instance, there has been a recent trend to use reinforcement learning models as a frontend model to inform changes in DDM parameters due to reward-based learning \autocite{mcdougle2021, miletic2021, fontanesi2019}.
Nevertheless, detailed frontend models are often challenging to develop, estimate, and compare.

Recently, we proposed an alternative approach that infers non-stationary parameter trajectories directly from the data, while imposing minimal constraints on how parameters change over time \autocite{schumacher2023}.
Our approach leverages a framework known as \textit{superstatistics} \autocite{beck2003, beck2004, mark2018}, which can be viewed from the lens of state space models and involves a superposition of multiple stochastic processes operating on distinct time scales (see Fig~\ref{fig:modeling_categories}\textbf{e}).
At its core, this model comprises a low-level observation model and a high-level transition model.
The former describes how data at a specific time point is generated, akin to the backend model.
Like the frontend approach (cf.~Fig \ref{fig:modeling_categories}\textbf{d}), the transition model characterizes how the parameters change over time.
However, the transition model in superstatistics is inherently a stochastic process, exemplified, for instance, by a Gaussian random walk, a regime switching process, or a mixture between smooth and abrupt transitions.

The superstatistics approach effectively addresses the limitations of prior methodologies.
Unlike stationary models, superstatistical models can readily generate non-stationary variations in the parameters of the low-level model, facilitating gradual or sudden transitions between different states.
Furthermore, parameter estimates are contingent on past data points, thereby treating the data no longer as IID.
In contrast to the trial-binning approach, models within the superstatistics framework leverage the entirety of available data, mitigating concerns about insufficient data points for parameter estimation. 
Different from GLM approaches, our superstatistics method imposes minimal assumptions on potential parameter trajectories, making it significantly less restrictive.

In contrast to frontend-backend models, superstatistics do not offer mechanistic explanations for parameter dynamics but provide greater flexibility in their estimation.
Although mechanistic explanations are central to psychological research, there are cases where suitable explanations are lacking or are applicable only to specific parameters.
Therefore, we consider these two approaches as complementary.
The superstatistical framework takes a bottom-up, exploratory approach, functioning as a tool for generating hypotheses.
In subsequent stages, one could potentially formulate plausible frontend models based on insights from parameter trajectories inferred with a superstatistical model.
Additionally, superstatistical models can serve as benchmarks for testing and validating competing frontend-backend models by comparing resulting parameter trajectories from both methods.

Having laid out the potential benefits of the superstatistics framework and its applicability in the realm of cognitive process models \autocite{schumacher2023}, a pivotal question arises: Do the inferred parameter trajectories genuinely reflect shifts in the cognitive constructs they aim to represent, or are they merely a modeling artefact? 
To address this inquiry, we perform an experimental validation study in which we manipulate the experimental context in a manner that allows us to confidently anticipate how individuals and, consequently, their inferred cognitive constructs, will respond. 
In other words, if the inferred parameter time series mirror the alterations in the experimental context, we garner substantial evidence that these trajectories indeed reflect changes in the psychological constructs.

Throughout, we employ the well-established $4$-parameter diffusion decision model \autocite[DDM;][]{ratcliff1978} as a low-level observation model.
The DDM is a mathematical model that simultaneously accounts for response time (RT) and choice data obtained from two-alternative decision tasks. 
Fundamentally, it posits that, in forced-choice binary decision tasks, individuals accumulate evidence for the decision alternatives until a certain threshold is met, triggering a decision.
Each of the DDM's four core parameters corresponds to a specific psychological construct: (i) the drift rate $v$ signifies the average speed of information uptake; (ii) the threshold $a$ serves as a proxy for decision caution; (iii) the relative starting point $\beta$ represents \textit{a priori} decision preferences; and (iv) the additional constant $\tau$ accounts for the duration of all processes taking place prior and following a decision, such as stimulus encoding or motor action \autocite[but see][]{verdonck2021}.

A primary reason for our choice of the DDM as the observation model lies in its rigorous prior validation \autocite{voss2004, lerche2019, arnold2015}.
These prior studies have convincingly demonstrated that the DDM's parameters are valid reflections of the intended psychological constructs. 
Moreover, the manipulation of experimental conditions leading to systematic alterations in specific DDM parameters is well-documented and comprehensively understood \autocite{ratcliff2008}.
For example, varying the difficulty of an experimental task alters the drift rate parameter, whereas providing verbal instructions to prioritize either speed or accuracy during task-solving leads to observable shifts in the threshold parameter and sometimes also in the non-decision time \autocite{lerche2018}.

In this study, we focus on the aforementioned experimental manipulations targeting the drift rate and the threshold parameters. We employed a \textit{color discrimination task}, which was also utilized in the validation study by \textcite{voss2004}. During this task, individuals must decide whether there are more blue or more orange pixels in a patch of pixels. The task difficulty can be easily manipulated by adjusting the ratio of blue and orange pixels. The farther the ratio is from 1:1, the easier the task becomes. Additionally, we manipulated the emphasis on speed or accuracy by verbally instructing participants to prioritize one over the other.

Systematic changes in cognitive model parameters can appear in different ways, ranging from changing slowly and gradually to more rapid and large shifts.
In our experiment, we focus on two different types.
Firstly, task difficulty changes frequently to the next easier or harder level, imitating gradual changes.
Secondly, the speed-accuracy emphasis changes less regularly after each trial block, resembling sudden shifts.
The primary aim of our experiment is to investigate whether the parameter trajectories inferred with a non-stationary DDM (NSDDM) match these changing patterns of the experimental conditions.
Specifically, we expect the drift rate parameter to mirror the gradual changes in the task difficulty.
Additionally, the threshold parameter should show sudden shifts when the priority switches between speed and accuracy.
It is crucial to understand that in this application, the NSDDM does not have information about the experimental context and has to infer the parameter trajectory solely from the behavioral data.

When dealing with various types of fluctuations, a crucial question arises: What kind of transition model is most suitable for capturing the expected dynamics?
To address this, we implemented different NSDDMs that vary only in their transition model for the drift rate and threshold parameter.
Specifically, we compare four distinct transition models: a Gaussian random walk; a mixture of a Gaussian random walk and uniformly distributed regime changes; a Lévy flight; and a regime-switching function, where parameters either remain constant from the previous time step or shift uniformly.
These four transition models represent different prior assumptions about plausible parameter trajectories.
They vary in complexity (i.e., the number of high-level parameters and functional expressiveness) and their ability to account for different types of temporal shifts.

Performing Bayesian model comparison and parameter estimation with superstatistical models can be computationally challenging \autocite{schumacher2023}. Therefore, we employ simulation-based inference \autocite[SBI,][]{cranmer2020} as implemented in the \texttt{BayesFlow} framework \autocite{radev2023}. 
\texttt{BayesFlow} enables us to carry out a principled Bayesian workflow utilizing simulation-based calibration \autocite[SBC,][]{talts2020, sailynoja2022} and other validation methods \autocite{schad2021toward, gelman2020bayesian} that would otherwise be excessively time-consuming.
The contributions of the present study can be summarized as follows:
\begin{enumerate}
    \item We perform an experimental validation of different non-stationary instantiations of the diffusion decision model.
    \item We propose an amortized method for Bayesian model comparison of non-stationary models via deep ensembles.
    \item We showcase the potential of amortized Bayesian inference for increasing the aspirations of cognitive modeling.
\end{enumerate}

\section*{Materials and Methods}
\label{sec:materials-methods}

\subsection*{Participants}
\label{sec:participants}
A total of $14$ participants ($9$ female, $5$ male) were recruited for the experiment.
The participants had an average age of $23.14$ years ($\text{SD} = 1.29$, $\text{Range} = [22, 26]$).
Every individual provided informed consent to participate in the study, and the research protocol received approval from the local ethics committee.
The entire study was conducted in accordance with the ethical principles outlined in the Helsinki Declaration.

\subsection*{Task}
\label{sec:task}
The participants completed a total of $800$ trials in a color discrimination task, including $32$ practice trials. In each trial, individuals were presented with a rectangular patch containing blue and orange pixels and had to determine whether there were more blue or orange pixels. Prior to the patch presentation, a fixation cross was displayed for $300$ ms. 
All stimuli were presented on a gray background.

Task difficulty was manipulated by varying the proportion of blue/orange pixels in the patch. The following ratios were utilized: $50.5$:$49.5$; $52.25$:$47.75$; $53.5$:$46.5$; and $55$:$45$. Half of the trials featured orange as the dominant color, while the other half featured blue. The difficulty level remained constant for either $8$ or $16$ trials before transitioning to the next level of difficulty.

In addition to manipulating task difficulty, participants received two types of instructions which changed every $48$th trial. In the ``accuracy'' condition, individuals were instructed to prioritize accuracy in their responses. Conversely, in the ``speed'' condition, participants were directed to emphasize speed while maintaining a reasonable level of accuracy. Feedback was provided after each trial to make participants aware of their performance: a green cross for correct responses, a red minus for incorrect responses, and a red clock for responses slower than 700 ms in the speed condition.

\subsection*{Superstatistics Framework}
\label{sec:superstatistics}
To represent non-stationary changes in DDM parameters, we adopt a superstatistics framework \autocite{beck2003, mark2018}.
Within this framework, each generative model comprises (at least) a \textit{low-level observation model} $\mathcal{G}$ characterized by time-dependent local parameters $\theta_t \in \mathbb{R}^K$ that vary according to a \textit{high-level transition model} $\mathcal{T}$ with static high-level parameters $\eta \in \mathbb{R}^D$.
These models simulate parameters and observable data $x_t \in \mathcal{X}$ according to the following general recurrent system
\begin{equation}
\begin{aligned}\label{eq:gen}
    \theta_t &= \mathcal{T}(\theta_{0:t-1}, \eta, \xi_t) \quad\, \text{with}\quad \xi_t \sim p(\xi \given \eta),\,\,\theta_0 \sim p(\theta) \\
    x_t &= \mathcal{G}(x_{1:t-1}, \theta_t, z_t) \quad \text{with}\quad z_t \sim p(z \given \theta_t),
\end{aligned}
\end{equation}
where $\mathcal{T}$ represents an arbitrary high-level transition function parameterized by $\eta$, and $\mathcal{G}$ is a (non-linear) transformation that encapsulates the functional assumptions of the low-level model. The random variates $\xi_t$ and $z_t$ govern the stochastic nature of the two model components through noise outsourcing. 
The initial parameter configuration $\theta_0$ adheres to a prior distribution $\theta_0 \sim p(\theta)$ encoding the available information about feasible starting parameter values.

The above formulation is very abstract and general, highlighting the flexibility of the superstatistics framework.
Moreover, it does not assume that the corresponding transition or likelihood densities, given by
\begin{align}
    \mathbb{T}(\theta_t \given \eta, \theta_{0:t-1}) &= \int  p(\theta_t, \xi \given \eta, \theta_{0:t-1})\,d\xi\\
    p(x_t \given \theta_t, x_{1:t-1}) &= \int  p(x_t, z \given \theta_t, x_{1:t-1})\,dz,
\end{align}
are tractable or available in closed-form, situating our approach in the context of simulation-based inference \autocite[SBI,][]{cranmer2020}.
Here, we build on SBI with neural networks \autocite{ardizzone2018, greenberg2019automatic,  radev2020bayesflow} as a principled approach to perform fully Bayesian inference by using only samples from the generative system defined by \autoref{eq:gen}.
Importantly, our estimation methods overcome key limitations of previous approaches related to the curse of dimensionality \autocite{mark2018}.

\subsection*{Low-Level Model}
\label{sec:low-level}
In this work, we use the same standard DDM implementation as a low-level observation model $\mathcal{G}$ for all NSDDMs. 
The low-level dynamics of the evidence accumulation process are described by the following stochastic ordinary differential equation:
\begin{align}
    \mathrm{d}x_n = v\mathrm{d}t_s + z \sqrt{\mathrm{d}t_s}
    \quad\text{with}\quad
    z\sim\mathcal{N}(0, 1).
\end{align}
Accordingly, the evidence $x_n$ on a given trial $n$ follows a random walk with drift $v$ and Gaussian noise $z$, where $t_s$ represents time on a continuous time scale.
The core assumption of the DDM is that evidence is accumulated with a fixed rate $v$ until one of two thresholds, $a$ or $0$, is reached, and the corresponding decision $D_n$ is made:
\begin{align}
    D_n = 
    \begin{cases}
        1, & \text{if } x_n \geq a\\
        0, & \text{if } x_n \leq 0
    \end{cases}.
\end{align}
Furthermore, the DDM incorporates an additive constant $\tau$, which represents the time allocated to all non-decisional processes (i.e., stimulus encoding and motor action).
Consequently, the DDM encompasses three distinct free parameters, namely $\theta = (v, a, \tau)$.
We fixed the starting point of the evidence accumulation process at $a/2$ since, in our case, the two boundaries of the accumulation process correspond to correct and incorrect responses, respectively.
Thus, it is unwarranted to estimate any potential \textit{a priori} bias towards either of these boundaries \autocite{voss2013diffusion}.

\begin{figure*}[t]
\centering
\includegraphics[width=0.99\textwidth]{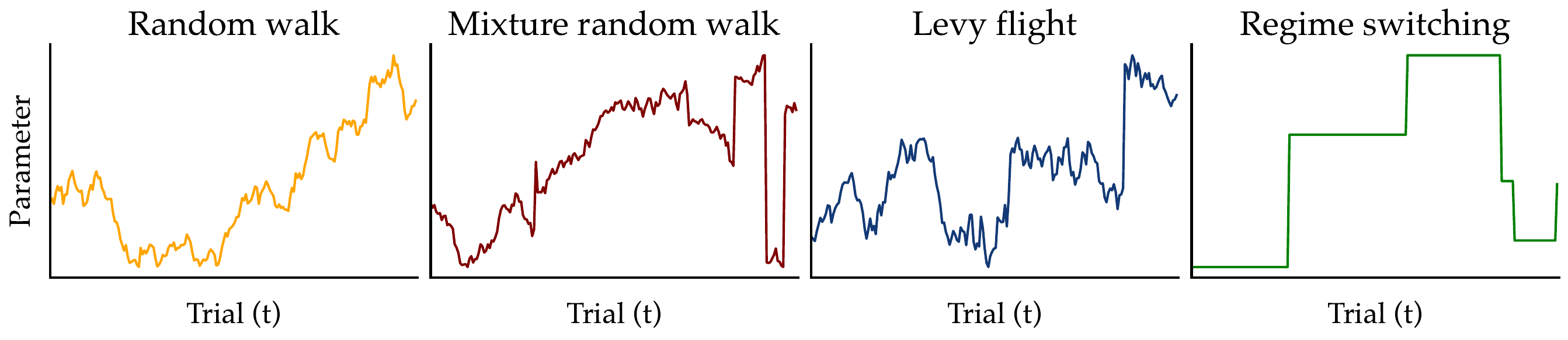}
\caption{An example illustration of the four high-level (transition) models considered in our study, governing the temporal variation of a hypothetical low-level model parameter.}
\label{fig:exemplar_parameter_trajectories}
\end{figure*}

\subsection*{High-Level Models}
\label{sec:high-level}
We formulate and compare four different high-level transition models, denoted as $\mathbb{T}_1$, ..., $\mathbb{T}_4$, which govern the trial-by-trial changes in local DDM parameters $\theta_{1:T}$.
These transition models vary in flexibility in allowing changes to the low-level parameters and their underlying complexity, including the number of high-level parameters involved (see Fig~\ref{fig:exemplar_parameter_trajectories} for an exemplar trajectory generated by each transition model).
To ensure that the low-level parameters remain within plausible ranges, we impose lower and upper bounds on their trajectories.\footnote{To facilitate gradient-based training we transformed the parameters to an unbounded space via scaling.}
Specifically, we set the upper bounds for the parameters $v$, $a$, and $\tau$ to $8$, $6$, and $4$, respectively. 
Additionally, since negative parameter values are not meaningful for our DDM specification, we set the lower bounds for all parameters to $0$.
For simplicity, our transition models do not assume dependencies between the trajectories of the local DDM parameters \textit{a priori}. We note that DDM parameters are typically found to be correlated \autocite{boehm2018}, and thus priors with less entropy (e.g., correlated Gaussian random walk) are also plausible.

\subparagraph*{\textit{Random Walk}}
The first transition model ($\mathbb{T}_1$) convolves the low-level model's parameters with a Gaussian distribution, resulting in a gradual change that follows a random walk:
\begin{align}\label{eq:random_walk}
    \mathbb{T}_1(\theta_{k, t} \given \theta_{k, t-1}, \sigma_k) = \mathcal{N}(\theta_{k, t} \given \theta_{k, t-1}, \sigma_k),
\end{align}
where $k$ denotes the individual model parameters.
According to this transition model, the current value of each parameter $\theta_{k, t}$ is only influenced by its previous value $\theta_{k, t-1}$, generating more or less auto-correlated and gradual changes.

\subparagraph*{\textit{Mixture Random Walk}}
The second transition model ($\mathbb{T}_2$) corresponds to a mixture distribution between a random walk (cf. \autoref{eq:random_walk}) and uniformly distributed shifts:

\begin{equation}\label{eq:mixture_random_walk}
    \mathbb{T}_2(\theta_{k, t} \given \theta_{k, t-1}, \rho_k, \sigma_k, a_k, b_k) = 
    \rho_k\,\mathcal{N}(\theta_{k, t} \given \theta_{k, t-1}, \sigma_k) + (1 - \rho_k)\,\mathcal{U}(a_k, b_k),
\end{equation}
where $\rho$ indicates the probability of the type of change (gradual change or shift) as a mixing coefficient for the two states.
The upper and lower bounds of the uniform distribution, denoted as $a$ and $b$, are set to cover plausible parameter ranges and are not treated as free parameters themselves.

\subparagraph*{\textit{Lévy-Flight}}
The Lévy flight transition model ($\mathbb{T}_3$) is similar to the Gaussian random walk.
However, instead of assuming normally distributed noise, it assumes an alpha-stable transition for each component of $\theta$:
\begin{equation}\label{eq:Lévy_walk}
    \mathbb{T}_3(\theta_{k, t} \given \theta_{k, t-1}, \sigma_k, \alpha_k) = 
    \text{Alpha-Stable}(\theta_{k, t} \given \theta_{k, t-1}, \sigma_k, \beta = 0, \alpha_k),
\end{equation}
where $0 < \alpha \le 2$ governs the heaviness of the noise distribution's tails.
If $\alpha_k = 2$ then the distribution is equivalent to a Gaussian distribution.
Notably, as the value of $\alpha$ decreases, the distribution's tails get heavier, allowing for larger shifts in the parameter values.
When simulating from the Lévy flight transition model, we use a scale of $\sigma_k / \sqrt{2}$, such that the corresponding Gaussian distribution for $\alpha_k = 2$ has a standard deviation of $\sigma_k$.

\subparagraph*{\textit{Regime Switching}} Finally, the regime switching transition model ($\mathbb{T}_4$) is a simpler version of the mixture random walk. The parameter's trajectory adheres to one of two possibilities: it either maintains its previous value or undergoes a uniform shift:
\begin{equation}\label{eq:regime_switching}
    \mathbb{T}_4(\theta_{k, t} \given \theta_{k, t-1}, \rho_k, a_k, b_k) = 
    \rho_k\,\delta(\theta_{k, t} - \theta_{k, t-1}) + (1 - \rho_k)\,\mathcal{U}(a_k, b_k),
\end{equation}
where $\delta(\cdot)$ is the Dirac delta distribution indicating that the parameter either does not change at all with probability $\rho$ or undergoes a sudden change with probability $1 - \rho$.

Strictly speaking, some of the above transition models can effectively be transformed into others by employing specific high-level parameter configurations.
For instance, the mixture random walk with $\sigma = 0$ reduces to the regime switching transition function. 
Conversely, when $\rho = 1$ it reduces to a simple Gaussian random walk.
Also, the Lévy flight transition model with $\alpha = 2$ turns into a random walk transition function.
The mixture random walk and the Lévy flight transition function have two high-level parameters and can thus be regarded as more complex and more flexible than the other two transition models, which only have a single high-level parameter.
Notably, the random walk transition model is the only one that cannot generate relatively large sudden shifts in parameter values.

\subsection*{Model Comparison Setup}

One of the major aims of this study is to compare four NSDDMs sharing the same low-level diffusion model but differing in their assumptions about the type of stochastic variation of the drift rate ($v$) and threshold ($a$) parameters. 
All four NSDDMs employ the same Gaussian random walk model $\mathbb{T}_1$ for the non-decision time parameter ($\tau$).
We base this decision on previous research \autocite{schumacher2023} and the rationale of our experimental manipulations, which should not imply sudden large shifts in the $\tau$ parameter.
For $\mathcal{M}_1$, the drift rate and threshold parameter also follow a Gaussian random walk, resulting in three high-level parameters, $\eta = (\sigma_v, \sigma_a, \sigma_{\tau})$.
In $\mathcal{M}_2$, both $v$ and $a$ follow a mixture between a Gaussian random walk and uniform shifts ($\mathbb{T}_2$), which results in a total of five high-level parameters, $\eta = (\sigma_v, \sigma_a, \sigma_{\tau}, \rho_v, \rho_a)$. 
In contrast, $\mathcal{M}_3$ introduces a trajectory for the drift rate and threshold parameters characterized by a Lévy flight ($\mathbb{T}_3$), which has five free high-level parameters, $\eta = (\sigma_v, \sigma_a, \sigma_{\tau}, \alpha_{v}$, $\alpha_{a})$.
Lastly, for $\mathcal{M}_4$, the two parameters $v$ and $a$ either remain the same as in the previous time point or shift uniformly ($\mathbb{T}_4$). This model has three high-level parameters, $\eta = (\sigma_{\tau}, \rho_v, \rho_a)$.
A listing of the weakly informative prior distributions assigned to the model parameters can be found in the \textbf{Appendix}.

It is noteworthy that these transition models not only differ in their parameter counts, but also in the degree to which they can generate diverse parameter trajectories.
Thus, our Bayesian model comparison approach relies on a more general notion of model complexity as embodied by the \textit{prior predictive distribution} (i.e., marginal likelihood).
The next section discusses Bayesian model comparison from the lens of amortized Bayesian inference as an efficient approximation method.

\begin{figure*}[t]
\centering
\includegraphics[width=0.99\textwidth]{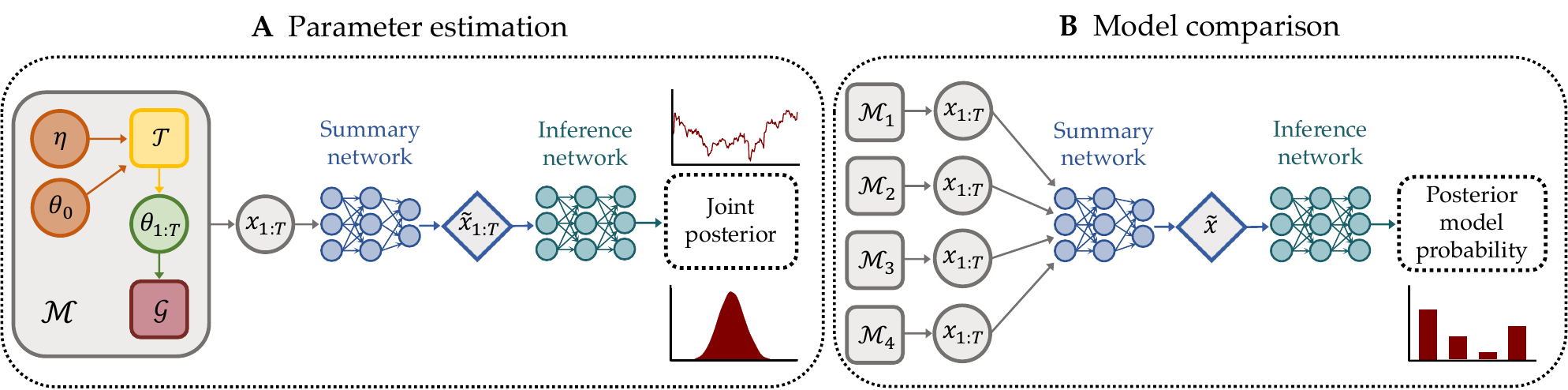}
\caption{A conceptual illustration of our amortized Bayesian inference training setup. \textbf{a} \textbf{Parameter estimation} A recurrent \textit{summary} network processes the synthetic time series $x_{1:T}$ and learns maximally informative temporal summary statistics $\Tilde{x}_{1:T}$. An \textit{inference network} (i.e., normalizing flow) learns to approximate the joint posterior distribution of time-varying low-level parameters $\theta_{1:T}$ and static high-level parameters $\eta$ given the learned summaries. \textbf{b} \textbf{Model comparison} A transformer \textit{summary} network consumes time series simulated from competing models and learns maximally informative summary vectors $\Tilde{x}$. An \textit{inference network} (i.e., a probabilistic classifier) learns to approximate posterior model probabilities (PMPs) given the summary vectors. Once trained, the networks can be efficiently validated using principled Bayesian methods and applied to the observed data.} 
\label{fig:sbi_flow}
\end{figure*}

\subsection*{Amortized Bayesian Inference}
    
Amortized Bayesian inference (ABI) is a flexible framework for estimating, comparing, and validating complex models through simulation-based training of specialized neural networks \autocite{radev2023}.
ABI consists of (i) a training phase where the networks learn a surrogate distribution, and (ii) an inference phase where the networks infer the target quantities (e.g., model parameters or model posterior probabilities) in real-time for any new data set supported by the model(s).
The neural networks are trained purely on simulations from the generative model and do not require an explicit likelihood or numerical integration. 
Thus, ABI re-casts expensive Bayesian inference into a neural network prediction task, such that sampling from the target posterior and model refits happen almost instantaneously.
In a previous study, we showed that ABI drastically outperforms traditional Bayesian methods for estimating time-varying parameters in terms of computation time \autocite{schumacher2023}.

\subparagraph{\textit{Amortized Parameter Estimation}}

Our deep learning approach for jointly estimating time-varying and static parameters follows \textcite{schumacher2023}, who extend ideas from ABI with static parameters \autocite{radev2020bayesflow, gonccalves2020training} to non-stationary Bayesian models.
Accordingly, our goal is not only to infer the trajectories of all three model parameters $\{\theta_t\}^{T}_{t=1}$, but also to estimate the posterior distribution for the static high-level parameters $\eta$ of the transition model.
Thus, we are interested in recovering the full joint posterior $p(\theta_{1:T}, \eta \given x_{1:T})$ from the observed time series $\{x_t\}_{t=1}^T$:

\begin{equation}\label{eq:jointpost}
    p(\theta_{1:T}, \eta \given x_{1:T}) \propto p(\eta, \theta_0)\,p(x_1 \given \theta_1)\,\times 
    \prod_{t=2}^T p(x_t \given \theta_t, x_{1:t-1})\,\prod_{t=1}^T \mathbb{T}(\theta_t \given \eta, \theta_{0:t-1}),
\end{equation}

\noindent where $p(\eta, \theta_0)$ is the joint prior over high-level parameters and initial low-level parameter values.
The joint prior typically factorizes as $p(\eta, \theta_0) = p(\eta)p(\theta_0)$, assuming that $\eta$ and $\theta_0$ are independent in the absence of any information.
Even though our SBI method is applicable to any model of the general form in Eq.~\ref{eq:jointpost}, our low-level (\textbf{Low-Level Model}) and high-level (\textbf{High-Level Models}) specifications lead to a simplified formulation
\begin{equation}
    p(\theta_{1:T}, \eta \given x_{1:T}) \propto p(\eta, \theta_0)\,\prod_{t=1}^T p(x_t \given \theta_t)\,\prod_{t=1}^T \mathbb{T}(\theta_t \given \eta, \theta_{t-1}).
\end{equation}
The simplified formulation follows from the fact that our transition models share the Markov property and the DDM likelihood depends on time only through the current parameter $\theta_t$ in the latent trajectory $\theta_{1:T}$.

Following the typical ABI offline training setting (see Fig~\ref{fig:sbi_flow}\textbf{a} for a conceptual illustration), we generate a \textit{data set of simulated data sets}, $\smash{\mathcal{D} = \{\eta^{(b)}, \theta^{(b)}_{1:T}, x^{(b)}_{1:T}\}_{b=1}^B}$, and use the simulated data to train a specialized neural network, $F_{\psi}(\theta_{1:T}, \eta;\,x_{1:T})$, which approximates the full joint posterior \autocite[i.e., a normalizing flow, see][]{papamakarios2021normalizing}.
In particular, we minimize the following loss in expectation over the full non-stationary generative model (i.e., the right hand-side of Eq.~\ref{eq:jointpost})
\begin{equation}
    \mathcal{L}(\psi) = \mathbb{E}_{(\eta, \theta_{1:T}, x_{1:T}) \sim \mathcal{D}}\left[-\log q_{\psi}(\theta_{1:T}, \eta \given x_{1:T}) \right],
\end{equation}
where we approximate the expectation over $p(\theta_0)\,p(\eta, \theta_{1:T}, x_{1:T})$ via our training set $\mathcal{D}$ and regularize against overfitting with standard techniques, such as dropout and weight decay. 
It is also possible to run the simulator(s) indefinitely and perform online training using on-the-fly simulation \autocite{radev2020bayesflow}. In fact, this approach should be preferred for \textit{fast simulators}, as it makes overfitting hardly possible. 
Thus, online learning is the approach we pursue for training the neural approximators.

In the context of dynamic Bayesian models, we have many choices on how to \textit{factorize} the joint posterior \autocite{sarkka2013}.
The two most common choices are to approximate the \textit{filtering distribution} or the \textit{smoothing distribution} \autocite{mark2018}.
The filtering distribution corresponds to an online analysis, where the low-level parameters $\theta_t$ at time step $t$ are only informed by past data points.
Differently, the smoothing distribution conditions the posterior of $\theta_t$ on all past and future data points, and provides potentially sharper estimates.
Thus, in this study, we exclusively target the approximate smoothing distribution due to its superior parameter recoverability in an offline analysis.\footnote{Note, that \textcite{schumacher2023} focused exclusively on the filtering distribution in their benchmarking experiments.}
In practice, we employ unidirectional or bidirectional long-short term memory (LSTM) networks \autocite{gers2000learning} with many-to-many input-output relationships as a backbone for approximating the filtering or smoothing distribution, respectively.
We then train four separate neural approximators, such that each network becomes an ``expert'' in inferring the smoothing distribution of the corresponding NSDDM.
The \textbf{Appendix} contains more details on the neural network settings and training hyperparameters.

\subparagraph{\textit{Amortized Model Comparison}}
To conduct a comparative analysis of the four NSDDMs, we focus on Bayes factors (BFs) and posterior model probabilities (PMPs). These measures can be classified as \textit{prior predictive}, since they depend on the \textit{marginal likelihood} (see below) as a proxy for a model's generative diversity, penalizing prior complexity \autocite{kass1995, mackay2003information}.
The efficacy of these measures has been demonstrated in a wide range of psychological modeling studies  \autocite{heck2023}. Nevertheless, an ongoing debate surrounds the preference between the two \autocite{vanravenzwaaij2022, tendeiro2019}. 
Since BFs and posterior odds (i.e., ratios between PMPs) are equivalent when all models are assumed to be equally likely \textit{a priori}, we estimate and analyse both quantities in our study.

Following the common Bayesian terminology \autocite{mackay2003information}, we can refer to the four competing models through an index set $\mathcal{M} = \{\mathcal{M}_1, \mathcal{M}_2, \mathcal{M}_3, \mathcal{M}_4\}$.
Prior predictive Bayesian model comparison aims to find the simplest most plausible model within $\mathcal{M}$.
To this end, we can compute PMPs for each of the competing models
\begin{equation}
\label{eq:pmp}
p(\mathcal{M}_j \given x_{1:T}) = \frac{p(x_{1:T} \given \mathcal{M}_j) \, p(\model_j)}{\mathbb{E}_{p(\mathcal{M})}\left[p(x_{1:T} \given \mathcal{M})\right]},
\end{equation}
where $p(\model)$ refers to the prior distribution over the discrete model space. The marginal likelihood $p(x_{1:T} \given \mathcal{M}_j)$ plays a crucial role in \autoref{eq:pmp}, and can be expressed by integrating out all parameters of the joint model,
\begin{equation}
\label{eq:likm}
p(x_{1:T} \given \mathcal{M}_j) = \int p(\eta, \theta_0)\,\prod_{t=1}^T p(x_t \given \theta_t, \mathcal{M}_j) 
\prod_{t=1}^T \mathbb{T}_j(\theta_t \given \eta,\theta_{t-1})\,d\eta\,d\theta_0,\dots,d\theta_T.
\end{equation}
Importantly, since the marginal likelihood averages the likelihood over the joint prior, it automatically incorporates a probabilistic Occam's razor, favoring models with constrained prior predictive flexibility.
When comparing a pair of competing models, $\mathcal{M}_j$ and $\mathcal{M}_{i}$, we can compute the ratio between their respective marginal likelihood,
\begin{equation}
\label{eq:bf}
\text{BF}_{ji} = \frac{p(x_{1:T} \given \mathcal{M}_j)}{p(x_{1:T} \given \mathcal{M}_i)}.
\end{equation}
This ratio is referred to as the Bayes factor (BF). Consequently, a $\text{BF}_{ji} > 1$ signifies a relative preference for model $j$ over model $i$ based on the given data $x_{1:T}$ \autocite{kass1995}.

Unfortunately, the marginal likelihood is notoriously hard to approximate \autocite{gronau2017tutorial} and even doubly intractable for mechanistic models with unknown or unnormalized likelihoods.
To circumvent this intractability, we follow the neural method of \textcite{radev2020, elsemüller2023} which enables amortized Bayesian model comparison for arbitrary computational models (see Fig~\ref{fig:sbi_flow}\textbf{b} for a graphical illustration).
This method involves the simultaneous training of two neural networks with different roles: a \textit{summary network} and an \textit{inference network}.
The summary network learns maximally informative summary statistics from the raw data (e.g., behavioral time series).
The inference network approximates the PMPs for the candidate models, $q_{\phi}(\mathcal{M} \given x_{1:T})$ given the outputs of the summary network.
Here, we subsume all trainable network parameters under $\phi$ and refer to the composition of the two networks as an \textit{evidential network}.

The training data for the evidential network consists of all simulations from the candidate models together with the corresponding model index, $\smash{\mathcal{D}(\mathcal{M}) = \{x_{1:T}^{(b)}, \mathcal{M}_j^{(b)}\}_{b=1}^{B'}}$, where $B'$ denotes the total number of simulations from all models.
Together, the two networks minimize the standard cross-entropy loss,
\begin{equation}
    \mathcal{L}(\phi) = \mathbb{E}_{(\mathcal{M}_j, x_{1:T}) \sim \mathcal{D}(\mathcal{M})}\Big[-\sum_{j=1}^J \mathbb{I}_{\mathcal{M}_j} \log q_{\phi}(\mathcal{M}_j \given x_{1:T}) \Big],
\end{equation}
and we approximate the expectation over $p(\eta, \theta_{1:T}, x_{1:T})$ by our training set $\mathcal{D}(\mathcal{M})$, and $\mathbb{I}_{\mathcal{M}_j}$ denotes an indicator function (i.e., one-hot encoding) for the true model index.
In principle, we could use online learning for amortized model comparison as well, but we found \textit{offline training} to yield sufficiently accurate results.

A key concern in amortized Bayesian model comparison is whether the network outputs truly reflect the underlying probabilities \autocite{guo2017calibration}. Ideally, a posterior probability estimate of $0.9$ for a given model suggests that a decision in favor of this model should be correct in $90\%$ of cases. However, if it is only correct in $80\%$ of cases, the estimate is \textit{overconfident}. This discrepancy is quantified using the Expected Calibration Error \autocite[ECE;][]{naeini2015obtaining}, which ranges from 0 (best) to 1 (worst). 
In practice, we estimate the ECE by averaging the deviations between predicted and true probabilities, calculated as relative frequencies within each probability bin.

More recently, \textcite{elsemuller2023a} demonstrated the importance of gauging the sensitivity of amortized neural approximators, especially in the context of model comparison. 
The authors suggest to train an ensemble of multiple evidential networks, instead of relying on a single network. 
Accordingly, we can measure the (lack of) agreement between ensemble members and obtain a hint at the robustness of the approximate PMPs.
Here, we trained an ensemble of ten evidential networks and computed the mean and standard deviation of the estimated PMPs across all ten networks.
For more details regarding the neural network architecture and training settings, we refer the reader to the \textbf{Appendix}.

\section*{Results}

\subsection*{Model Comparison}

\begin{figure*}[t]
\centering
\includegraphics[width=\textwidth]{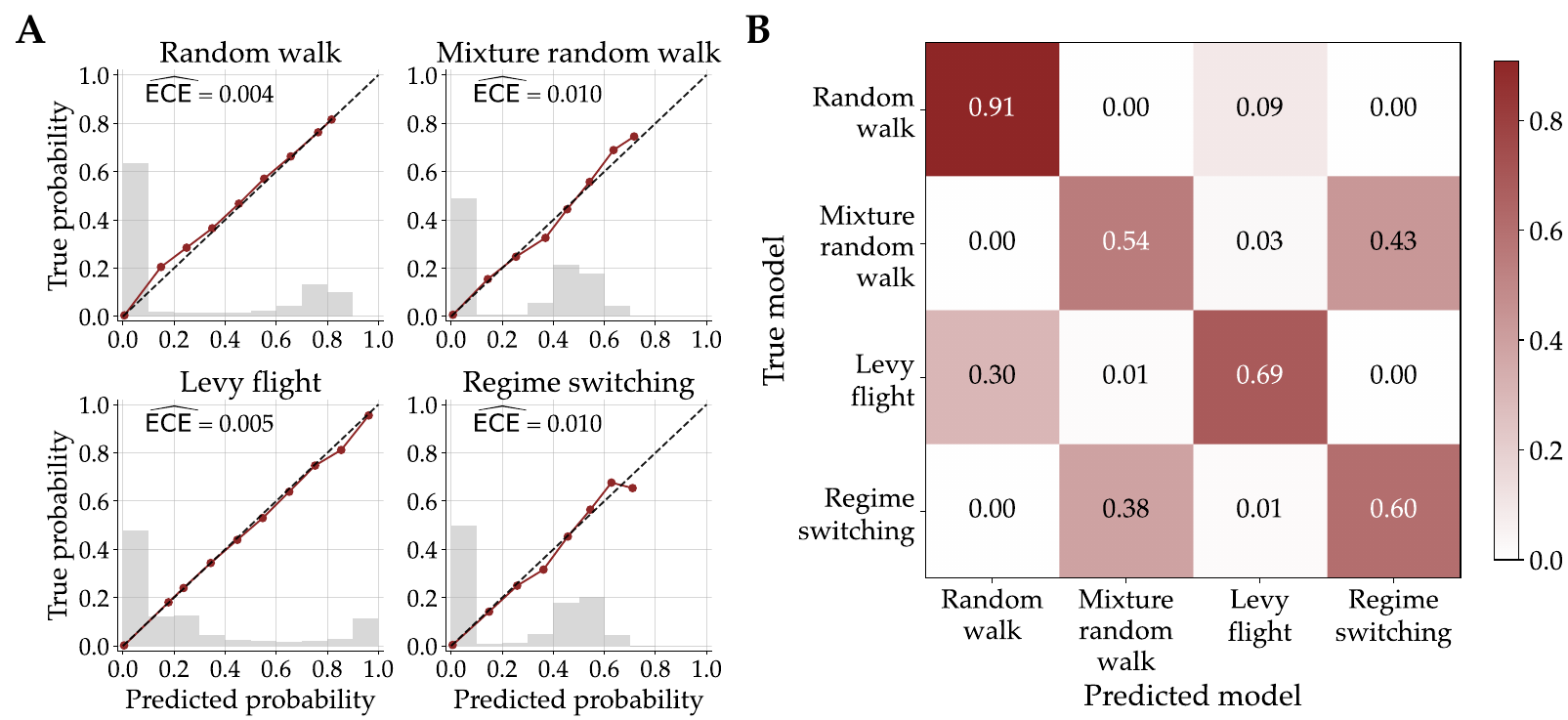}
\caption{In silico model comparison and sensitivity results. \textbf{a} Calibration curves of all four NSDDMs aggregated across the neural approximator ensemble. Additionally, the expected calibration error (\smash{$\widehat{{\mathrm{{ECE}}}}$}) is annotated within each subfigure. The gray histograms depict the relative frequencies of the predicted model probabilities. \textbf{b} Confusion matrix between true data generating model and predicted model. The proportion values were averaged across the ten neural approximator within the ensemble.}
\label{fig:model_comp_validation}
\end{figure*}

As a first step, we assess the closed-world (i.e., \textit{in silico}) performance of our model comparison method in terms of computational faithfulness and accuracy of model recovery.
To assess the former, we perform simulation-based calibration \autocite[SBC;][]{talts2020, sailynoja2022} based on $10\,000$ synthetic data sets each consisting of $800$ trials per model.
Fig~\ref{fig:model_comp_validation}\textbf{a} shows the calibration curves for each NSDDM averaged across the ten evidential networks in our deep ensemble.
We observe excellent calibration with very minimal expected calibration errors ($\widehat{{\mathrm{{ECE}}}}$) across all models. Thus, we conclude that the approximate posterior probabilities are well-calibrated in the closed-world setting.

Next, we assess the accuracy of our model comparison networks in terms of their ability to correctly identify the ground-truth data-generating model.
To this end, we apply the deep ensemble to the $40\,000$ synthetic data sets we have already simulated for assessing calibration. 
In Fig~\ref{fig:model_comp_validation}\textbf{b}, we present the resulting confusion matrix, which illustrates the agreement between true and predicted models averaged across the ten approximators.
Among the four models, the random walk DDM is the only one that rarely gets confused with the other models.
A possible explanation is that it is the only transition model not capable of generating sudden shifts in parameter values.
The remaining models are susceptible to more frequent misclassifications.
For example, the mixture random walk DDM is correctly identified only $54\%$ of the time, and it is often confused with the regime switching model, occurring $43\%$ of the time.
Notably, the Lévy flight DDM is prone to mimicry with the random walk DDM (on average $30\%$ of the time). 

It is essential to emphasize that these results do not imply a deficiency in our model comparison method, but rather underscore the fact that certain pairs of models, such as the mixture random walk and the regime switching DDM, can generate remarkably similar data patterns.
For instance, a significant portion of the prior distribution's mass for the $\alpha$ parameter of the Lévy flight transition model centers around $2$.
If $\alpha \approx 2$, then the Lévy alpha-stable distribution closely resembles a Gaussian distribution, with equality in the case of $\alpha = 2$.
Consequently, simulating the Lévy flight DDM would often yield data patterns that could have just as plausibly originated from the simpler random walk DDM.

Similarly, a substantial portion of the prior mass for the $\sigma$ priors of the mixture random walk transition model clusters around $0$, which subsequently transforms it into a regime switching transition model, resulting in large overlap in synthetic data sets.
Interestingly, the mixture random walk and the Lévy flight DDM are rarely confused, even though both models can produce subtle local changes and large sudden shifts.
This implies that these two transition models generate qualitatively similar but quantitatively easy to distinguish parameter trajectories.

To better understand the similarities between the transition models, we conducted a model misspecification analysis focusing on the mixture random walk and regime-switching models, as these two models exhibited the highest model mimicry. We cross-fitted the models to $100$ synthetic data sets, each consisting of $800$ trials. We then evaluated parameter recovery performance by computing the normalized root mean squared error (NRMSE) between true and estimated parameters for both the well-specified and misspecified scenarios. We found no notable difference in parameter recovery between the two scenarios (see Fig~\ref{fig:misspecification_closed_world} in the \textbf{Appendix}).

In summary, the observation of occasional model confusion is not a limitation of our method; rather, it underscores our method's effectiveness in discerning when two models generate highly similar data, making them less straightforward to differentiate from each other.
Moreover, the amortization property of our method enables us to easily conduct such simulation studies prior to analyzing real data -- estimating $40\,000$ posterior model probabilities would have been infeasible for any other method.

\begin{figure*}[t]
\centering
\includegraphics[width=\textwidth]{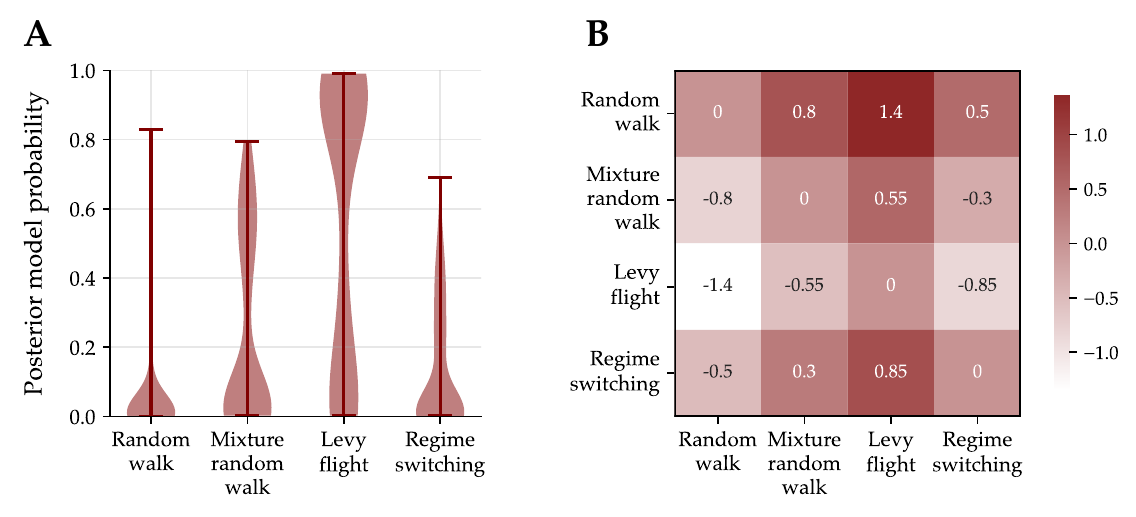}
\caption{Empirical model comparison results. \textbf{a} Distribution of posterior model probabilities (PMP) from all ensembles and $14$ individual participants. \textbf{b} Heatmap of average $\text{log}_{10}$ Bayes factors (BF). Both metrics agree on favoring the Lévy flight over the other transition models.}
\label{fig:model_probabilties}
\end{figure*}

After successfully validating our model comparison method, we apply the deep ensemble to the empirical data of the $14$ participants.
Each approximator in the ensemble was used to infer posterior model probabilities (PMP) for each model, considering each individual's data separately.
Subsequently, we displayed the distribution of the PMPs over all ensembles and individuals (Fig~\ref{fig:model_probabilties}\textbf{a}).
The analysis reveals that the Lévy flight DDM is the most plausible model with an average PMP of approximately $60\%$. It was the most plausible model for $9$ out of the $14$ participants.
In contrast, the mixture random walk model collects an average PMP of less than $30\%$.
Nevertheless, it was estimated to be the most plausible model for $5$ participants.
The random walk DDM and regime switching DDM were consistently less plausible than the other models and did not emerge as superior for any of the participants.

In addition to PMPs, we computed $\text{log}_{10}$ Bayes factors (BF).
Fig~\ref{fig:model_probabilties}\textbf{b} depicts a heatmap of BFs for all one-to-one comparisons between our four NSDDMs, averaged across the participants and the evidential networks of the ensemble.
Following \textcite{kass1995}, an absolute value of $\text{log}_{10}(\text{BF}) > 2$ indicates \textit{decisive} evidence, absolute values between $1$ to $2$ signify \textit{strong}, and between $0.5$ to $1$ \textit{substantial} evidence.
An absolute value of $\text{log}_{10}(\text{BF}) < 0.5$ is labeled as \textit{not worth more than a bare mention}. 
The BF patterns in Fig~\ref{fig:model_probabilties}\textbf{b} align with the PMP findings, implying \textit{strong} evidence for the Lévy flight DDM over the random walk DDM and \textit{substantial} evidence over the other NSDDMs.
Also, both the mixture random walk and the regime switching DDM have \textit{substantial} evidence over the random walk model. 
Interestingly, there is little evidence favoring the mixture random walk DDM over the regime-switching model, suggesting comparable performance.

These findings offer two substantive insights.
First, the ability of transition models to generate sudden shifts in parameters seems essential, as seen in the random walk DDM's lower plausibility.
Moreover, the regime switching DDM, allowing for occasional shifts, but neglecting small gradual changes, performed less effectively than the more complex models.
This result underscores the importance of accommodating both gradual as well as sharp changes in model parameters for achieving optimal fit.
Consequently, the more complex NSDDMs, particularly the mixture random walk DDM and Lévy flight DDM, emerged as more plausible than their simpler counterparts, despite the implicit penalty for prior complexity imposed by Bayesian model comparison.

\begin{figure*}[t]
\centering
\includegraphics[width=\textwidth]{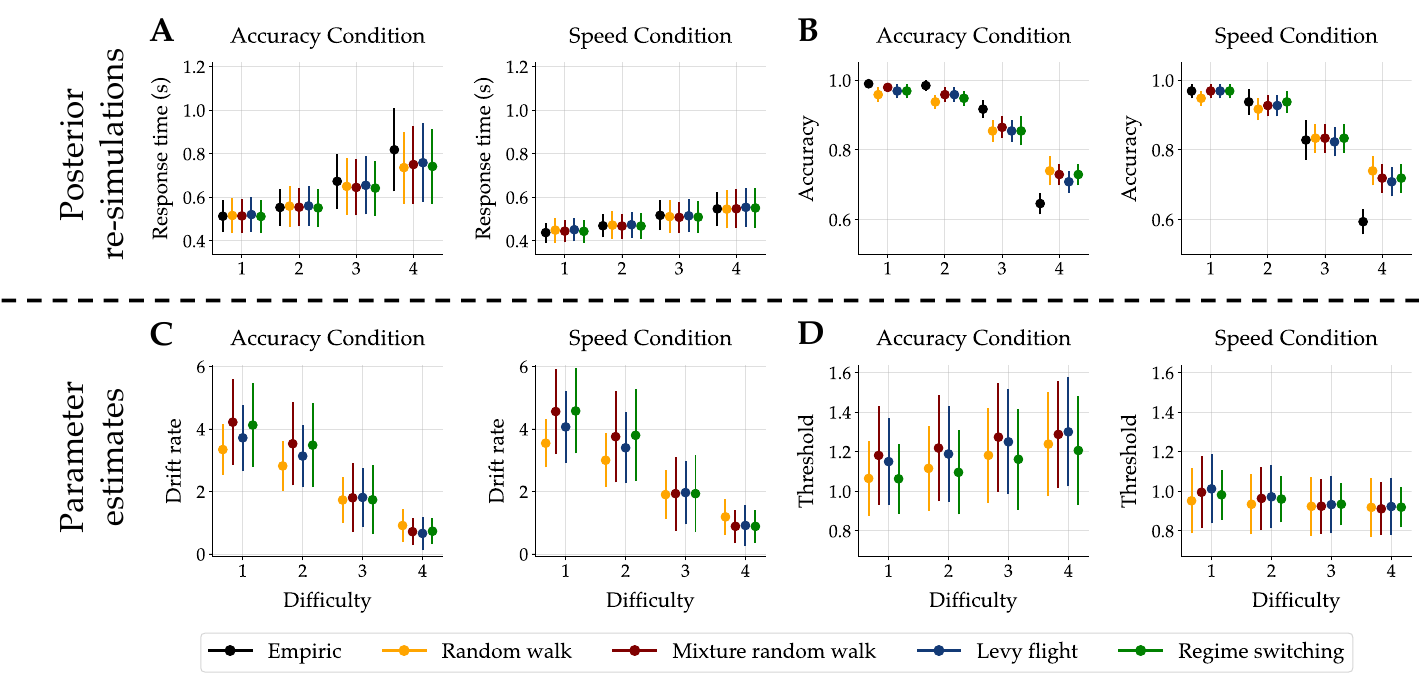}
\caption{Aggregated results from all models fitted to the empirical data. The top row illustrates posterior re-simulations as a measure of the model's generative performance and absolute goodness-of-fit to the data. The bottom row depicts parameter estimates of the drift rate and the threshold parameter from the non-stationary diffusion decision models (NSDDM). \textbf{a} Empirical and re-simulated RTs for each difficulty level and both conditions. \textbf{b} Empirical and re-simulated proportions of correct choices (accuracy) for each difficulty level and both conditions separately. \textbf{c} Posterior estimates of the drift rate parameter for each difficulty level and both conditions separately. \textbf{d} Posterior estimates of the threshold parameter for each difficulty level and both conditions separately. Points indicate medians and the error bars represent the median absolute deviations (MAD) across individuals and re-simulations.}
\label{fig:main_results}
\end{figure*}

\subsection*{Posterior Re-simulation}
Subsequently, we fit all four variants of the NSDDM to each of the $14$ data sets, evaluating the absolute goodness-of-fit of each model.
To achieve this, we conducted $500$ re-simulations with randomly sampled posterior parameter trajectories for each individual data set.
In Fig~\ref{fig:main_results}\textbf{a}, we present the median and median absolute deviation (MAD) of response times (RT) across all individuals and re-simulations.
We provide these aggregates for each NSDDM, categorized by task difficulty level and the two experimental conditions.
Notably, an initial observation reveals that the experimental manipulations were effective on average: empirical median RTs increased with task difficulty, and individuals tended to respond faster in the speed condition compared to the accuracy condition.
Remarkably, all four variants of the NSDDM demonstrated an outstanding fit to these empirical data patterns.
Solely, RTs in the accuracy condition with the highest task difficult level consistently are underestimated by all NSDDM variants.

The empirical and re-simulated proportion of correct choices (accuracy) are aggregated and presented in the same way as the RTs (see Fig~\ref{fig:main_results}\textbf{b}).
Again, the empirical data align with the anticipated patterns resulting from our experimental manipulations.
As expected, accuracy diminishes with increasing task difficulty.
Individuals are generally less accurate in the \textit{speed} condition compared to the \textit{accuracy} condition.
Although NSDDMs successfully reproduce the general patterns in the choice data, we observe notably worse re-simulation compared to that of the RTs data.
In both accuracy and speed conditions, re-simulated accuracies exhibit a less pronounced decline as a function of difficulty than observed in the empirical data.
Further, the difference in accuracy between the two experimental conditions is less pronounced in the re-simulated data compared to the behavioral data.
Notably, the random walk DDM underperforms relative to the other three NSDDMs in this analysis.

It is important to highlight that, unlike conventional approaches, the models did not receive any information regarding the specific experimental context an individual faced at any given moment.
From these analyses, we conclude that all NSDDM implementations successfully capture the general patterns in the empirical RT data.
Individual participant analyses, detailed in the \textbf{Appendix}, affirm the same conclusions.

In addition to analyzing the absolute model fit at the aggregate level, we evaluated the fit across the RT time series. For each participant, we generated $250$ posterior re-simulations for the first $700$ trials using the corresponding best-fitting NSDDM. The remaining $68$ data points were reserved for predictive analysis. For this analysis, we employed a \textit{one-step-ahead} prediction approach, where we iteratively forecasted the subsequent data point starting at time step $700$, followed by re-fitting the model for each of the remaining steps.

\begin{figure*}[t]
\centering
\includegraphics[width=\textwidth]{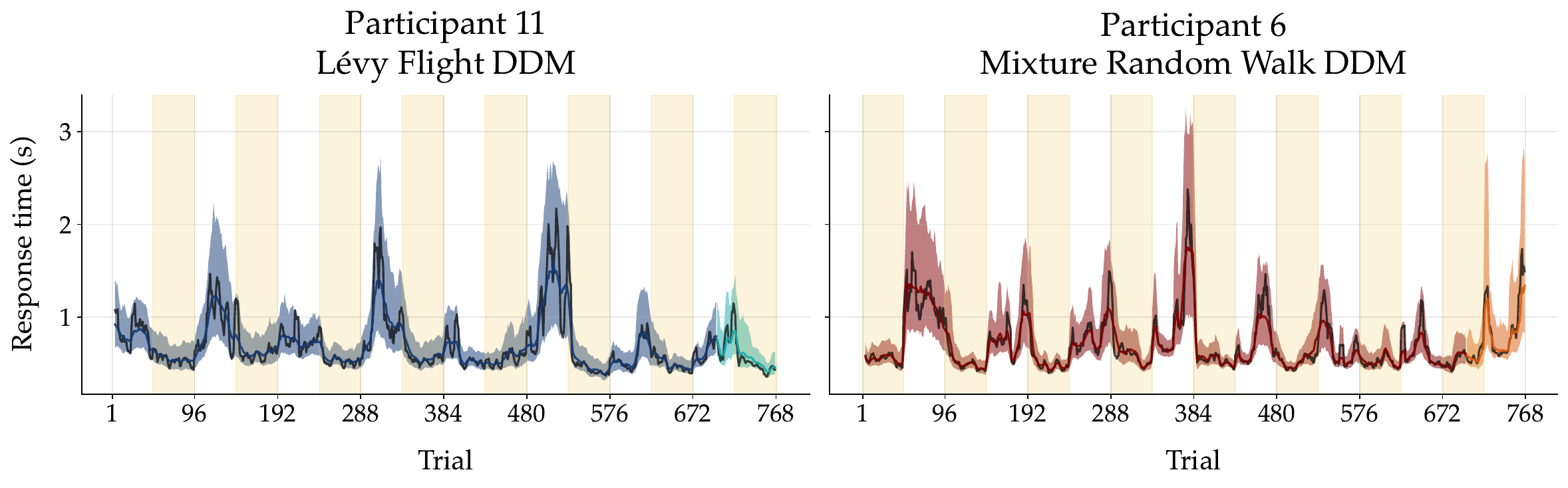}
\caption{Model fit to response time (RT) time series. The empirical RT time series of two exemplar individuals are shown in black. From trial $1$ to $700$, the posterior re-simulations (aka retrodictive checks) using the best fitting non-stationary diffusion decision model (NSDDM) for the specific individual are shown in blue and red, respectively. In this instance, the left column showcases results from a Lévy flight DDM, while the right column displays parameter trajectories from a mixture random walk DDM. For the remaining trials, one-step-ahead posterior predictions from the NSDDMs are depicted in cyan and orange, respectively. Solid lines correspond to the median and shaded bands to $90\%$ credibility intervals (CI). The empirical, re-simulated, and predicted RT time series were smoothed via a simple moving average (SMA) with a period of $5$. The yellow shaded regions indicate trials where speed was emphasised over accuracy, while blank white areas denote instances where the opposite emphasis was applied.}
\label{fig:rt_time_series_resim}
\end{figure*}

Fig~\ref{fig:rt_time_series_resim} illustrates the empirical and re-simulated RT time series for two exemplary participants. 
Results for the remaining $12$ participants can be seen in the \textbf{Appendix}.
The colored lines depict the median and the shaded bands represent to $90\%$ credibility intervals (CI) across the $250$ re-simulations.
Both the empirical data (solid black lines) and the re-simulated/predicted RTs were smoothed using a simple moving average (SMA) with a period of $5$.
Yellow shaded regions highlight trials where speed was emphasised over accuracy, whereas blank white areas denote instances where the opposite emphasis was applied.
Overall, RTs were slower and more variable in the accuracy condition.
Notably, the NSDDM not only closely replicated the empirical time series but also effectively predicted future data points. 
This suggests that the model does not overfit the data.

\subsection*{Parameter Estimates}

At the heart of the current validation study are the inferred parameters, prompting a crucial question: Do these parameter dynamics align with the sequence of experimental manipulations? We address this question by examining both the time-averaged and time-varying estimates. 

\subparagraph*{\textit{Aggregate Analysis}}
We initially examine the parameter estimates averaged across individuals for each difficulty level and condition separately.
This provides a comprehensive overview of average effects on model parameters in different experimental contexts, at first, without delving into the temporal aspect.
The bottom panel of Fig~\ref{fig:main_results} illustrates the posterior medians and MADs collapsed onto the different experimental contexts for the drift rate (Fig~\ref{fig:main_results}\textbf{c}) and threshold parameter (Fig~\ref{fig:main_results}\textbf{d}).

Analyzing the aggregated drift rate estimates reveal an anticipated pattern.
On average, the drift rate decreases as task difficulty increases, observed in both the accuracy and speed conditions.
Additionally, slightly higher overall values are estimated in the speed condition compared to the accuracy condition.
While all four NSDDMs yield fairly similar parameter values, the distinctions in average parameter values between difficulty levels are less pronounced when estimated with the random walk DDM.

With the second experimental manipulation - namely, the instruction to emphasize speed or accuracy - we aimed to manipulate the participants' decision caution, which is assumed to be captured by the threshold parameter.
Examining the aggregated estimates of the threshold parameter in Fig~\ref{fig:main_results}\textbf{d}, we observe generally increased values in the accuracy condition compared to the speed condition.
Interestingly, in the accuracy condition, the threshold parameter also slightly increases with growing task difficulty — a pattern not observed in the speed condition.
A comparison between the estimates of the four NSDDMs reveals that the mixture random walk DDM and the Lévy flight DDM yield higher threshold estimates in the accuracy condition compared to the other two NSDDMs.
Conversely, all four NSDDMs seem to converge in their threshold parameter estimates in the speed condition.

\begin{figure*}[t]
\centering
\includegraphics[width=0.99\textwidth]{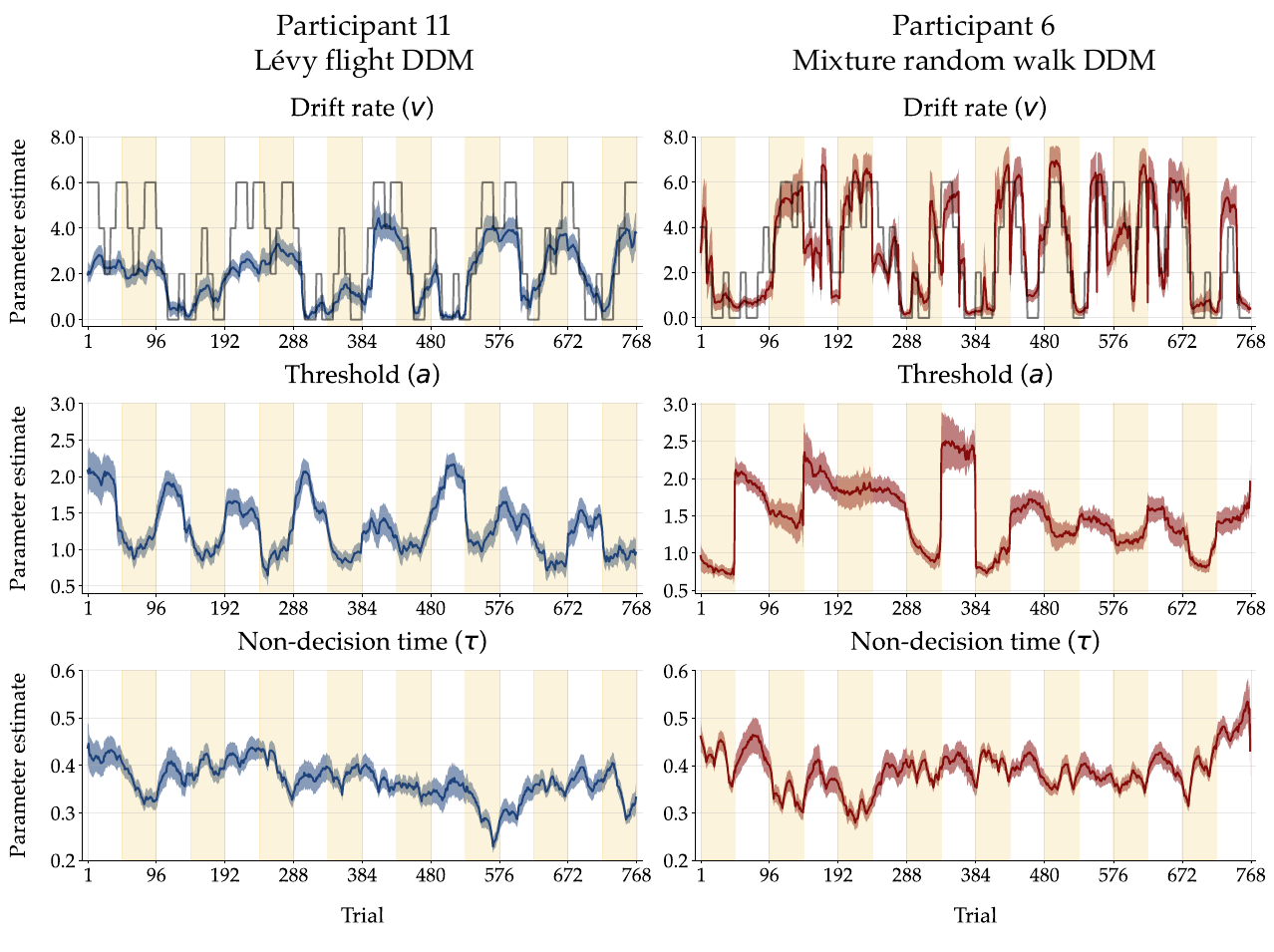}
\caption{Estimated parameter trajectories of two exemplar individuals corresponding to the respective best-fitting non-stationary diffusion decision model (NSDDM). In this instance, the left column showcases results from a Lévy flight DDM, while the right column displays parameter trajectories from a mixture random walk DDM. Each low-level parameter (drift rate, threshold, and non-decision time) is displayed on a separate row. The solid lines are color-coded (blue for the Lévy flight DDM and red for the mixture random walk DDM) to represent the posterior medians, while the shaded regions mark the median absolute deviation (MAD). The yellow shaded regions indicate trials where speed was emphasised over accuracy, while blank white areas denote instances where the opposite emphasis was applied. The sequences of task difficulty levels are depicted with black lines and overlaid with the drift rate in the top panels.}
\label{fig:parameter_trajectory}
\end{figure*}

\subparagraph*{\textit{Parameter Trajectories}}
For a more fine-grained analysis, particularly considering temporal aspects, we present the complete inferred parameter trajectories of the three low-level parameters of a NSDDM for two exemplary individuals in Fig~\ref{fig:parameter_trajectory}.
The \textbf{Appendix} contains the inferred parameter trajectories of the remaining $12$ participants.
Each participant's trajectory is depicted with the posterior median (solid lines) and the median absolute deviation (MAD, shaded bands) across all $768$ experimental trials, estimated with the model with the highest posterior model probability for that specific individual.
The trajectory of participant $11$ corresponds to a Lévy flight DDM, whereas the trajectory of participant $6$ comes from a mixture random walk DDM.
Shaded blocks along the timeline denote the experimental condition at a given trial, with yellow indicating an emphasis on speed.

The top panel illustrates the estimated trajectories of the drift rate parameter alongside the sequences of task difficulty levels (depicted by black line).
Here, $0$ corresponds to the most difficult level, while $6$ represents the easiest.
It is important to note that the absolute values of the difficulty conditions hold no intrinsic meaning.
As observed, the drift rates of both participants align with the overarching trend of the difficulty condition sequence.
They decrease when the difficulty is high and increase as the task becomes easier.

Regarding the trajectory of the threshold parameter (middle panel), we anticipated that a shift from an accuracy instruction to a speed instruction would lead to a decrease in the threshold parameter, and \textit{vice versa}.
This hypothesized pattern is clearly evident when examining the estimated threshold parameter trajectories of the two participants in the middle panel of Fig~\ref{fig:parameter_trajectory}.
For instance, the threshold parameter estimated for participant $11$ oscillates around an approximate value of $1$ in the speed condition.
Moreover, it consistently rises whenever a switch in the accuracy condition takes place.
Intriguingly, the parameter's value during accuracy emphasis is not as uniform compared to the speed condition.
In some blocks, it fluctuates around $2$, while in others, it hovers around $1.5$ or even lower.
Similarly, participant $6$ displays pronounced shifts in the threshold parameter when a change in the condition occurs, with these shifts being more pronounced in the first half of the experiment and diminishing in the second half.

Finally, the bottom panel of Fig~\ref{fig:parameter_trajectory} illustrates the trajectory of the non-decision time parameter.
Although our experimental manipulations did not systematically target the dynamics of this parameter, it is sometimes assumed that the manipulation of speed and accuracy instructions may also influence it \autocite{voss2004, arnold2015}.
While both individuals exhibit some fluctuations in $\tau$, no systematic differences between the two conditions are apparent.

Upon reviewing the parameter trajectories of the remaining participants in the \textbf{Appendix}, similar patterns emerge.
In summary, both the inferred means and trajectories of the drift rate and threshold parameters align with the sequence of experimental manipulations, as predicted by our design.
Moreover, our NSDDMs were able to estimate these trajectories directly from the behavioral data, getting no explicit information whatsoever about the experimental context.
Thus, our validation study suggests that NSDDMs can detect genuine changes in cognitive constructs.

\section*{Discussion}

Psychology and cognitive science are witnessing a growing interest in incorporating dynamic aspects into mechanistic models that seek to formalize and explain cognitive processes.
In a previous study, we explored a method to estimate plausible trajectories of cognitive process model parameters directly from behavioral data \autocite{schumacher2023}.
Nevertheless, an empirical validation of this modeling approach was lacking.
Thus, the current study sought to bridge this gap by experimentally examining the validity of the inferred diffusion decision model (DDM) parameter dynamics.

\subsection*{Experimental Validation}

The present study posed the following core question: Can non-stationary DDMs (NSDDM) effectively detect experimentally induced changes in cognitive constructs from behavioral data alone?
If so, our findings can provide the first substantial evidence for the validity of the superstatistics framework as applied to cognitive models.
Notably, our results demonstrated that the NSDDMs indeed reliably identified the sequence of two experimental manipulations, despite the absence of any contextual information.
Moreover, posterior re-simulation revealed good fit to the general data pattern, both on an aggregate level as well as on the level of the raw time series.
This performance stands as compelling evidence supporting the validity of NSDDMs.

Nevertheless, we observed some misfits in accuracy in the most difficult condition, regardless of the emphasis on speed or accuracy.
Identifying the exact reasons for these discrepancies is challenging, but several factors may have contributed to the difficulty in achieving an accurate fit.
First and foremost, our models did not receive information about the condition of the current trial.
We aimed to validate whether the inferred parameter trajectories could capture changes in conditions, but this made it significantly harder to fit specific data patterns accurately.
Additionally, some prior assumptions might have made it unlikely to fit the behavior in the most difficult condition accurately.
Furthermore, we did not perform any pretreatment of the data, such as excluding trials with very short or long response times.
The goal of this study was to experimentally validate the parameter trajectories, not to achieve perfect data fitting.
Therefore, we did not conduct a detailed analysis to improve our modeling decisions to address these misfits.

Despite these challenges, the inferred parameter trajectories provided valuable insights into how the model responded to varying task conditions. The trajectory of the drift rate parameter for all individuals closely mirrored the sequence of the task difficulty manipulation.
Specifically, the drift rate parameter decreased when task difficulty increased, and conversely, increased as task difficulty decreased.
This not only confirms the anticipated impact of the manipulation, but also highlights the NSDDMs' ability to discern these variations directly from the behavioral data, agnostic to additional contextual information.

Interestingly, drift rates increased throughout the experiment, although this was not the case for trials with the highest task difficulty.
This observation suggests a practice effect among participants, where task performance generally improved with experience, except under the most challenging condition.
Practice effects are a widely recognized phenomenon in various decision-making and memory paradigms \autocite{healey2014, healey2016, forstmann2008, wagenmakers2008, wynton2017}.
In fact, practice effects have been studied with various dynamic cognitive modeling approaches \autocite{kahana2018, gunawan2022, evans2019, evans2018}.
A notable contribution to this field comes from \textcite{gunawan2022}, who conducted a comprehensive re-analysis of three datasets derived from widely cited articles.
Their study compared three dynamic models: (i) a smooth polynomial trend, (ii) a non-smooth autoregressive process, and (iii) a regime switching model instantiated by a hidden Markov model (HMM) with two different states.

In their study, \textcite{gunawan2022} employed a low-level model similar to the DDM, namely the linear ballistic accumulator model \autocite[LBA;][]{brown2008}.
However, their transition models, specifically the polynomial trend and the autoregressive process, differed in that they allowed LBA parameters to change only from block to block, neglecting trial-to-trial parameter fluctuations (except for the HMM).
Their findings indicated that the HMM outperformed the other two dynamic model instantiations.
This superiority can possibly be attributed to the model's capacity to flexibly capture parameter changes from trial to trial, in contrast to changes occurring only from block to block.
Even though the trial-by-trial specification of the HMM captures the microstructure of the decision-making process, it is still less flexible than the models we examined in the current study.
HMMs assume a pre-defined number of possible states, whereas this is not the case with the implementation of our regime switching model.
The advantage of not fixing the number of distinct states beforehand is particularly evident when the exact latent quantity is unknown prior to investigation.
Moreover, results from our model comparison clearly favored transition models that account for both, gradual changes as well as sudden shifts. This suggests that regime-switching models may fall short in certain fields of application.
Nevertheless, both models have their merits, and the choice between them should be guided by the specific research question at hand and formal model comparison.

As our study focused on experimentally validating parameter trajectories estimated with NSDDMs, we deliberately refrained from further analysing practice effects.
However, we suggest that our flexible framework could be a promising alternative for investigating practice effects.
Unlike pure regime switching models, it can reveal a mixture of practice-related changes, ranging from abrupt shifts to gradual changes.
When exploring substantive research questions, such as practice effects, with superstatistical models, it is imperative to depart from the approach taken in the current study. That is, one should always incorporate contextual information from the experimental setting when estimating parameter trajectories.
Here the question arises, how to incorporate this information?
In a previous study, we simply assumed separate low-level parameters for each experimental condition \autocite{schumacher2023}.
This approach is particularly appropriate when conditions randomly change from trial-to-trial.
However, future research could explore alternative ways of including experimental context information with the goal of further informing the parameters.

Concerning the second experimental manipulation, that is, the emphasis on speed or accuracy, their effect on the threshold parameter is more diverse across individuals.
While a majority of participants demonstrated shifts in the threshold parameter in response to instructional changes, the consistency and magnitude of these changes varied significantly among individuals.
Some participants exhibited only a few adjustments in the threshold parameters, seemingly overlooking the change in instruction on certain occasions.
In contrast, others consistently heightened their threshold parameter during accuracy-focused tasks, followed by a subsequent decrease when transitioning to speed-oriented conditions. 
Meanwhile, some participants displayed rather unsystematic changes in decision caution, suggesting that these participants reacted differently to the speed-accuracy manipulation.

\textcite{kucharsky2021} introduced a dynamic LBA incorporating a hidden Markov transition model with two states, akin to the model proposed by \textcite{gunawan2022}.
Their focus centered on scrutinizing the speed-accuracy trade-off, exploring the hypothesis that individuals dynamically switch between different operating states under varying instruction conditions.
By fitting their model to previously collected data, they provided evidence that individuals tend to oscillate between two stable states: a deliberative, stimulus-driven mode emphasizing accuracy and sacrificing speed, and a guessing mode characterized by random and relatively faster choices.

However, our approach for estimating parameter trajectories reveals a more intricate scenario, challenging the assumed binary operational shift.
Contrary to expectations, individuals manifest more than two discernible states.
At times, they exhibit an extreme adaptation to a change in condition, while at other times, they display little or no reaction to the altered condition.
This complexity underscores the necessity for more flexible transition models, as employed in our study.
Failing to utilize such adaptive models could potentially obscure the complex unfolding of individuals' cognition and behavior over time.


\subsection*{Model Comparison}

When implementing non-stationary models, a modeler encounters a myriad of options, ranging from various transition models to decisions about which parameter follows which transition model.
In this study, we limited our choices to a small subset of the possibility space.
Based on our experimental manipulations, we anticipated that the DDM parameters, particularly the threshold parameter, would not only undergo gradual changes, but also more abrupt shifts in response to changing conditions.
Consequently, we tested different implementations accommodating such shifts (mixture random walk, Lévy flight, regime-switching) against a transition model that does not, namely, the simple Gaussian random walk.

The inferred posterior model probabilities (PMPs) and Bayes factors (BFs) consistently favored the Lévy flight and, occasionally, the mixture random walk transition models.
However, in terms of the absolute goodness-of-fit, as assessed through posterior re-simulations, the performance of all four NSDDMs showed remarkable similarity.
This leads to two notable conclusions. First, even the models with lower PMPs demonstrated a good fit to the data, likely owing to the inherent flexibility of the superstatistical framework. 
Second, our Bayesian model comparison method could reliably detect the most favorable model even when the absolute differences were marginal.

\subsection*{Comparison to Time-Variant Models}

Throughout this article, we have focused exclusively on the DDM as a specific example within the broader class of evidence accumulation models. The DDM assumes that certain parameters, such as the threshold parameter, remain constant during a single trial, but we allowed them to vary across different trials.

The literature has introduced models like the collapsing bounds DDM \autocite{bowman2012, shadlen2013} and the urgency-gating model \autocite{ditterich2006}, which allow for time-variance within a trial. The collapsing bounds DDM posits that the threshold decreases throughout of the decision-making process, while the urgency-gating model incorporates both the leaky integration of evidence samples and an ``urgency signal'' that prevents excessive delays in decision-making.

These models have been developed primarily to explain specific empirical data patterns, such as slow errors, without relying on the assumption of random trial-to-trial variability in the core parameters of the DDM \autocite[but see][]{hawkins2015}. This raises the question of whether these models could offer a simpler explanation than our NSDDMs.

While the collapsing bounds DDM and the urgency-gating model can account for certain data patterns without assuming trial-to-trial variability, they cannot identify or explain systematic changes in the underlying constructs caused by factors such as learning, fatigue, motivation, or sudden insights. Uncovering such changes is precisely the goal of NSDDMs.

Nonetheless, the collapsing bounds DDM and the urgency-gating model could be valuable to explore within a superstatistical framework. It has been suggested that these models may provide more plausible explanations for tasks in which stimuli or conditions are dynamic within a single trial \autocite{palestro2018, evans2017a}. With the superstatistical framework it would be straightforward to implement the collapsing bounds DDM and the urgency-gating model, resulting in models that allow for within trial dynamics but also aim to uncover systematic changes over the course of an experiment.

\subsection*{Limitations}

Psychological research is usually interested in group or overall estimate of parameters.
Thus, it would have been informative to compute and inspect ``average'' parameter trajectories.
Unfortunately, our experiment was designed in a way that the difficulty and the speed-accuracy instruction manipulation was randomized across participants.
This made it impossible to average the individual trajectories directly.
Instead, we collapsed the estimates by the different experimental conditions and provided an aggregate view across individuals.
Although this is certainly a limitation of this study, we argue that the current analysis is sufficient to address our specific research question.

Moreover, despite using many default settings from the \texttt{BayesFlow} software \autocite{radev2023}, the configuration and training of neural approximators for both parameter inference and model comparison for non-stationary models can still be a challenge.
A basic understanding of deep learning principles and simulation-based inference is an essential prerequisite.
These requirements may pose obstacles to the adoption of our method, highlighting the necessity for improved software and tutorials addressing these intricacies.

\subsection*{Outlook}

Going forward, our superstatistics framework offers numerous opportunities for future research. It could become a powerful tool in the methodological toolkit of researchers interested in temporal changes in cognitive constructs. As a general framework, it provides significant flexibility to uncover potential parameter dynamics in a data-driven manner. While we have focused on a specific model describing the evidence accumulation process during speeded binary decision-making, many other cognitive process models stand to benefit from such an approach. For instance, parameters in reinforcement learning models, such as the learning rate or the softmax temperature parameter, are likely to change over time \autocite{li2023, ger2024}.

Furthermore, even when the temporal evolution of cognitive parameters is not the central research question, employing non-stationary models can offer advantages over stationary models \autocite{schumacher2023}. Our analysis of estimated trajectories demonstrates clear changes in parameters, highlighting how assuming stationarity could result in misleading conclusions. Exploring these dynamics more deeply could provide valuable insights and drive advancements in various areas of cognitive science.
For instance, researchers could attempt to link non-stationary parameter trajectories to additional neurophysiological time series data, such as EEG or eye movements. This approach could be beneficial in two ways: incorporating additional data as a constraint might improve the precision of cognitive parameter trajectory estimates, and it could enhance our understanding of the relationship between brain activity and specific cognitive constructs.

With great flexibility comes a plethora of choices.
In this study, we compared different transition models guided by the contrast between gradual and sudden changes.
However, there are more degrees of freedom when implementing superstatistical models, or Bayesian models in general \autocite{gelman2020bayesian}.
\textcite{elsemuller2023a} advocates for the crucial role of sensitivity analysis, illustrating a potent methodology to facilitate informed decisions regarding factors such as the type and shape of prior distributions, neural network architectures, and other pivotal elements.
We believe that using such an approach in the context of superstatistics could provide better guidelines for their implementation.

Up to this point, we focused on the estimates of the low-level parameter trajectories.
Yet, it is crucial to note that we also obtain posterior distributions for the static high-level parameters.
These estimates can also yield valuable insights into individuals' behavior and cognition.
Depending on the chosen transition model, these estimates can offer indications of the frequency with which individuals transition between distinct operational states or the variability inherent in their cognitive constructs.
Thus, analyzing these high-level parameters could constitute a compelling avenue for future research.

\subsection*{Conclusion}

In conclusion, the experimental validation of non-stationary diffusion decision models presented in this study represents a significant step forward in the field of cognitive modeling. Our results provide compelling evidence that the estimated parameter trajectories genuinely reflect tangible changes in the targeted psychological constructs. We hope that our validation opens the door to widespread applications of non-stationary models in future modeling endeavors, offering a more nuanced understanding of cognitive processes across varying time scales.

\section*{Data and Code Availability}
All models, data, and scripts for reproducing the results of this work are publicly available in the project's GitHub repository \url{https://github.com/bayesflow-org/Non-Stationary-DDM-Validation}. The neural superstatistics method is implemented in the \texttt{BayesFlow} Python library for amortized Bayesian workflows \autocite{radev2023}.

\newpage

\printbibliography

\clearpage
\newpage

\appendix
\begin{center}
    \bfseries\Large Appendix
\end{center}

\section{Prior Distributions}\label{secA1}
In the following we list the prior distributions we used for all four NSDDM's.

\subsection*{DDM Starting Values}
For the starting values of the parameter trajectories we used half-normal distributions with a mean $\mu$ and a standard deviation $\sigma$ denoted as $\mathcal{HN}(\mu, \sigma)$:

\begin{equation*}
    \begin{aligned}
        v_0 &\sim \mathcal{HN}(2.0, 2.0) \\
        a_0 &\sim \mathcal{HN}(2.0, 1.5) \\
        \tau_0 &\sim \mathcal{HN}(0.3, 1.0)
    \end{aligned}
\end{equation*}

\subsection*{Random Walk Transition Model}
Half-normal distributions were used for the standard deviations of the Gaussian random walk transition model:

\begin{equation*}
    \begin{aligned}
        \sigma_v &\sim \mathcal{HN}(0.0, 0.1) \\
        \sigma_a &\sim \mathcal{HN}(0.0, 0.1) \\
        \sigma_{\tau} &\sim \mathcal{HN}(0.0, 0.01)
    \end{aligned}
\end{equation*}

We decided to use a relatively narrower prior on $\sigma_\tau$ because the non-decision time parameter is not expected to fluctuate as heavily as the other two parameters.

\subsection*{Mixture Random Walk Transition Model}
The mixture random walk transition model used the same prior for the Gaussian random walk as described above. Additionally, Uniform distributions denoted as $\mathcal{U}$ were used for the mixture proportion parameter $\rho$:

\begin{equation*}
    \begin{aligned}
        \rho_v &\sim \mathcal{U}(0.0, 0.2) \\
        \rho_a &\sim \mathcal{U}(0.0, 0.1) \\
    \end{aligned}
\end{equation*}

\subsection*{Lévy Flight Transition Model}
The Lévy flight transition model uses an alpha stable distribution instead of a Gaussian distribution for the transition. We used the same priors for the standard deviations as in the random walk and the mixture random walk. The alpha stable distribution has an additional parameter $\alpha$, which determines the fatness of the tails. This parameter is bound between $1$ and $2$. Therefore, we used a Beta distribution denoted as $\text{B}$ and added $1$ to the sampled values:

\begin{equation*}
    \begin{aligned}
        \tilde{\alpha}_v &\sim \text{B}(1.5, 1.5) \\
        \tilde{\alpha}_a &\sim \text{B}(2.5, 1.5) \\
        \alpha_v &= \tilde{\alpha}_v + 1 \\
        \alpha_a &= \tilde{\alpha}_a + 1 \\
    \end{aligned}
\end{equation*}

\subsection*{Regime Switching Transition Model}
The same prior distributions as for the mixture random walk were used for the mixture probabilities of the regime switching transition model.

\clearpage
\newpage

\section{Neural Network Architectures and Training Setups}
In the following, we outline our implementation of the neural approximators and the training setup used for model comparison and parameter estimation.

\subsection*{Model Comparison}
For model comparison we trained an ensemble of ten neural approximators.
Each approximator consists of a summary network and an inference network.
The summary network is a many-to-one transformer architecture for time series encoding \autocite{wen2023}.
The time series transformer has $128$ template and $64$ summary dimensions.
For inference, we use a network that approximates posterior model probabilities (PMPs) as employed in \textcite{elsemüller2023}.

We performed offline training for each of the ten neural approximators separately.
The training data consisted of $25\,000$ simulations per model.
Training was performed with $25$ epochs and a batch size of $16$ starting with an initial learning rate of $0.0005$.
The learning rate was adjusted with a cosine decay from its initial value to $0$. 

\subsection*{Parameter Estimation}
For parameter estimation we trained one neural approximator for each of the four NSDDM implementations.
Each approximator consists of a hierarchical summary network as employed in \textcite{elsemüller2023} and two inference networks.
Three bidirectional long-short term memory (LSTM) networks were used for the hierarchical summary network.
The number of hidden units were $512$, $256$, and $128$ respectively.

For inference, we use a composition of two invertible neural networks \autocite{radev2020bayesflow}, one for the low-level and one for the high-level parameters.
The network for the low-level parameters has $8$ coupling layers with an interleaved \textit{affine} and \textit{spline} internal coupling design.
The network for the high-level parameters only differs from the former in its number of coupling layers which is $6$.

Since our simulators can be run fast, the training of the four neural approximators was performed online, with $75$ epochs, $1\,000$ iterations per epoch, and a batch size of $16$. Thus, each approximator was trained on $N = 1\,200\,000$ simulated data sets. The initial learning rate was set to $0.0005$ and was reduced with a cosine decay function to $0$.

\clearpage
\newpage

\section{Model Misspecification}

The validation of our model comparison workflow indicated that the mixture random walk DDM and the regime switching DDM are often confused with each other. 
Therefore, we took a closer look at the comparison of their trajectories. First, we investigated their performance in the \textit{closed world} by cross-fitting them on the basis of synthetic data. Second, we directly compared the parameter trajectories from both models inferred for each of the $14$ participants separately (open world).

\subsection{Closed World}

We simulated $100$ synthetic datasets, each consisting of $T = 800$ trials, using the mixture random walk DDM and the regime-switching model separately. We then fitted both models to both types of datasets (cross-fitting). To evaluate the models' parameter recovery performance, we calculated the normalized root mean squared error (NRMSE) between the true and estimated parameters across all time steps for both scenarios (i.e., well-specified and misspecified). The NRMSE is given by:
\begin{equation*}
    \begin{aligned}
        \text{NRMSE}(\hat{\theta}, \theta) = \frac{\sqrt{\frac{1}{n} \sum_{i=1}^{n} (\theta_i - \hat{\theta}_i)^2}}{\hat{\theta}_{\text{max}} - \hat{\theta}_{\text{min}}},
    \end{aligned}
\end{equation*}
where $\theta_i$ and $\hat{\theta}_i$ represent the true and estimated univariate parameter, respectively, and $\hat{\theta}_{\text{max}}$ and $\hat{\theta}_{\text{min}}$ are the maximum and minimum estimated parameter values used for normalization.

\begin{figure*}[hbt!]
\label{fig:misspecification_closed_world}
\centering
\includegraphics[width=0.95\textwidth]{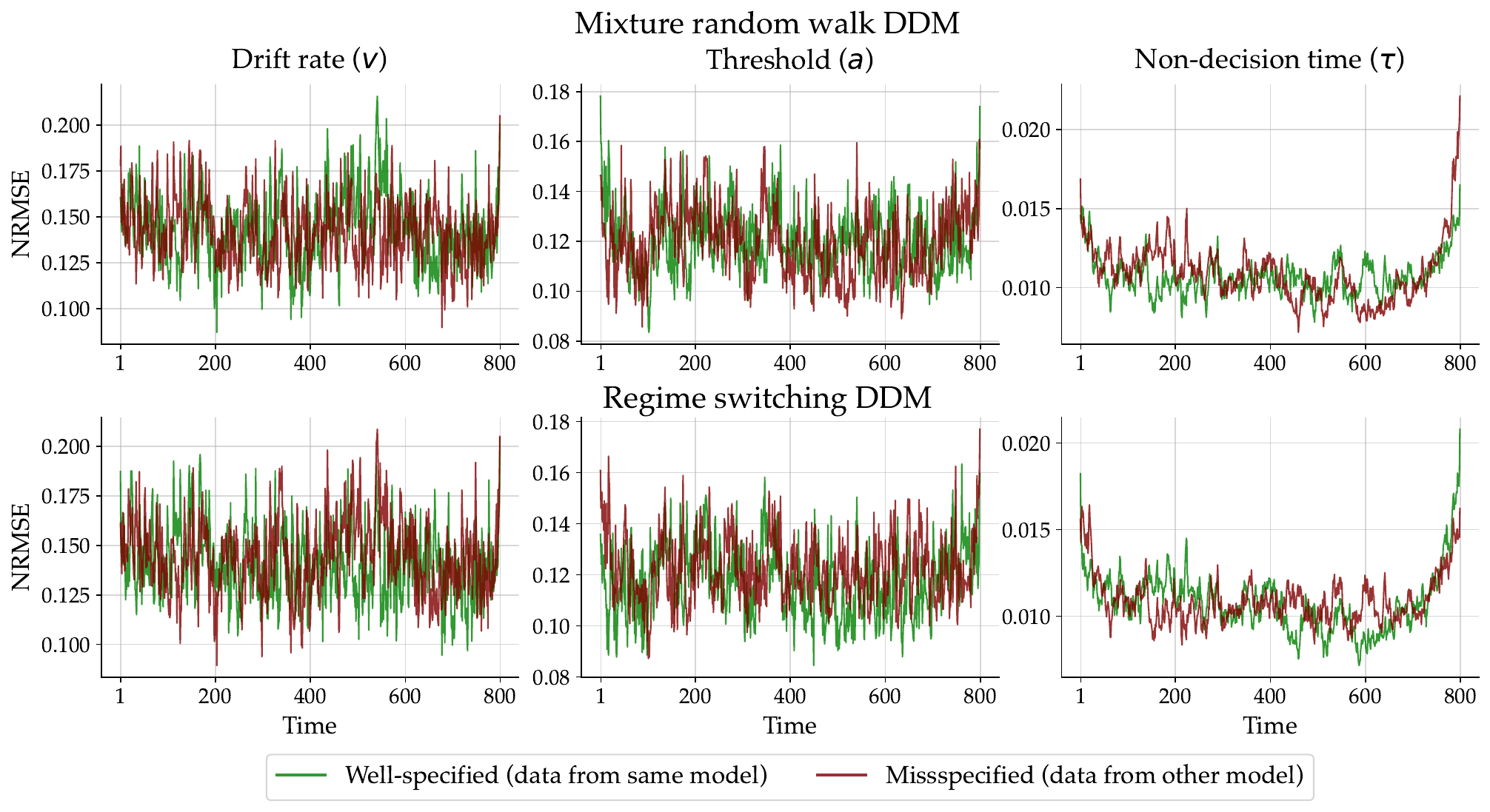}
\caption{
   The average parameter recovery performance for the mixture random walk DDM and the regime switching DDM measured by the normalized root mean squared error (NRSME) between true data-generating and estimated parameters. The green lines indicate the results in the well-specified scenarios where a model is fitted to data generated by the relative model. The red lines indicate the results in the missspecified scenarios where a model is fitted to the data simulated by the other model.
}
\end{figure*}

Fig~\ref{fig:misspecification_closed_world} shows the NRMSE for each model and parameter separately. We observed no notable difference between fitting a model to self-generated data (well-specified) and fitting it to data generated by the other model (misspecified). This suggests that both transition models can reproduce each other's trajectories quite well, making them difficult to distinguish in data space.

\clearpage

\subsection{Open World}

\begin{figure*}[h!]
\label{fig:misspecification_open_world}
\centering
\includegraphics[width=0.8\textwidth]{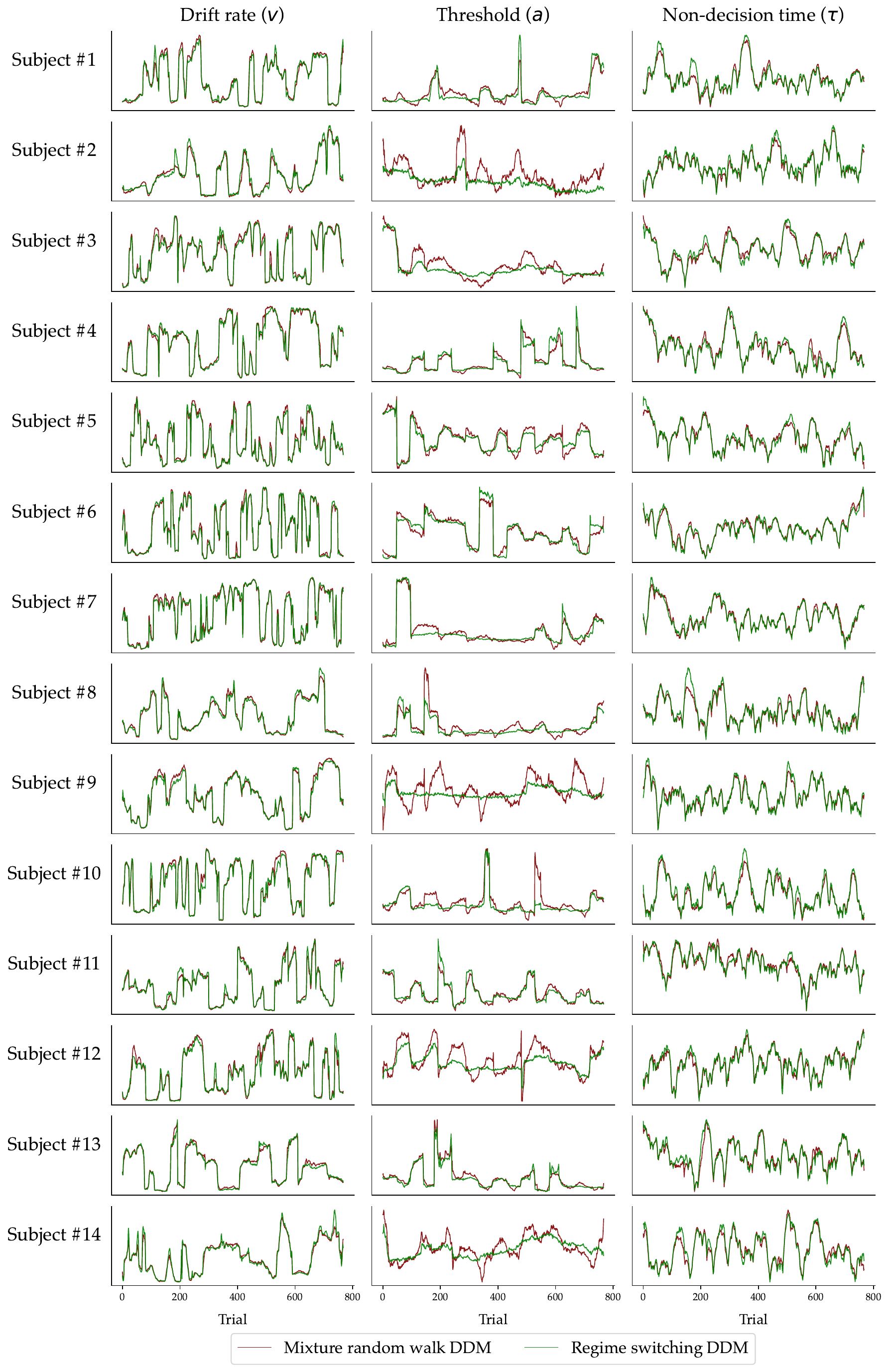}
\caption{
    The median parameter trajectories inferred with the mixture random walk DDM (red) and the regime switching DDM (green) for each of the $14$ subject and core DDM parameter (drift rate, threshold, non-decision time) separately.
}
\end{figure*}

Fig~\ref{fig:misspecification_open_world} shows the parameter trajectories inferred using the mixture random walk DDM (red lines) and the regime-switching model (green lines) for each subject and core DDM parameter. The estimates for the drift rate and non-decision time are nearly identical between the models. Similarly, the threshold parameter trajectories are consistent for most subjects. However, for a few subjects (e.g., subjects $2$ and $9$), the models diverge significantly. The cause of this discrepancy remains unclear, and is likely due to the superior expressiveness of the mixture random walk DDM on the real data.

\clearpage
\newpage

\section{Ablation Study}

To assess the stability of our results, we conducted an ablation study.
We fitted our models to subsets of the data ($1/8, 1/4$ and  $1/2$ of the total trials per person).
We then performed posterior re-simulations and evaluated the absolute fit to the data at an aggregated level, replicating Fig~\ref{fig:main_results}.

\begin{figure*}[h!]
\label{fig:main_results_n96}
\centering
\includegraphics[width=\textwidth]{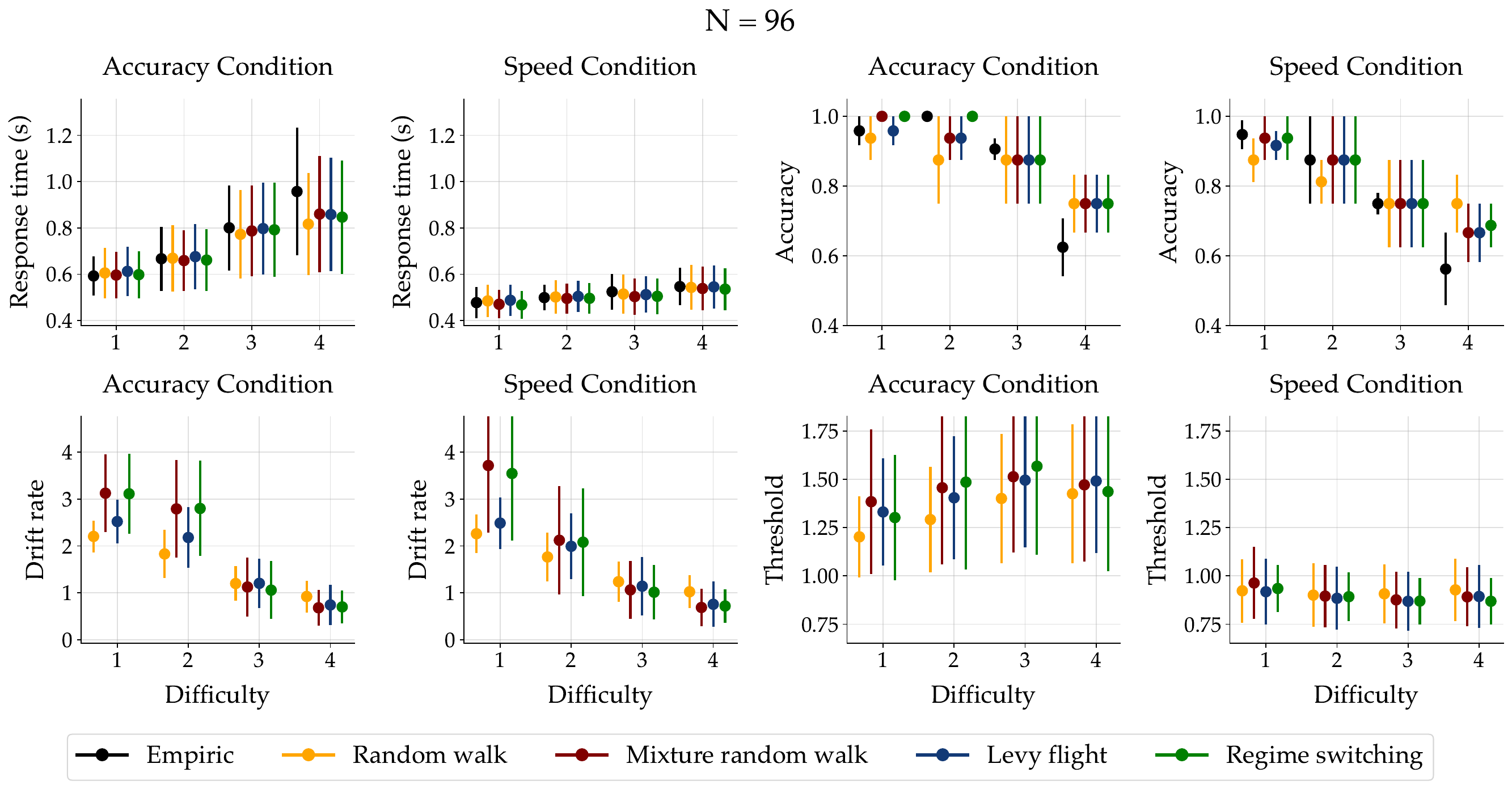}
\caption{Aggregated results from all models fitted to $1/8$ of the total trials of the empirical data. The top row illustrates posterior re-simulations as a measure of the model's generative performance and absolute goodness-of-fit to the data. The bottom row depicts parameter estimates of the drift rate and the threshold parameter from the non-stationary diffusion decision models (NSDDM). \textbf{a} Empirical and re-simulated RTs for each difficulty level and both conditions. \textbf{b} Empirical and re-simulated proportions of correct choices (accuracy) for each difficulty level and both conditions separately. \textbf{c} Posterior estimates of the drift rate parameter for each difficulty level and both conditions separately. \textbf{d} Posterior estimates of the threshold parameter for each difficulty level and both conditions separately. Points indicate medians and the error bars represent the median absolute deviations (MAD) across individuals and re-simulations.}
\end{figure*}

The absolute goodness-of-fit remains relatively strong even with as few as approximately $T = 100$ trials per person (Fig~\ref{fig:main_results_n96}). As expected, the uncertainty in posterior re-simulations and parameter estimates increases as the number of trials decreases. Nevertheless, the patterns remain robust with about $100$ trials.

\begin{figure*}[t]
\centering
\includegraphics[width=\textwidth]{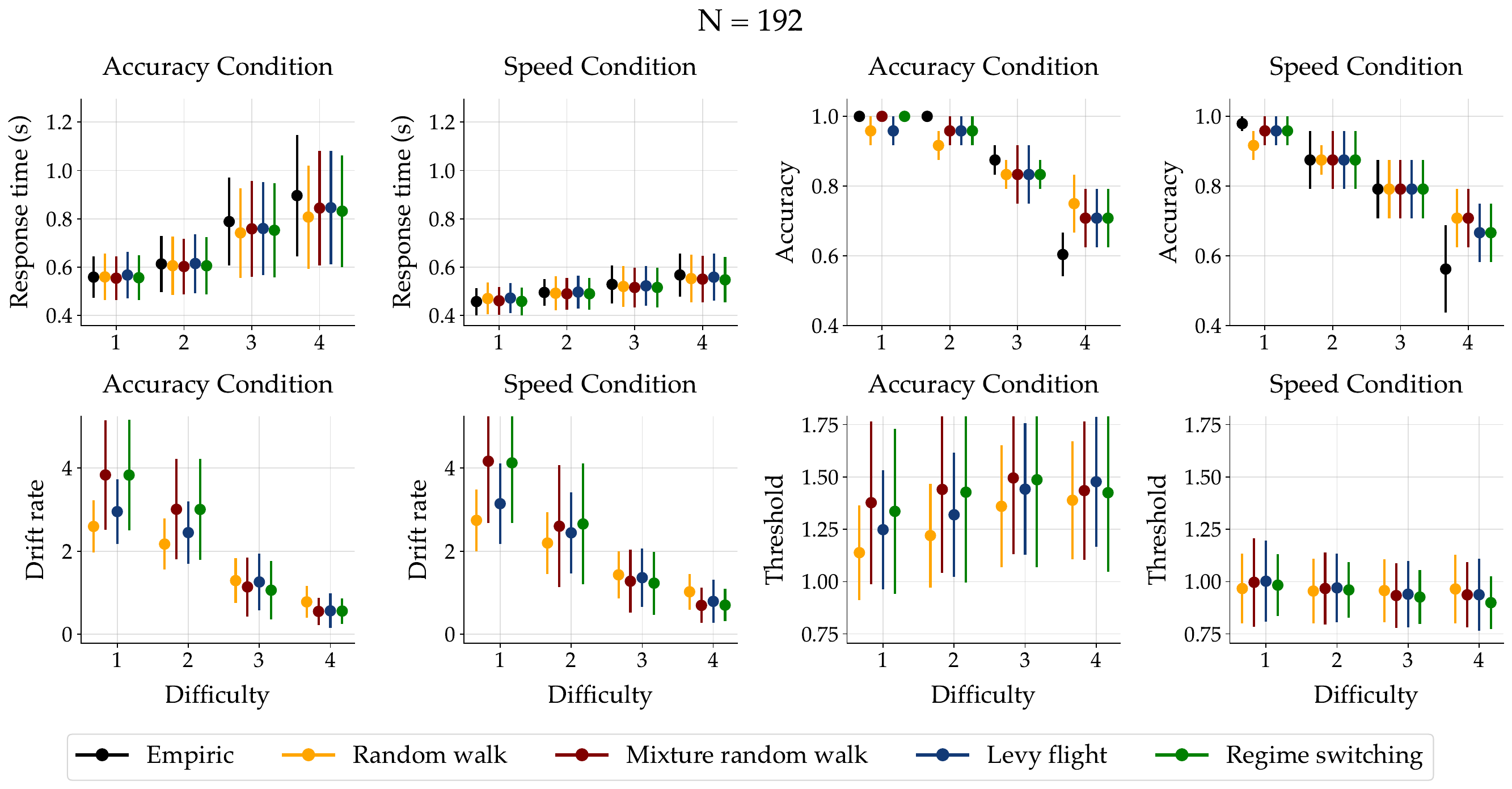}
\caption{Aggregated results from all models fitted to $1/4$ of the total trials of the empirical data. The top row illustrates posterior re-simulations as a measure of the model's generative performance and absolute goodness-of-fit to the data. The bottom row depicts parameter estimates of the drift rate and the threshold parameter from the non-stationary diffusion decision models (NSDDM). \textbf{a} Empirical and re-simulated RTs for each difficulty level and both conditions. \textbf{b} Empirical and re-simulated proportions of correct choices (accuracy) for each difficulty level and both conditions separately. \textbf{c} Posterior estimates of the drift rate parameter for each difficulty level and both conditions separately. \textbf{d} Posterior estimates of the threshold parameter for each difficulty level and both conditions separately. Points indicate medians and the error bars represent the median absolute deviations (MAD) across individuals and re-simulations.}
\end{figure*}

\begin{figure*}[t]
\centering
\includegraphics[width=\textwidth]{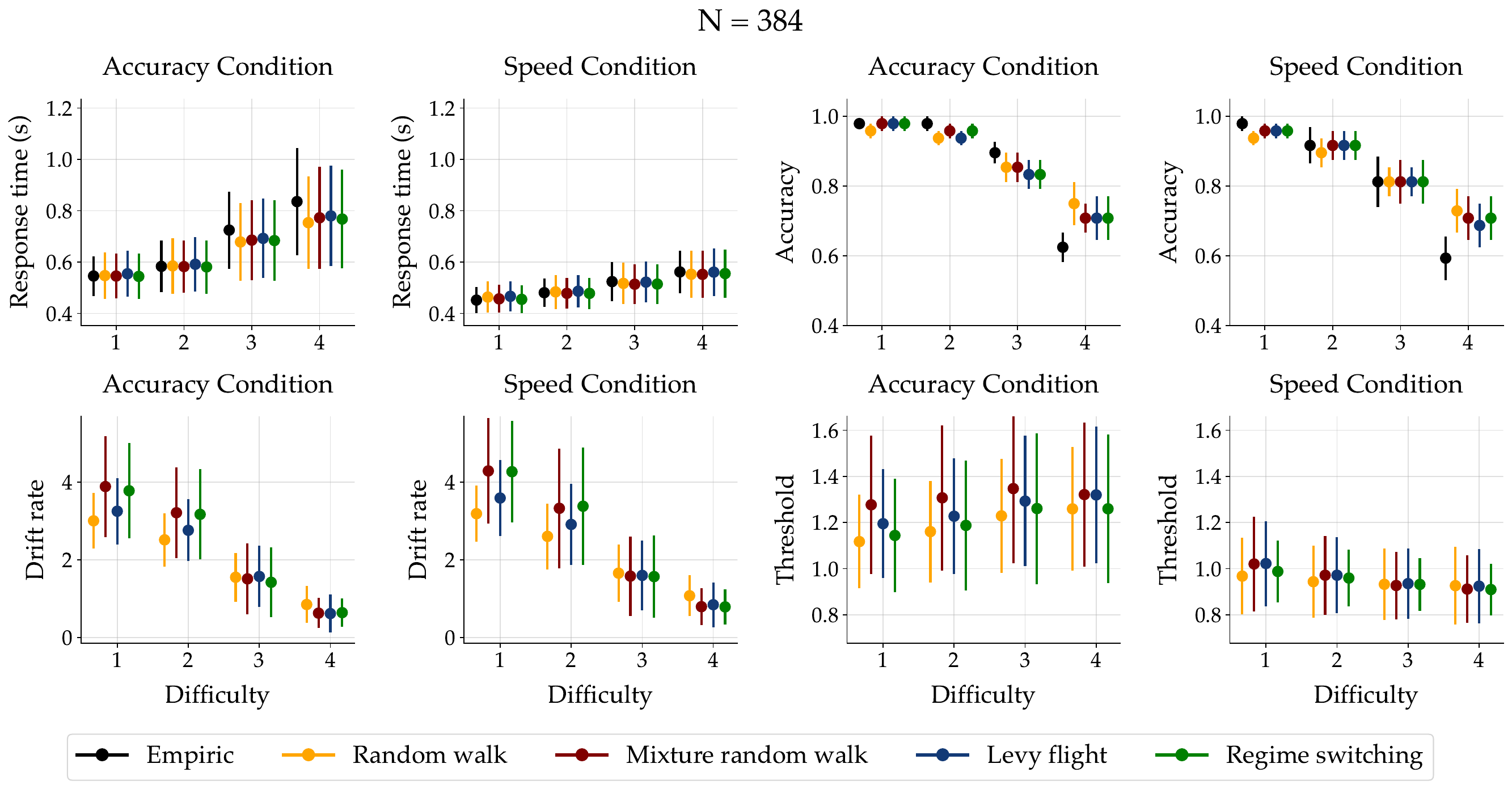}
\caption{Aggregated results from all models fitted to $1/2$ of the total trials of the empirical data. The top row illustrates posterior re-simulations as a measure of the model's generative performance and absolute goodness-of-fit to the data. The bottom row depicts parameter estimates of the drift rate and the threshold parameter from the non-stationary diffusion decision models (NSDDM). \textbf{a} Empirical and re-simulated RTs for each difficulty level and both conditions. \textbf{b} Empirical and re-simulated proportions of correct choices (accuracy) for each difficulty level and both conditions separately. \textbf{c} Posterior estimates of the drift rate parameter for each difficulty level and both conditions separately. \textbf{d} Posterior estimates of the threshold parameter for each difficulty level and both conditions separately. Points indicate medians and the error bars represent the median absolute deviations (MAD) across individuals and re-simulations.}
\end{figure*}

\clearpage
\newpage

\section{Individual Analysis}

The following section shows the individual specific posterior re-simulations and parameter estimates for each difficulty level and both conditions separately. The visualizations are constructed in the vain of Fig~\ref{fig:main_results} in the main text.

\begin{figure*}[h!]
\centering
\includegraphics[width=\textwidth]{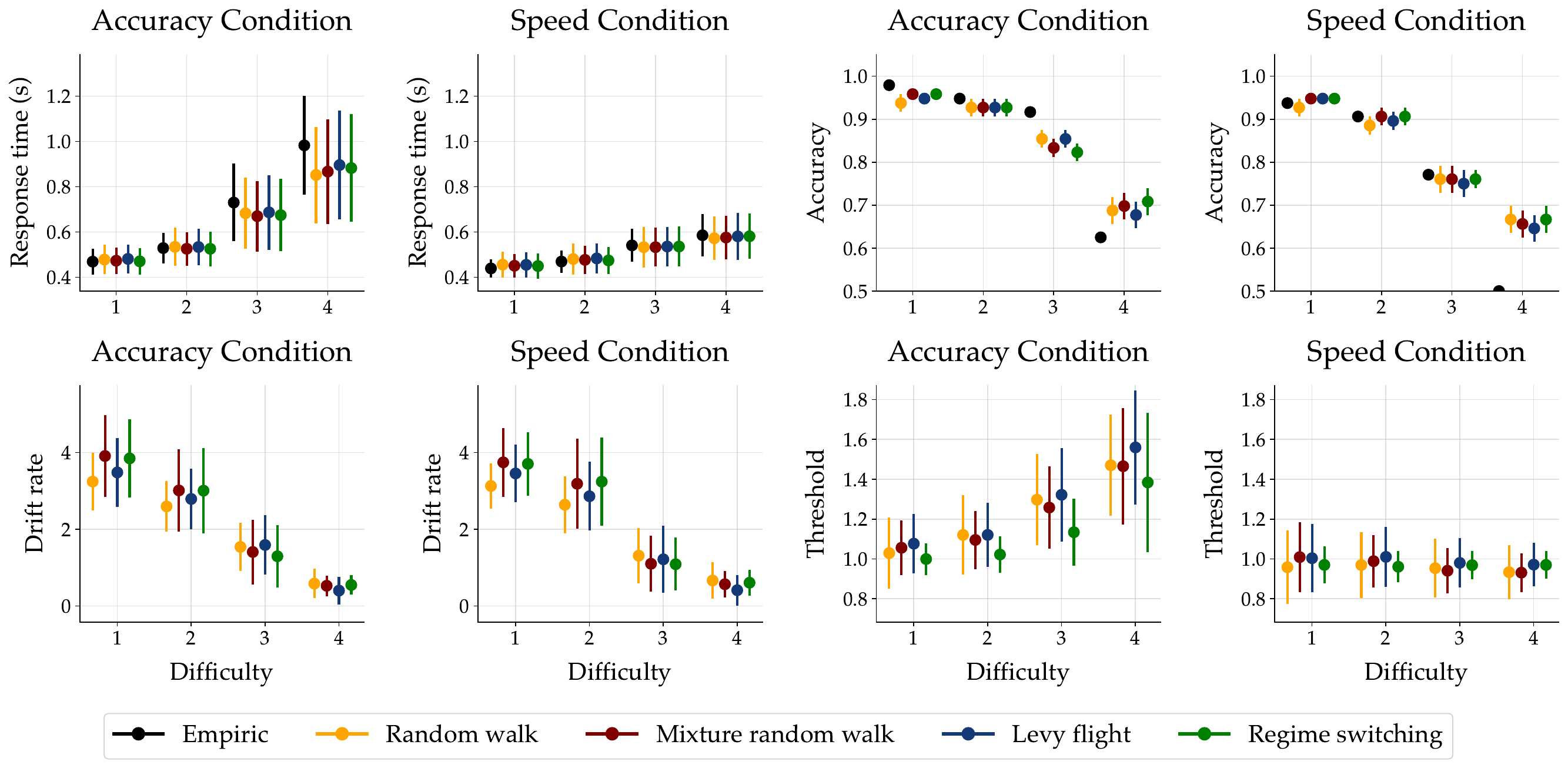}
\caption{Aggregate results from all models fitted to the data from participant $1$. The top row illustrates posterior re-simulations as a measure of the model's generative performance and absolute goodness-of-fit to the data. The bottom row depicts parameter estimates of the drift rate and the threshold parameter from the non-stationary diffusion decision models (NSDDM). \textbf{a} Empirical and re-simulated response times for each difficulty level and both conditions. \textbf{b} Empirical and re-simulated proportions of correct choices (accuracy) for each difficulty level and both conditions separately. \textbf{c} Posterior estimates of the drift rate parameter for each difficulty level and both conditions separately. \textbf{d} Posterior estimates of the threshold parameter for each difficulty level and both conditions separately. Points indicate medians and the error bars represent the median absolute deviations (MAD) across individual data and re-simulations.}
\end{figure*}

\FloatBarrier

\begin{figure*}[h!]
\centering
\includegraphics[width=\textwidth]{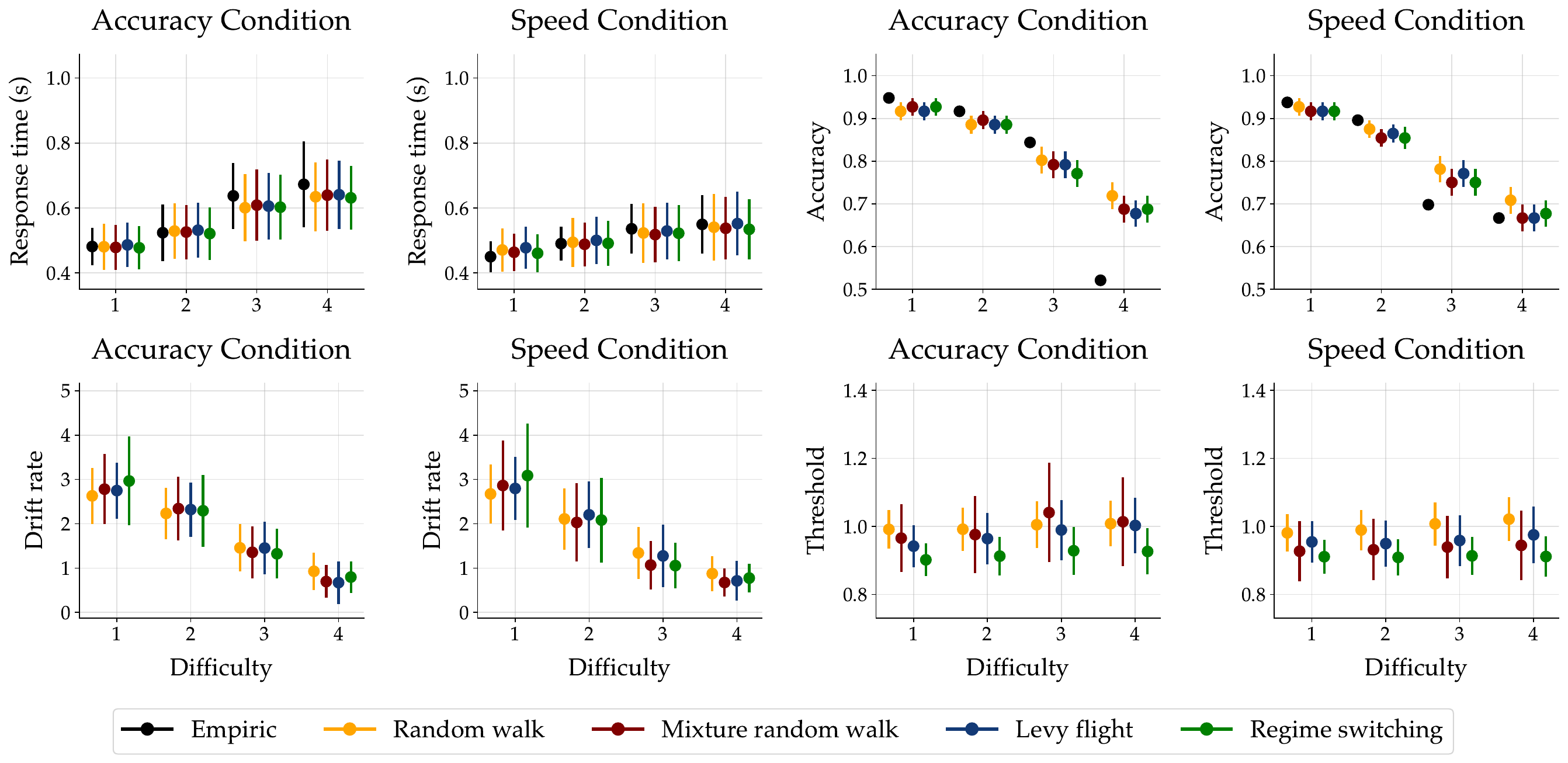}
\caption{Aggregate results from all models fitted to the data from participant $2$. The top row illustrates posterior re-simulations as a measure of the model's generative performance and absolute goodness-of-fit to the data. The bottom row depicts parameter estimates of the drift rate and the threshold parameter from the non-stationary diffusion decision models (NSDDM). \textbf{a} Empirical and re-simulated response times for each difficulty level and both conditions. \textbf{b} Empirical and re-simulated proportions of correct choices (accuracy) for each difficulty level and both conditions separately. \textbf{c} Posterior estimates of the drift rate parameter for each difficulty level and both conditions separately. \textbf{d} Posterior estimates of the threshold parameter for each difficulty level and both conditions separately. Points indicate medians and the error bars represent the median absolute deviations (MAD) across individual data and re-simulations.}
\end{figure*}

\FloatBarrier

\begin{figure*}[h!]
\centering
\includegraphics[width=\textwidth]{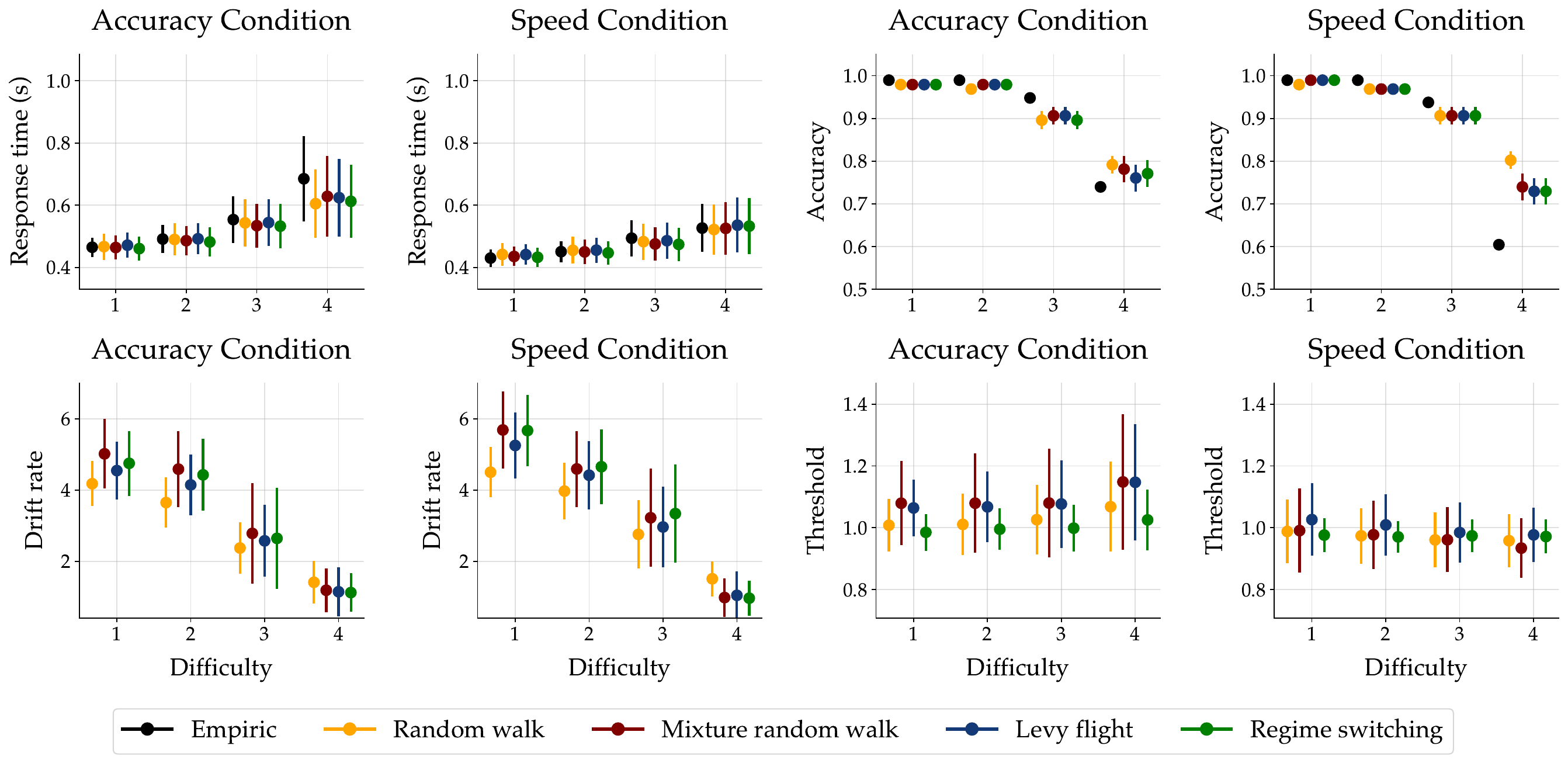}
\caption{Aggregate results from all models fitted to the data from participant $3$. The top row illustrates posterior re-simulations as a measure of the model's generative performance and absolute goodness-of-fit to the data. The bottom row depicts parameter estimates of the drift rate and the threshold parameter from the non-stationary diffusion decision models (NSDDM). \textbf{a} Empirical and re-simulated response times for each difficulty level and both conditions. \textbf{b} Empirical and re-simulated proportions of correct choices (accuracy) for each difficulty level and both conditions separately. \textbf{c} Posterior estimates of the drift rate parameter for each difficulty level and both conditions separately. \textbf{d} Posterior estimates of the threshold parameter for each difficulty level and both conditions separately. Points indicate medians and the error bars represent the median absolute deviations (MAD) across individual data and re-simulations.}
\end{figure*}

\FloatBarrier

\begin{figure*}[h!]
\centering
\includegraphics[width=\textwidth]{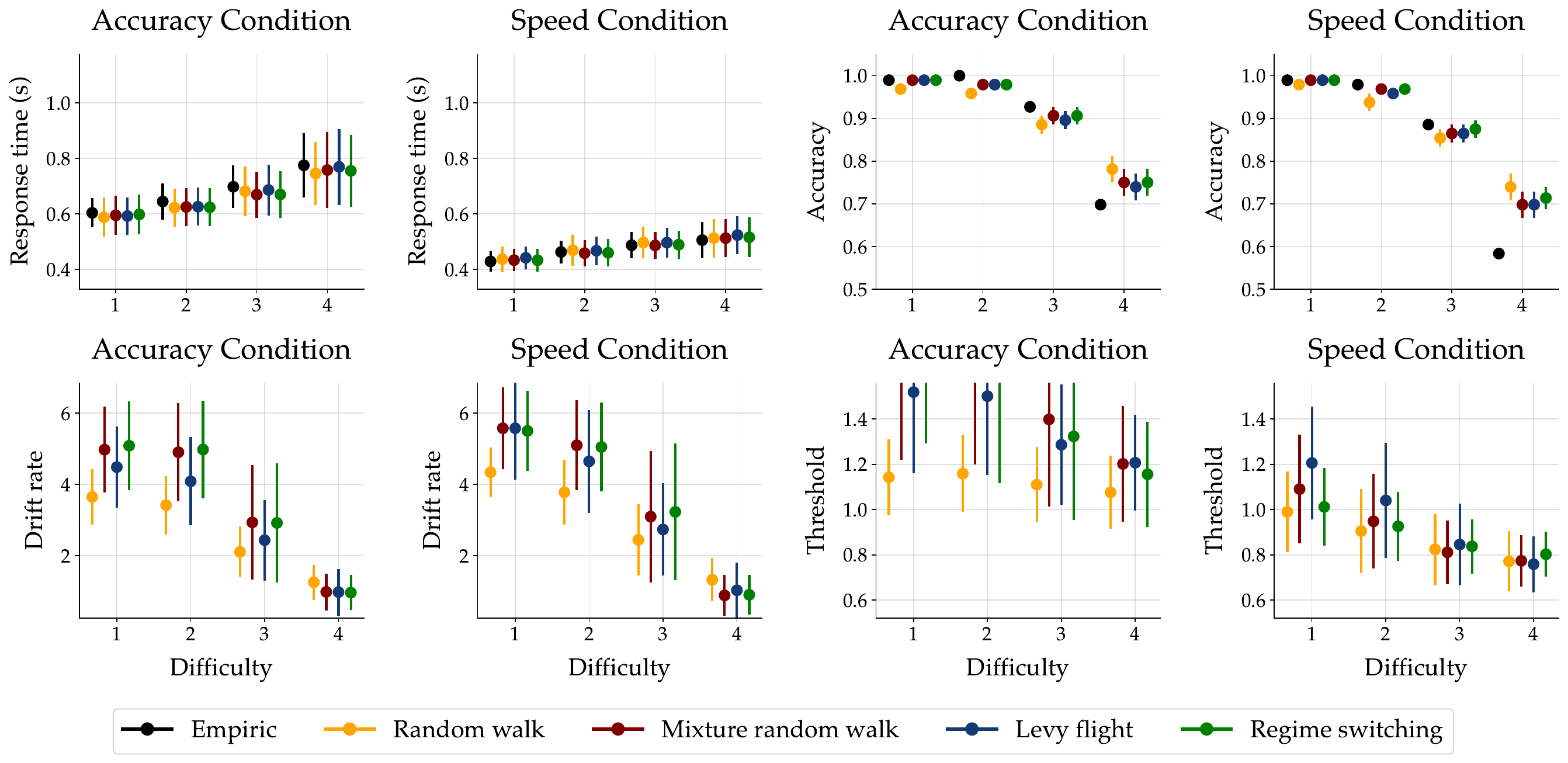}
\caption{Aggregate results from all models fitted to the data from participant $4$. The top row illustrates posterior re-simulations as a measure of the model's generative performance and absolute goodness-of-fit to the data. The bottom row depicts parameter estimates of the drift rate and the threshold parameter from the non-stationary diffusion decision models (NSDDM). \textbf{a} Empirical and re-simulated response times for each difficulty level and both conditions. \textbf{b} Empirical and re-simulated proportions of correct choices (accuracy) for each difficulty level and both conditions separately. \textbf{c} Posterior estimates of the drift rate parameter for each difficulty level and both conditions separately. \textbf{d} Posterior estimates of the threshold parameter for each difficulty level and both conditions separately. Points indicate medians and the error bars represent the median absolute deviations (MAD) across individual data and re-simulations.}
\end{figure*}

\FloatBarrier

\begin{figure*}[h!]
\centering
\includegraphics[width=\textwidth]{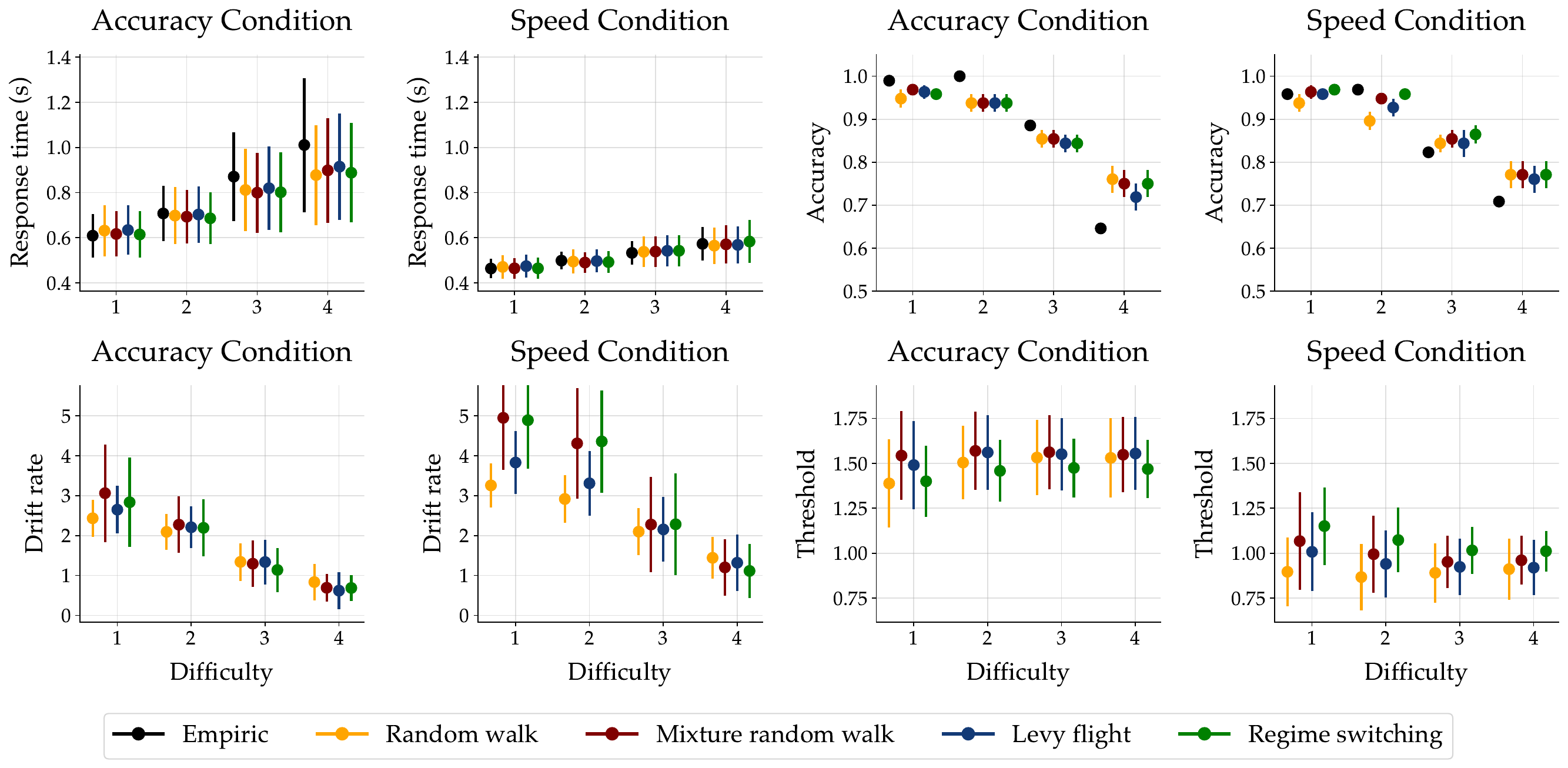}
\caption{Aggregate results from all models fitted to the data from participant $5$. The top row illustrates posterior re-simulations as a measure of the model's generative performance and absolute goodness-of-fit to the data. The bottom row depicts parameter estimates of the drift rate and the threshold parameter from the non-stationary diffusion decision models (NSDDM). \textbf{a} Empirical and re-simulated response times for each difficulty level and both conditions. \textbf{b} Empirical and re-simulated proportions of correct choices (accuracy) for each difficulty level and both conditions separately. \textbf{c} Posterior estimates of the drift rate parameter for each difficulty level and both conditions separately. \textbf{d} Posterior estimates of the threshold parameter for each difficulty level and both conditions separately. Points indicate medians and the error bars represent the median absolute deviations (MAD) across individual data and re-simulations.}
\end{figure*}

\FloatBarrier

\begin{figure*}[h!]
\centering
\includegraphics[width=\textwidth]{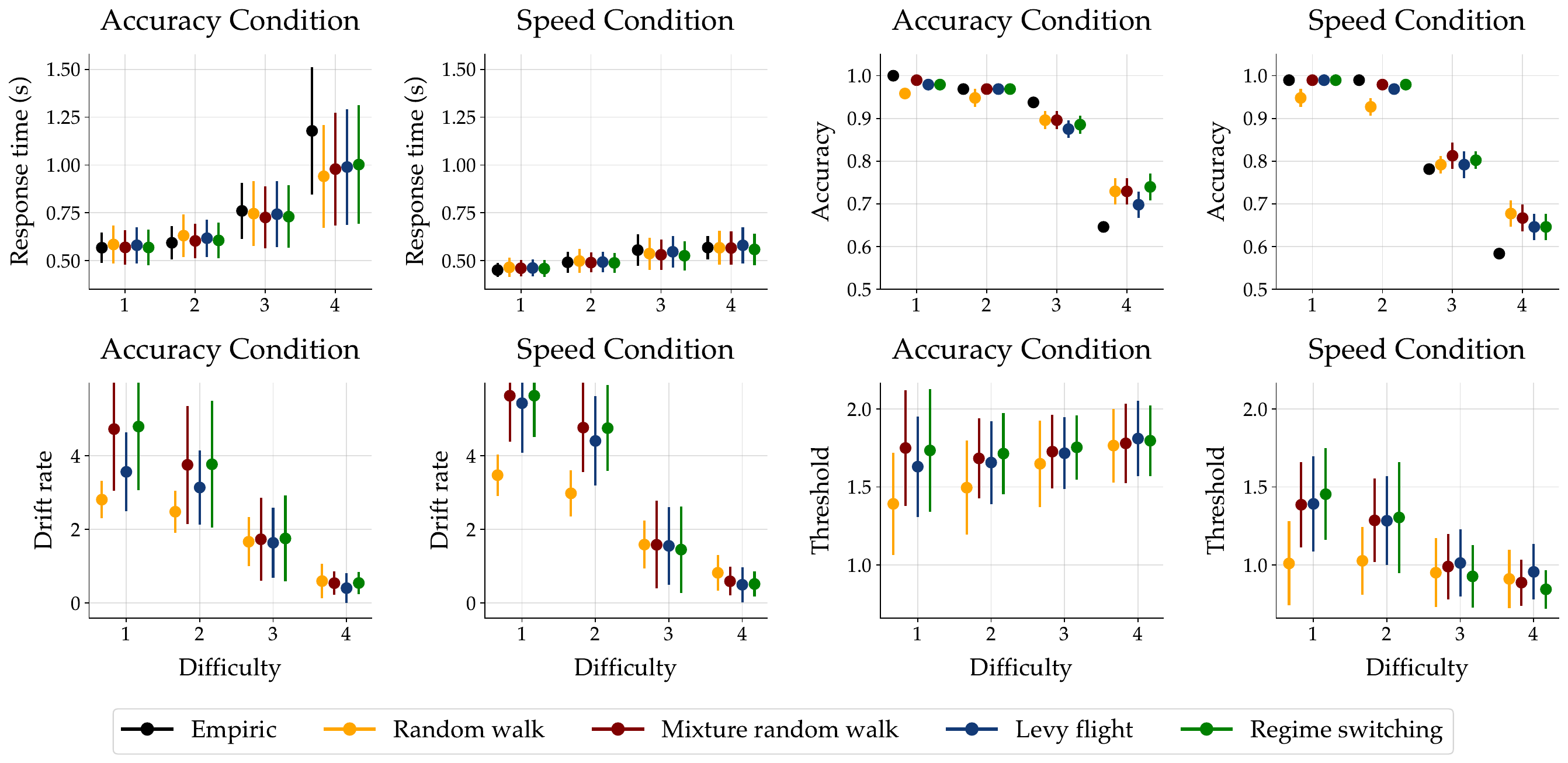}
\caption{Aggregate results from all models fitted to the data from participant $6$. The top row illustrates posterior re-simulations as a measure of the model's generative performance and absolute goodness-of-fit to the data. The bottom row depicts parameter estimates of the drift rate and the threshold parameter from the non-stationary diffusion decision models (NSDDM). \textbf{a} Empirical and re-simulated response times for each difficulty level and both conditions. \textbf{b} Empirical and re-simulated proportions of correct choices (accuracy) for each difficulty level and both conditions separately. \textbf{c} Posterior estimates of the drift rate parameter for each difficulty level and both conditions separately. \textbf{d} Posterior estimates of the threshold parameter for each difficulty level and both conditions separately. Points indicate medians and the error bars represent the median absolute deviations (MAD) across individual data and re-simulations.}
\end{figure*}

\FloatBarrier

\begin{figure*}[h!]
\centering
\includegraphics[width=\textwidth]{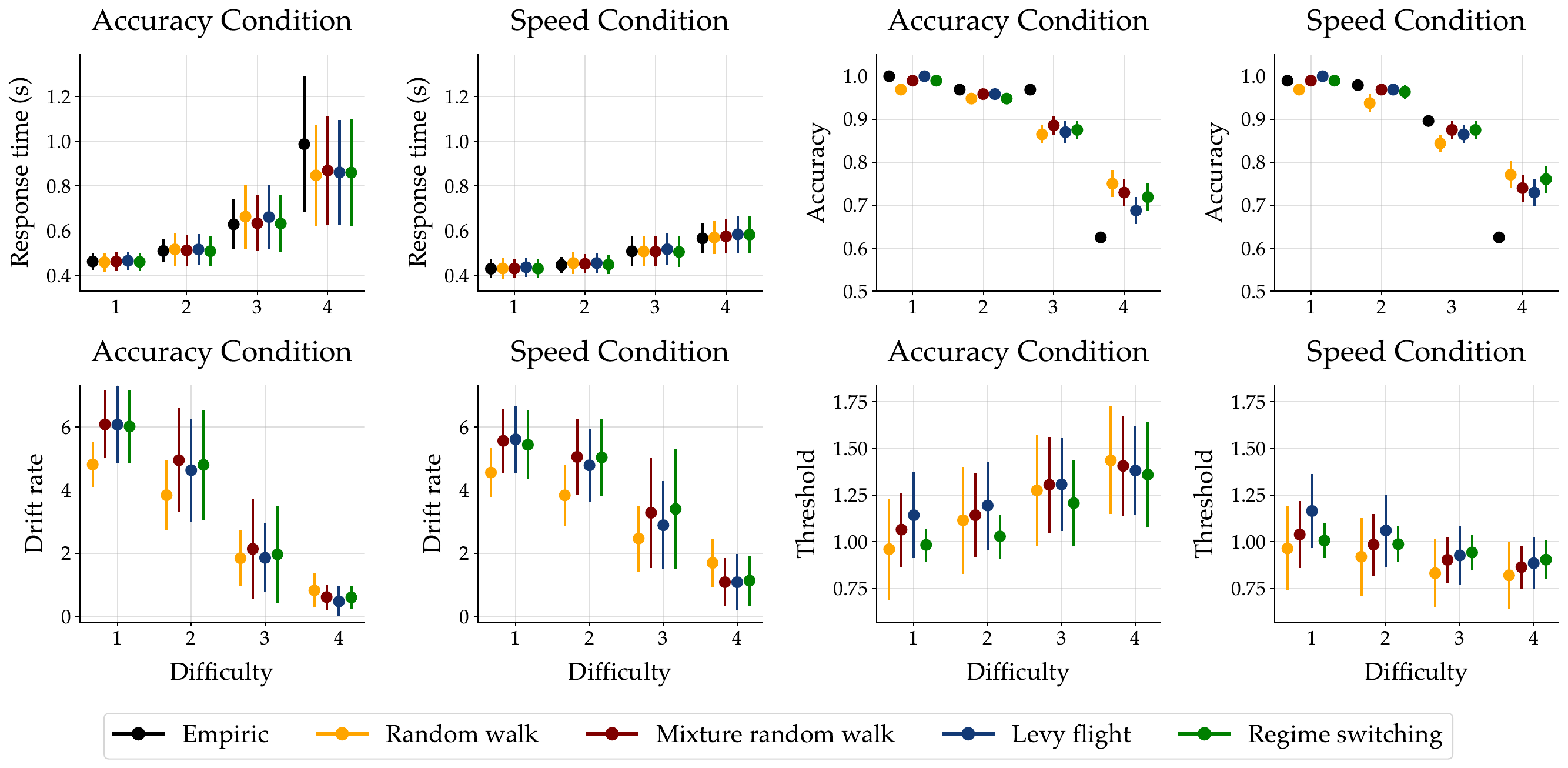}
\caption{Aggregate results from all models fitted to the data from participant $7$. The top row illustrates posterior re-simulations as a measure of the model's generative performance and absolute goodness-of-fit to the data. The bottom row depicts parameter estimates of the drift rate and the threshold parameter from the non-stationary diffusion decision models (NSDDM). \textbf{a} Empirical and re-simulated response times for each difficulty level and both conditions. \textbf{b} Empirical and re-simulated proportions of correct choices (accuracy) for each difficulty level and both conditions separately. \textbf{c} Posterior estimates of the drift rate parameter for each difficulty level and both conditions separately. \textbf{d} Posterior estimates of the threshold parameter for each difficulty level and both conditions separately. Points indicate medians and the error bars represent the median absolute deviations (MAD) across individual data and re-simulations.}
\end{figure*}

\FloatBarrier

\begin{figure*}[h!]
\centering
\includegraphics[width=\textwidth]{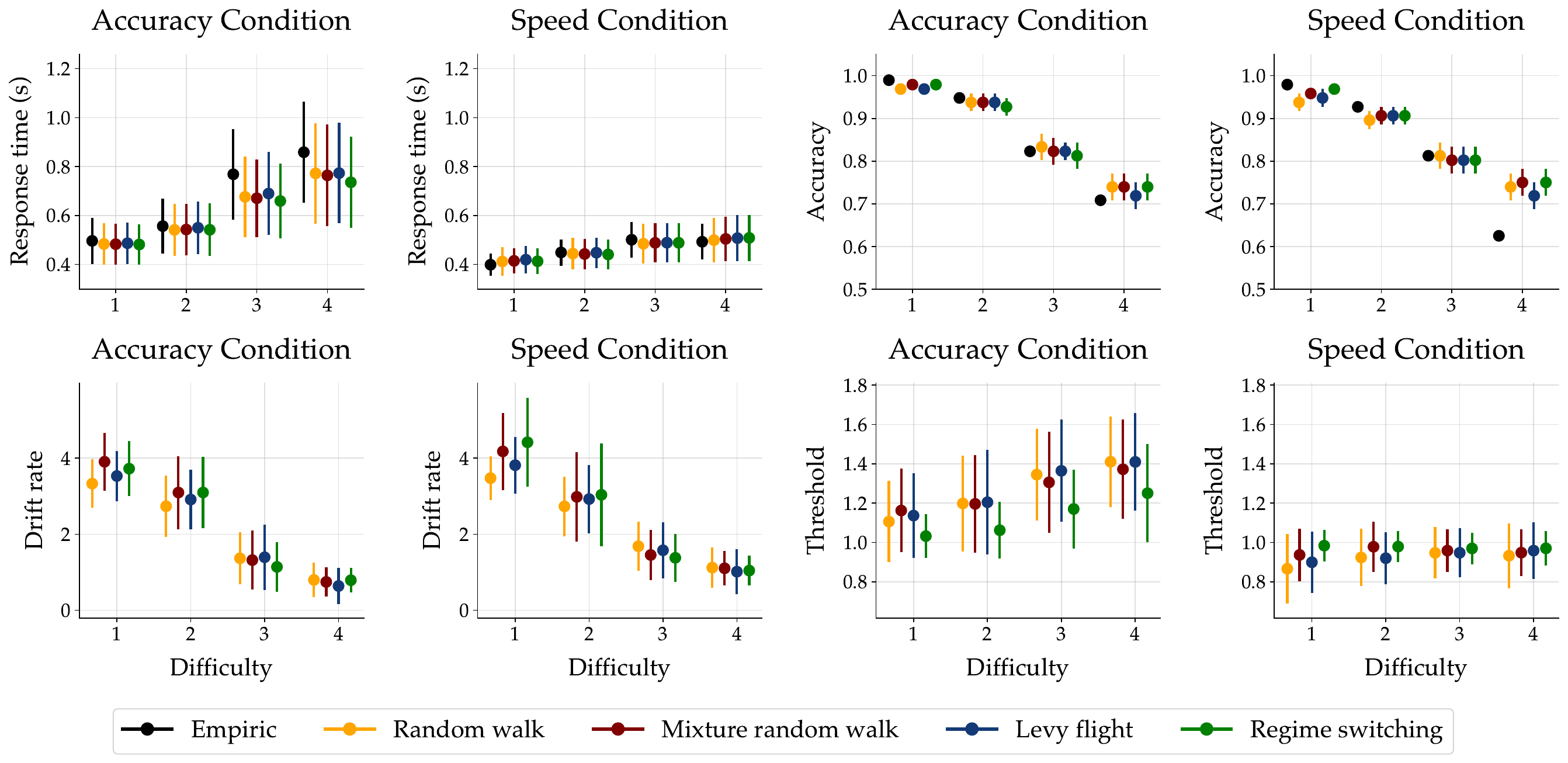}
\caption{Aggregate results from all models fitted to the data from participant $8$. The top row illustrates posterior re-simulations as a measure of the model's generative performance and absolute goodness-of-fit to the data. The bottom row depicts parameter estimates of the drift rate and the threshold parameter from the non-stationary diffusion decision models (NSDDM). \textbf{a} Empirical and re-simulated response times for each difficulty level and both conditions. \textbf{b} Empirical and re-simulated proportions of correct choices (accuracy) for each difficulty level and both conditions separately. \textbf{c} Posterior estimates of the drift rate parameter for each difficulty level and both conditions separately. \textbf{d} Posterior estimates of the threshold parameter for each difficulty level and both conditions separately. Points indicate medians and the error bars represent the median absolute deviations (MAD) across individual data and re-simulations.}
\end{figure*}

\FloatBarrier

\begin{figure*}[h!]
\centering
\includegraphics[width=\textwidth]{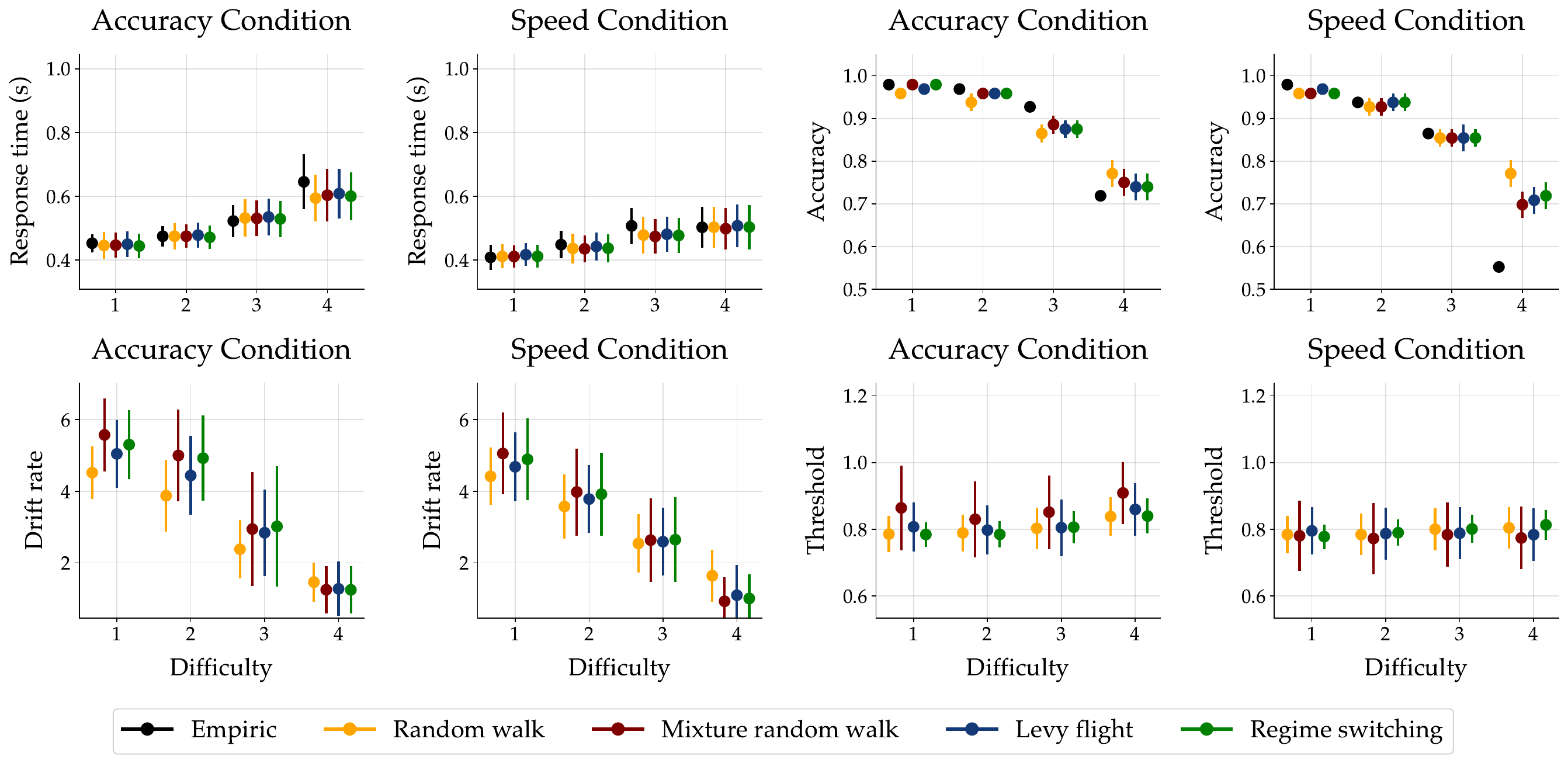}
\caption{Aggregate results from all models fitted to the data from participant $9$. The top row illustrates posterior re-simulations as a measure of the model's generative performance and absolute goodness-of-fit to the data. The bottom row depicts parameter estimates of the drift rate and the threshold parameter from the non-stationary diffusion decision models (NSDDM). \textbf{a} Empirical and re-simulated response times for each difficulty level and both conditions. \textbf{b} Empirical and re-simulated proportions of correct choices (accuracy) for each difficulty level and both conditions separately. \textbf{c} Posterior estimates of the drift rate parameter for each difficulty level and both conditions separately. \textbf{d} Posterior estimates of the threshold parameter for each difficulty level and both conditions separately. Points indicate medians and the error bars represent the median absolute deviations (MAD) across individual data and re-simulations.}
\end{figure*}

\FloatBarrier

\begin{figure*}[h!]
\centering
\includegraphics[width=\textwidth]{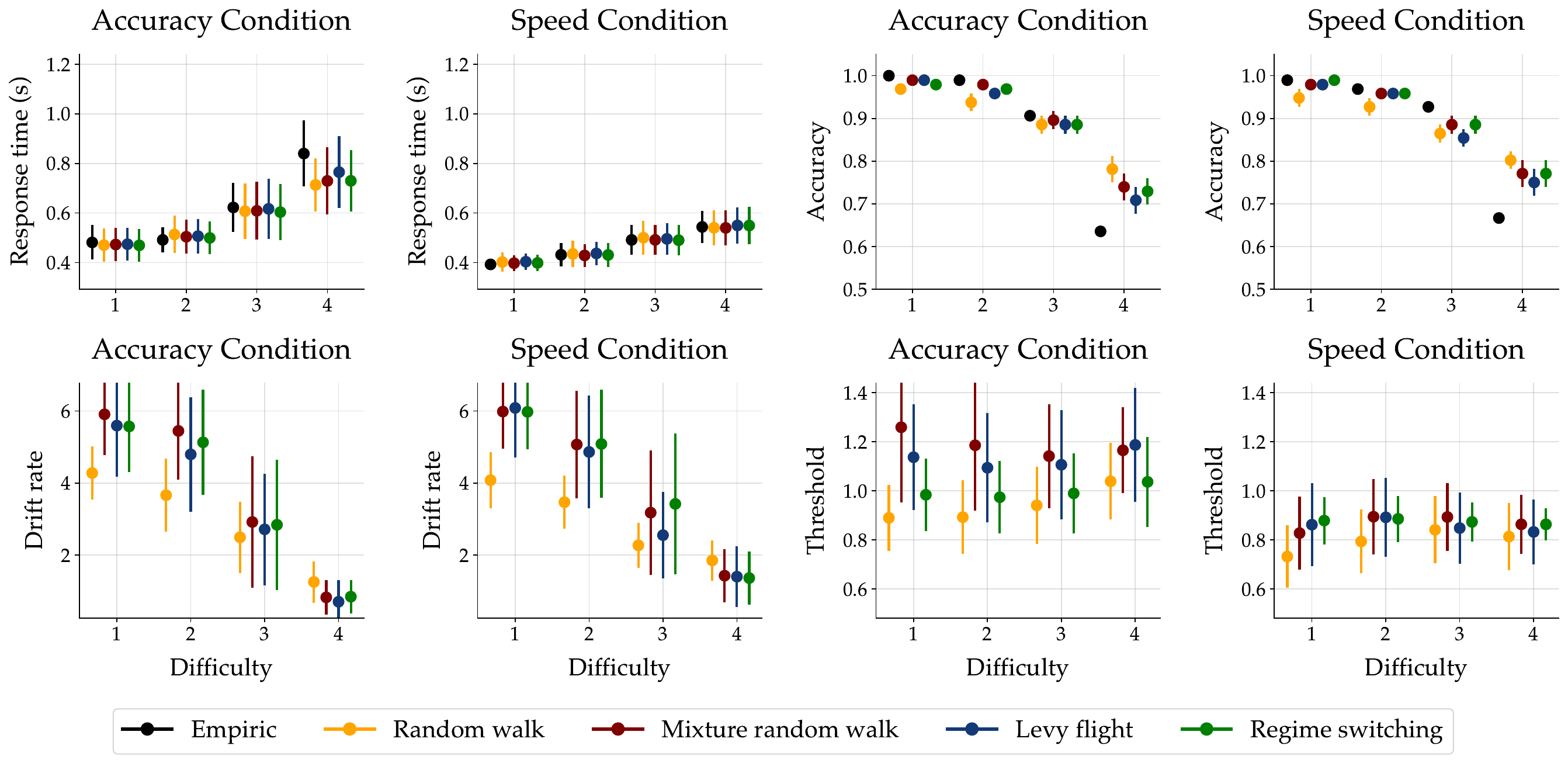}
\caption{Aggregate results from all models fitted to the data from participant $10$. The top row illustrates posterior re-simulations as a measure of the model's generative performance and absolute goodness-of-fit to the data. The bottom row depicts parameter estimates of the drift rate and the threshold parameter from the non-stationary diffusion decision models (NSDDM). \textbf{a} Empirical and re-simulated response times for each difficulty level and both conditions. \textbf{b} Empirical and re-simulated proportions of correct choices (accuracy) for each difficulty level and both conditions separately. \textbf{c} Posterior estimates of the drift rate parameter for each difficulty level and both conditions separately. \textbf{d} Posterior estimates of the threshold parameter for each difficulty level and both conditions separately. Points indicate medians and the error bars represent the median absolute deviations (MAD) across individual data and re-simulations.}
\end{figure*}

\FloatBarrier

\begin{figure*}[h!]
\centering
\includegraphics[width=\textwidth]{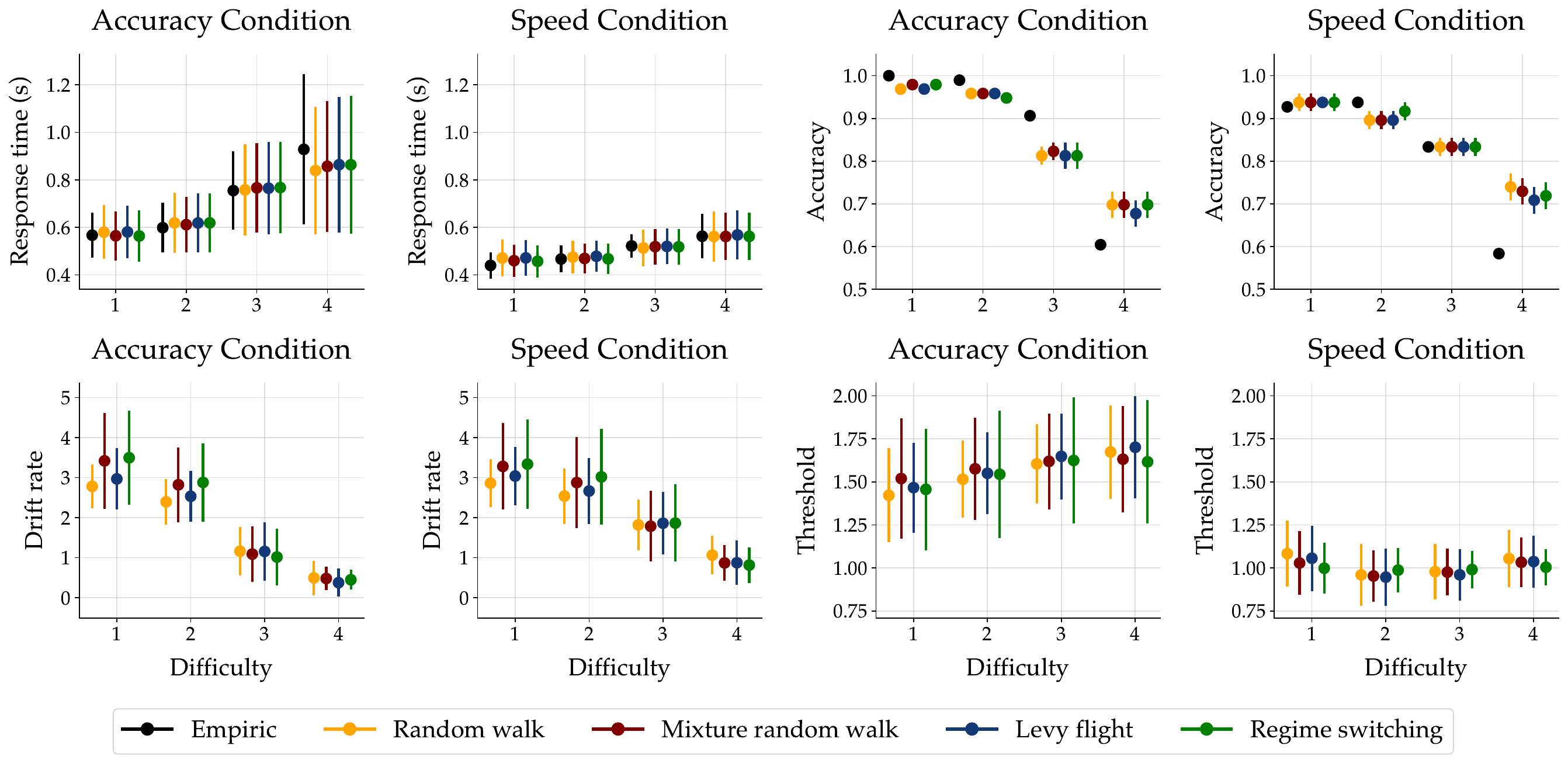}
\caption{Aggregate results from all models fitted to the data from participant $11$. The top row illustrates posterior re-simulations as a measure of the model's generative performance and absolute goodness-of-fit to the data. The bottom row depicts parameter estimates of the drift rate and the threshold parameter from the non-stationary diffusion decision models (NSDDM). \textbf{a} Empirical and re-simulated response times for each difficulty level and both conditions. \textbf{b} Empirical and re-simulated proportions of correct choices (accuracy) for each difficulty level and both conditions separately. \textbf{c} Posterior estimates of the drift rate parameter for each difficulty level and both conditions separately. \textbf{d} Posterior estimates of the threshold parameter for each difficulty level and both conditions separately. Points indicate medians and the error bars represent the median absolute deviations (MAD) across individual data and re-simulations.}
\end{figure*}

\FloatBarrier

\begin{figure*}[h!]
\centering
\includegraphics[width=\textwidth]{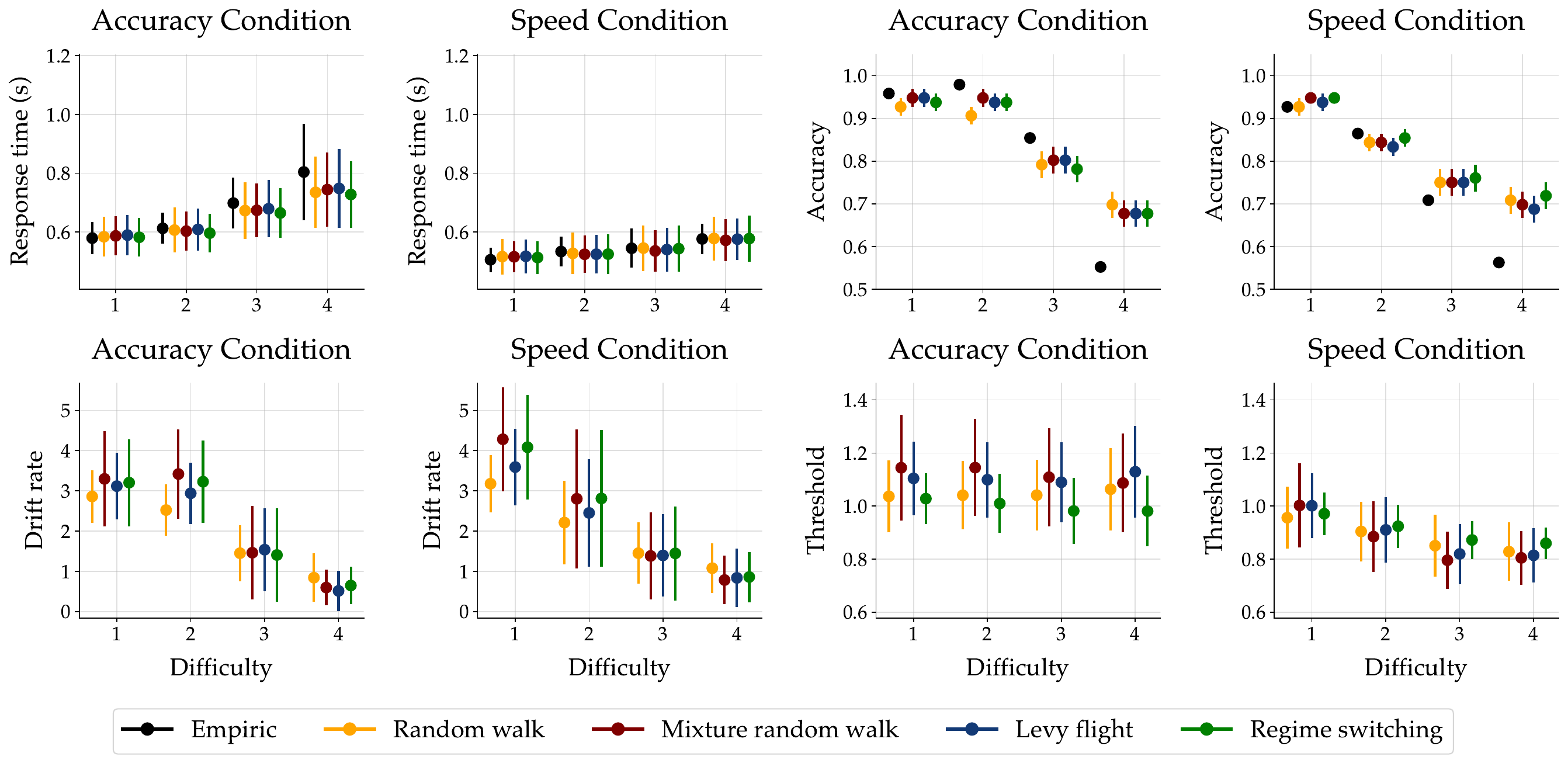}
\caption{Aggregate results from all models fitted to the data from participant $12$. The top row illustrates posterior re-simulations as a measure of the model's generative performance and absolute goodness-of-fit to the data. The bottom row depicts parameter estimates of the drift rate and the threshold parameter from the non-stationary diffusion decision models (NSDDM). \textbf{a} Empirical and re-simulated response times for each difficulty level and both conditions. \textbf{b} Empirical and re-simulated proportions of correct choices (accuracy) for each difficulty level and both conditions separately. \textbf{c} Posterior estimates of the drift rate parameter for each difficulty level and both conditions separately. \textbf{d} Posterior estimates of the threshold parameter for each difficulty level and both conditions separately. Points indicate medians and the error bars represent the median absolute deviations (MAD) across individual data and re-simulations.}
\end{figure*}

\FloatBarrier

\begin{figure*}[h!]
\centering
\includegraphics[width=\textwidth]{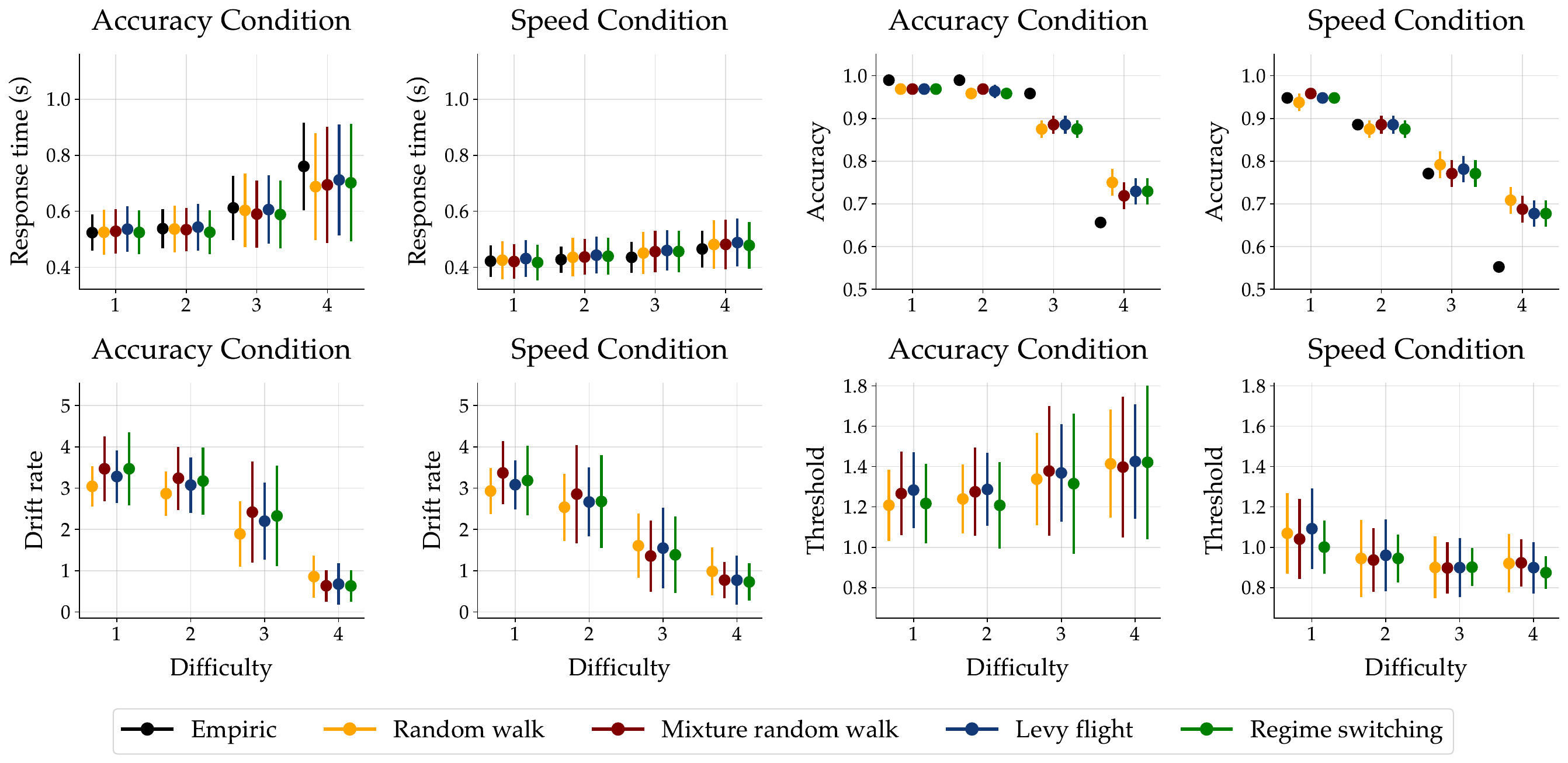}
\caption{Aggregate results from all models fitted to the data from participant $13$. The top row illustrates posterior re-simulations as a measure of the model's generative performance and absolute goodness-of-fit to the data. The bottom row depicts parameter estimates of the drift rate and the threshold parameter from the non-stationary diffusion decision models (NSDDM). \textbf{a} Empirical and re-simulated response times for each difficulty level and both conditions. \textbf{b} Empirical and re-simulated proportions of correct choices (accuracy) for each difficulty level and both conditions separately. \textbf{c} Posterior estimates of the drift rate parameter for each difficulty level and both conditions separately. \textbf{d} Posterior estimates of the threshold parameter for each difficulty level and both conditions separately. Points indicate medians and the error bars represent the median absolute deviations (MAD) across individual data and re-simulations.}
\end{figure*}

\FloatBarrier

\begin{figure*}[h!]
\centering
\includegraphics[width=\textwidth]{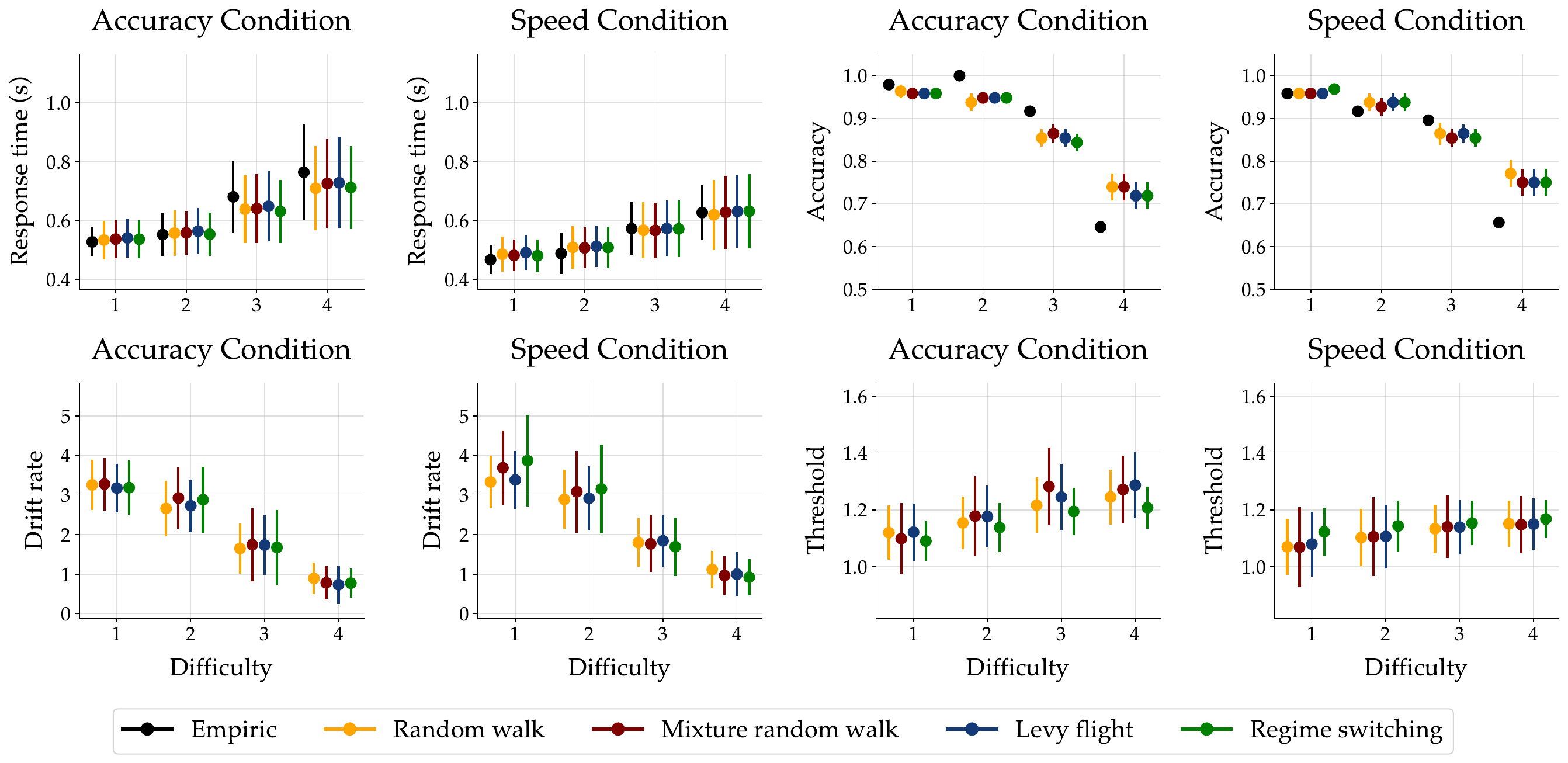}
\caption{Aggregate results from all models fitted to the data from participant $14$. The top row illustrates posterior re-simulations as a measure of the model's generative performance and absolute goodness-of-fit to the data. The bottom row depicts parameter estimates of the drift rate and the threshold parameter from the non-stationary diffusion decision models (NSDDM). \textbf{a} Empirical and re-simulated response times for each difficulty level and both conditions. \textbf{b} Empirical and re-simulated proportions of correct choices (accuracy) for each difficulty level and both conditions separately. \textbf{c} Posterior estimates of the drift rate parameter for each difficulty level and both conditions separately. \textbf{d} Posterior estimates of the threshold parameter for each difficulty level and both conditions separately. Points indicate medians and the error bars represent the median absolute deviations (MAD) across individual data and re-simulations.}
\end{figure*}

\FloatBarrier
\clearpage
\newpage

\section{Response Time Time Series}
In the following, we present the model fit to the whole response time time series for the remaining $12$ participants.

\begin{figure*}[h!]
\centering
\includegraphics[width=\textwidth]{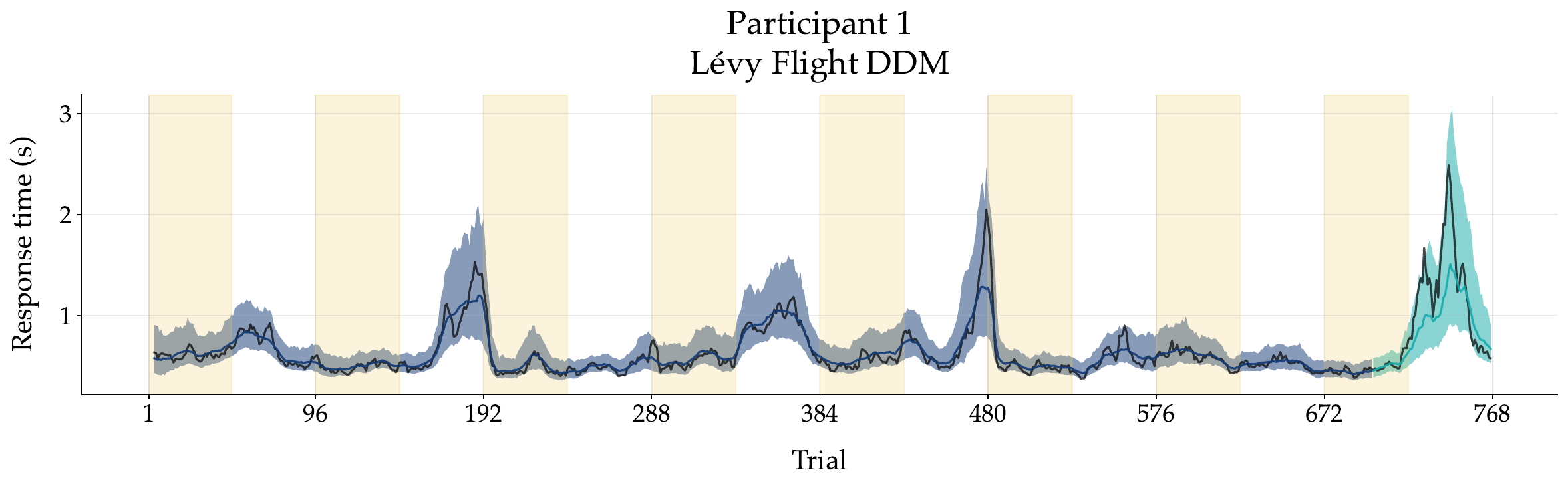}
\caption{Model fit to response time (RT) time series. The empirical RT time series of participant $1$ is shown in black. From trial $1$ to $700$, the posterior re-simulation (aka retrodictive check) using the best fitting non-stationary diffusion decision model (NSDDM) for this specific individual are shown in blue. In this instance, the results stem from a Lévy flight DDM. For the remaining trials, one-step-ahead predictions are depicted in cyan. Solid lines correspond to the median and shaded bands to $90\%$ credibility intervals (CI). The empirical, re-simulated, and predicted RT time series were smoothed via a simple moving average (SMA) with a period of $5$. The yellow shaded regions indicate trials where speed was emphasised over accuracy, while blank white areas denote instances where the opposite emphasis was applied.}
\end{figure*}

\begin{figure*}[h!]
\centering
\includegraphics[width=\textwidth]{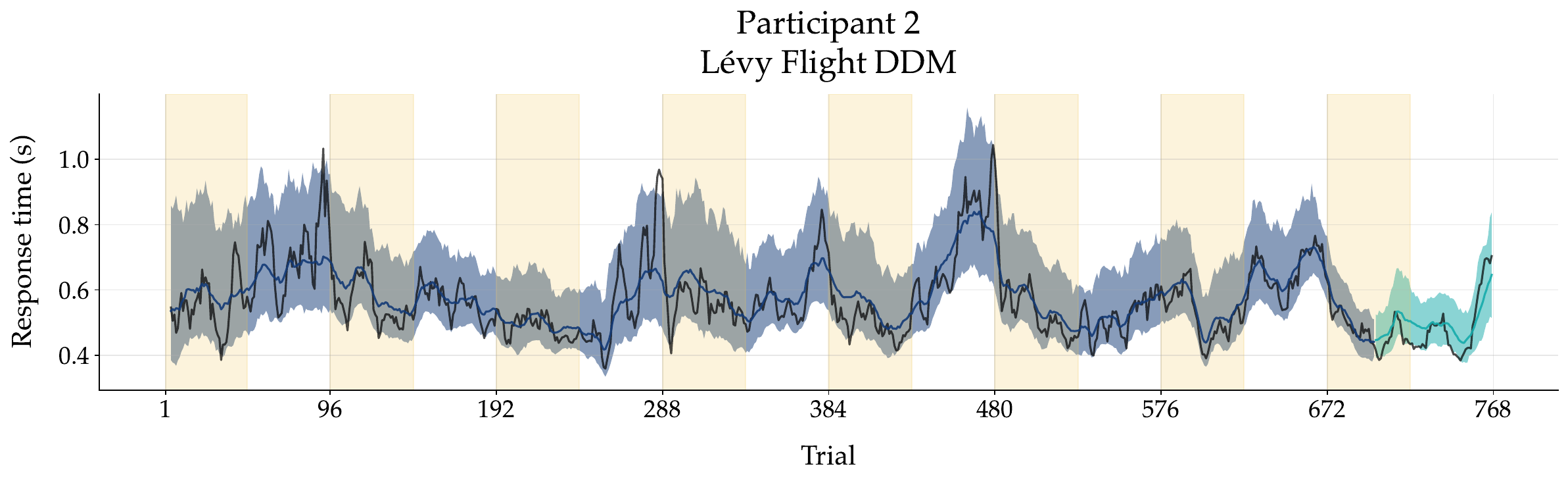}
\caption{Model fit to response time (RT) time series. The empirical RT time series of participant $2$ is shown in black. From trial $1$ to $700$, the posterior re-simulation (aka retrodictive check) using the best fitting non-stationary diffusion decision model (NSDDM) for this specific individual are shown in blue. In this instance, the results stem from a Lévy flight DDM. For the remaining trials, one-step-ahead predictions are depicted in cyan. Solid lines correspond to the median and shaded bands to $90\%$ credibility intervals (CI). The empirical, re-simulated, and predicted RT time series were smoothed via a simple moving average (SMA) with a period of $5$. The yellow shaded regions indicate trials where speed was emphasised over accuracy, while blank white areas denote instances where the opposite emphasis was applied.}
\end{figure*}

\begin{figure*}[h!]
\centering
\includegraphics[width=\textwidth]{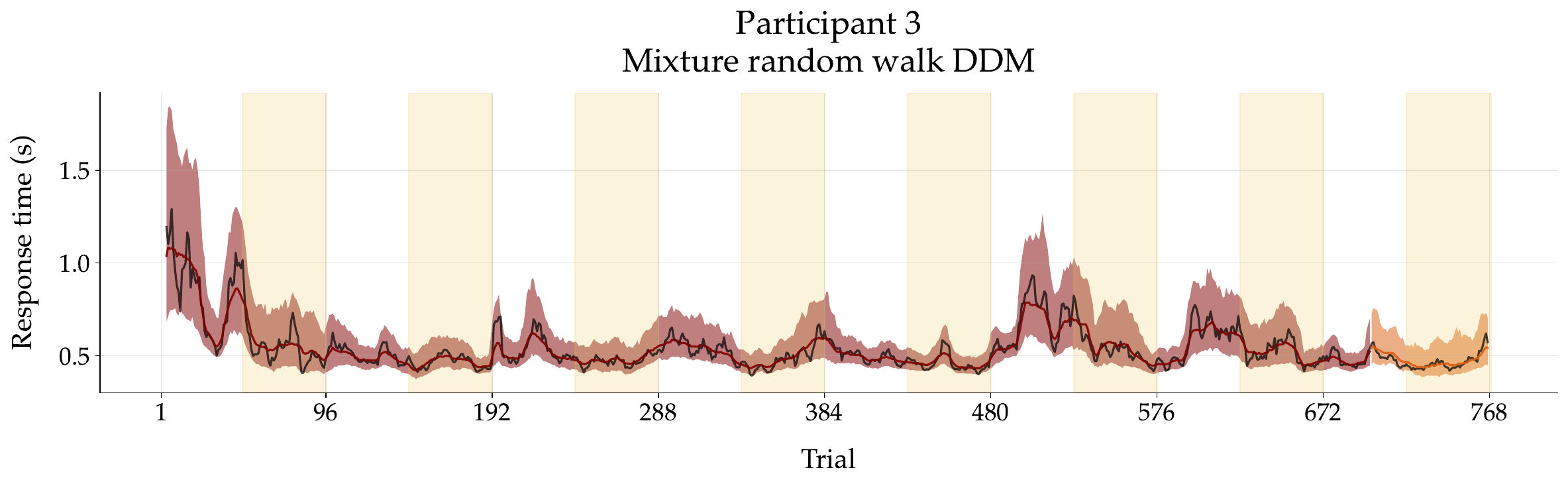}
\caption{Model fit to response time (RT) time series. The empirical RT time series of participant $3$ is shown in black. From trial $1$ to $700$, the posterior re-simulation (aka retrodictive check) using the best fitting non-stationary diffusion decision model (NSDDM) for this specific individual are shown in red. In this instance, the results stem from a mixture random walk DDM. For the remaining trials, one-step-ahead predictions are depicted in orange. Solid lines correspond to the median and shaded bands to $90\%$ credibility intervals (CI). The empirical, re-simulated, and predicted RT time series were smoothed via a simple moving average (SMA) with a period of $5$. The yellow shaded regions indicate trials where speed was emphasised over accuracy, while blank white areas denote instances where the opposite emphasis was applied.}
\end{figure*}

\begin{figure*}[h!]
\centering
\includegraphics[width=\textwidth]{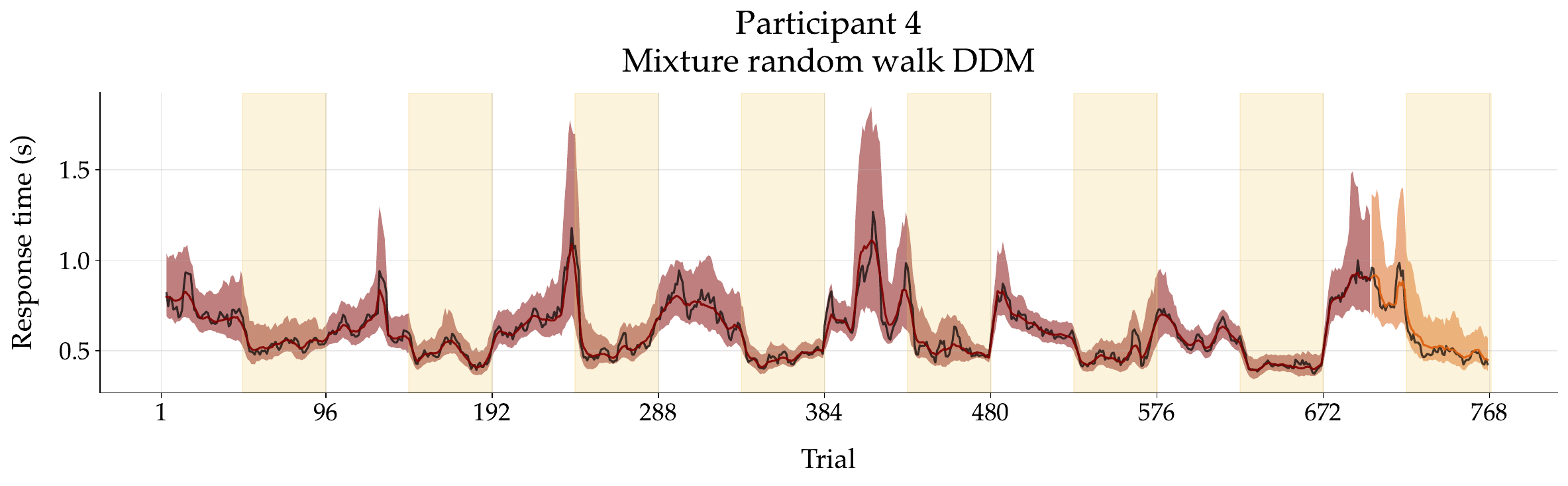}
\caption{Model fit to response time (RT) time series. The empirical RT time series of participant $4$ is shown in black. From trial $1$ to $700$, the posterior re-simulation (aka retrodictive check) using the best fitting non-stationary diffusion decision model (NSDDM) for this specific individual are shown in red. In this instance, the results stem from a mixture random walk DDM. For the remaining trials, one-step-ahead predictions are depicted in orange. Solid lines correspond to the median and shaded bands to $90\%$ credibility intervals (CI). The empirical, re-simulated, and predicted RT time series were smoothed via a simple moving average (SMA) with a period of $5$. The yellow shaded regions indicate trials where speed was emphasised over accuracy, while blank white areas denote instances where the opposite emphasis was applied.}
\end{figure*}

\begin{figure*}[h!]
\centering
\includegraphics[width=\textwidth]{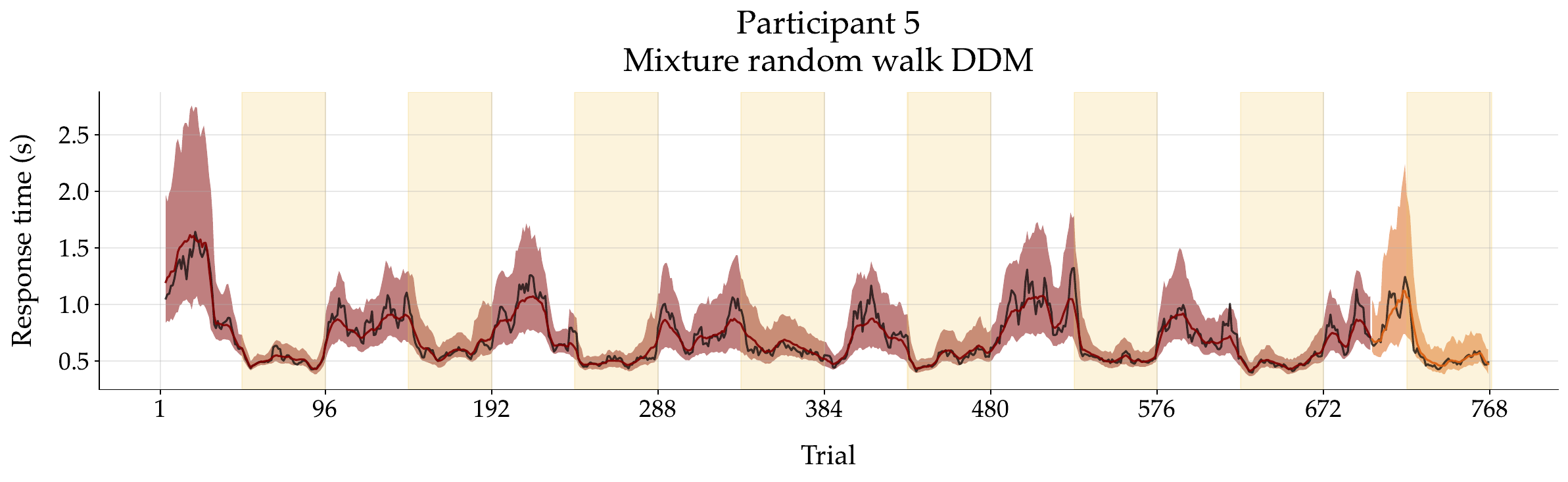}
\caption{Model fit to response time (RT) time series. The empirical RT time series of participant $5$ is shown in black. From trial $1$ to $700$, the posterior re-simulation (aka retrodictive check) using the best fitting non-stationary diffusion decision model (NSDDM) for this specific individual are shown in red. In this instance, the results stem from a mixture random walk DDM. For the remaining trials, one-step-ahead predictions are depicted in orange. Solid lines correspond to the median and shaded bands to $90\%$ credibility intervals (CI). The empirical, re-simulated, and predicted RT time series were smoothed via a simple moving average (SMA) with a period of $5$. The yellow shaded regions indicate trials where speed was emphasised over accuracy, while blank white areas denote instances where the opposite emphasis was applied.}
\end{figure*}

\begin{figure*}[h!]
\centering
\includegraphics[width=\textwidth]{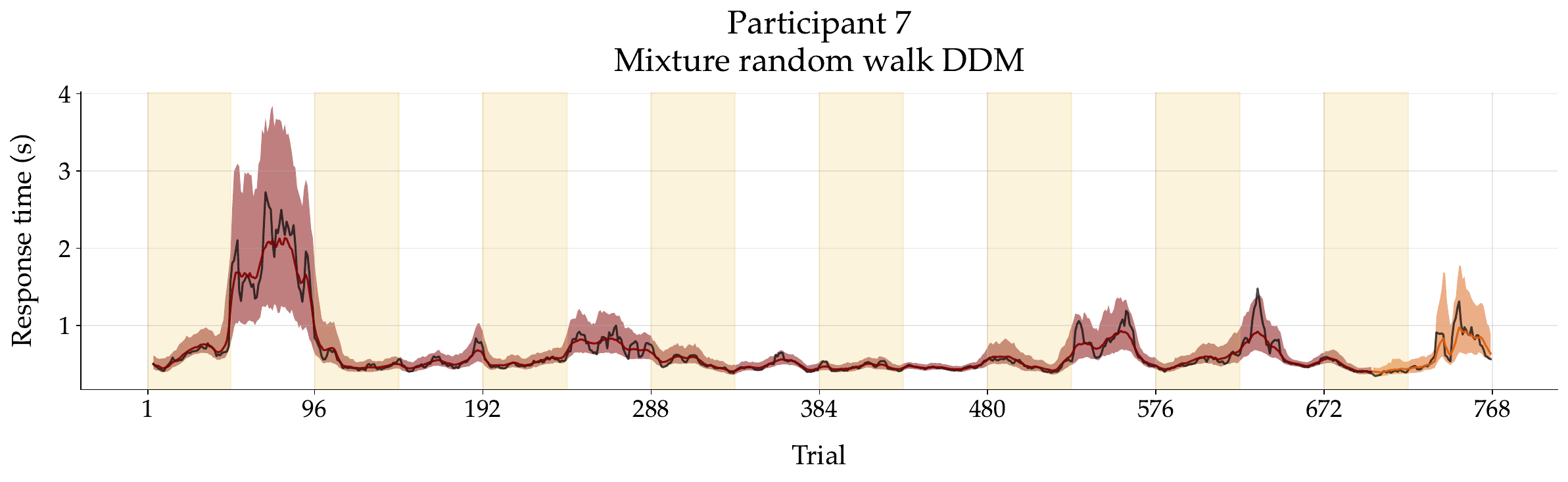}
\caption{Model fit to response time (RT) time series. The empirical RT time series of participant $7$ is shown in black. From trial $1$ to $700$, the posterior re-simulation (aka retrodictive check) using the best fitting non-stationary diffusion decision model (NSDDM) for this specific individual are shown in red. In this instance, the results stem from a mixture random walk DDM. For the remaining trials, one-step-ahead predictions are depicted in orange. Solid lines correspond to the median and shaded bands to $90\%$ credibility intervals (CI). The empirical, re-simulated, and predicted RT time series were smoothed via a simple moving average (SMA) with a period of $5$. The yellow shaded regions indicate trials where speed was emphasised over accuracy, while blank white areas denote instances where the opposite emphasis was applied.}
\end{figure*}

\begin{figure*}[h!]
\centering
\includegraphics[width=\textwidth]{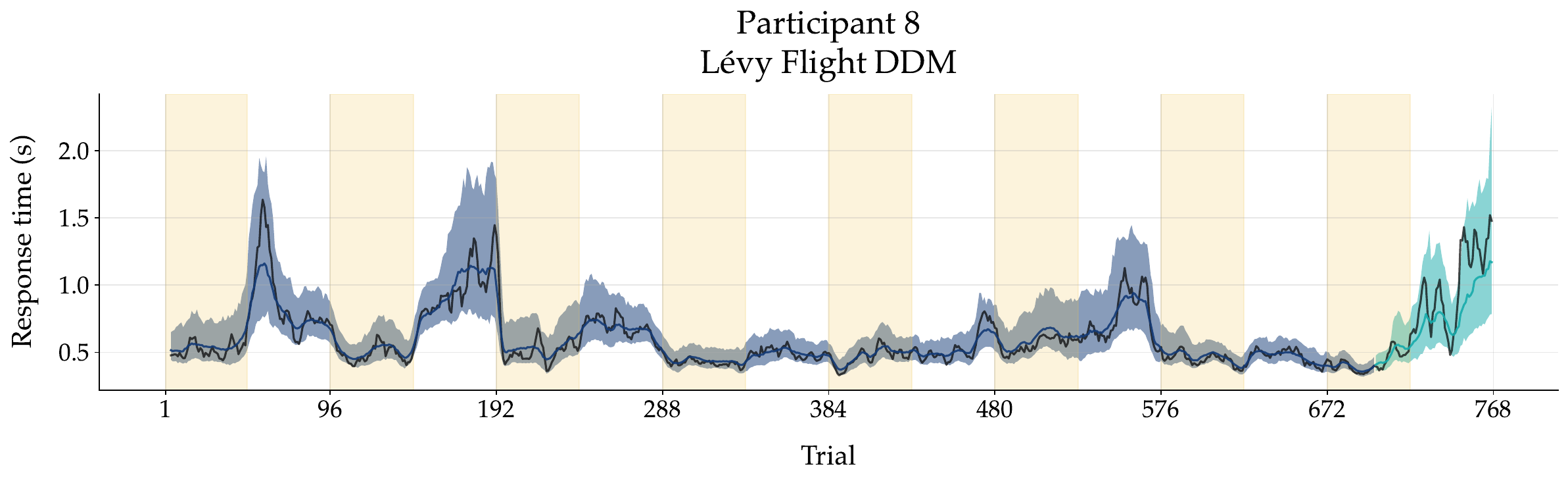}
\caption{Model fit to response time (RT) time series. The empirical RT time series of participant $8$ is shown in black. From trial $1$ to $700$, the posterior re-simulation (aka retrodictive check) using the best fitting non-stationary diffusion decision model (NSDDM) for this specific individual are shown in blue. In this instance, the results stem from a Lévy flight DDM. For the remaining trials, one-step-ahead predictions are depicted in cyan. Solid lines correspond to the median and shaded bands to $90\%$ credibility intervals (CI). The empirical, re-simulated, and predicted RT time series were smoothed via a simple moving average (SMA) with a period of $5$. The yellow shaded regions indicate trials where speed was emphasised over accuracy, while blank white areas denote instances where the opposite emphasis was applied.}
\end{figure*}

\begin{figure*}[h!]
\centering
\includegraphics[width=\textwidth]{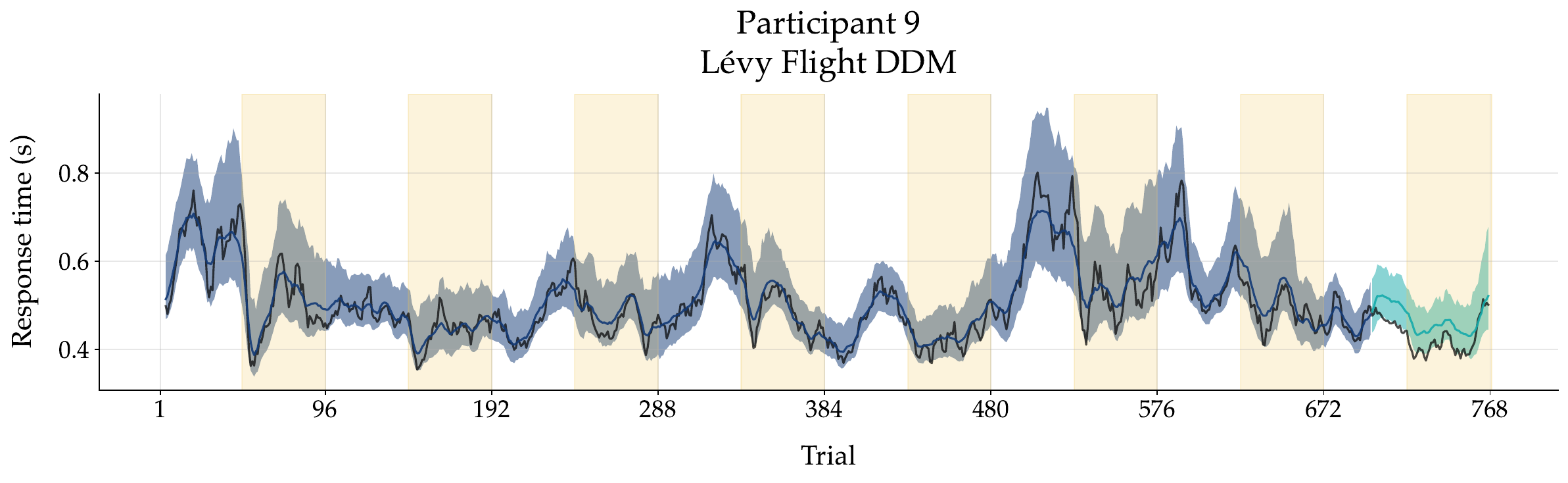}
\caption{Model fit to response time (RT) time series. The empirical RT time series of participant $9$ is shown in black. From trial $1$ to $700$, the posterior re-simulation (aka retrodictive check) using the best fitting non-stationary diffusion decision model (NSDDM) for this specific individual are shown in blue. In this instance, the results stem from a Lévy flight DDM. For the remaining trials, one-step-ahead predictions are depicted in cyan. Solid lines correspond to the median and shaded bands to $90\%$ credibility intervals (CI). The empirical, re-simulated, and predicted RT time series were smoothed via a simple moving average (SMA) with a period of $5$. The yellow shaded regions indicate trials where speed was emphasised over accuracy, while blank white areas denote instances where the opposite emphasis was applied.}
\end{figure*}

\begin{figure*}[h!]
\centering
\includegraphics[width=\textwidth]{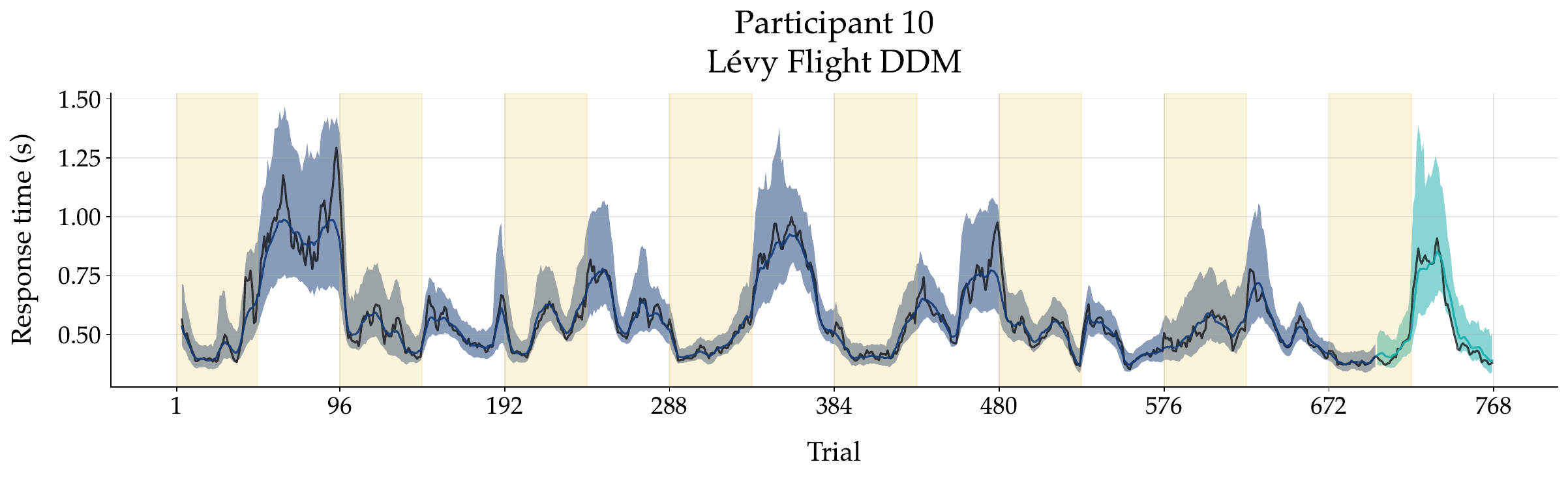}
\caption{Model fit to response time (RT) time series. The empirical RT time series of participant $10$ is shown in black. From trial $1$ to $700$, the posterior re-simulation (aka retrodictive check) using the best fitting non-stationary diffusion decision model (NSDDM) for this specific individual are shown in blue. In this instance, the results stem from a Lévy flight DDM. For the remaining trials, one-step-ahead predictions are depicted in cyan. Solid lines correspond to the median and shaded bands to $90\%$ credibility intervals (CI). The empirical, re-simulated, and predicted RT time series were smoothed via a simple moving average (SMA) with a period of $5$. The yellow shaded regions indicate trials where speed was emphasised over accuracy, while blank white areas denote instances where the opposite emphasis was applied.}
\end{figure*}

\begin{figure*}[h!]
\centering
\includegraphics[width=\textwidth]{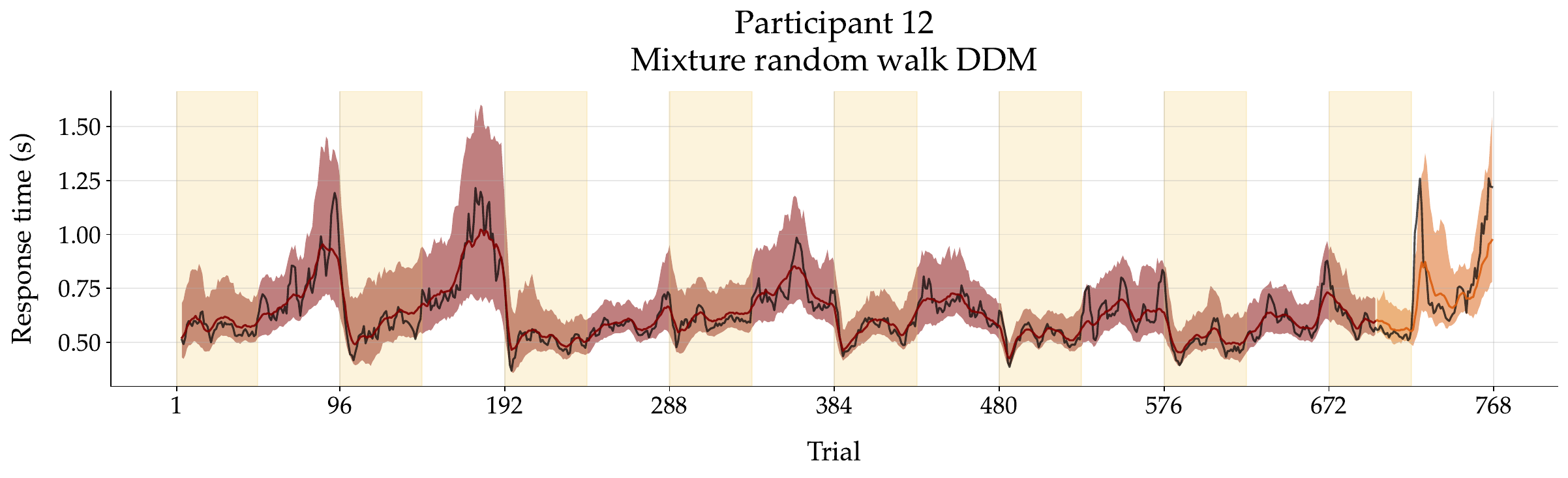}
\caption{Model fit to response time (RT) time series. The empirical RT time series of participant $7$ is shown in black. From trial $1$ to $700$, the posterior re-simulation (aka retrodictive check) using the best fitting non-stationary diffusion decision model (NSDDM) for this specific individual are shown in red. In this instance, the results stem from a mixture random walk DDM. For the remaining trials, one-step-ahead predictions are depicted in orange. Solid lines correspond to the median and shaded bands to $90\%$ credibility intervals (CI). The empirical, re-simulated, and predicted RT time series were smoothed via a simple moving average (SMA) with a period of $5$. The yellow shaded regions indicate trials where speed was emphasised over accuracy, while blank white areas denote instances where the opposite emphasis was applied.}
\end{figure*}

\begin{figure*}[h!]
\centering
\includegraphics[width=\textwidth]{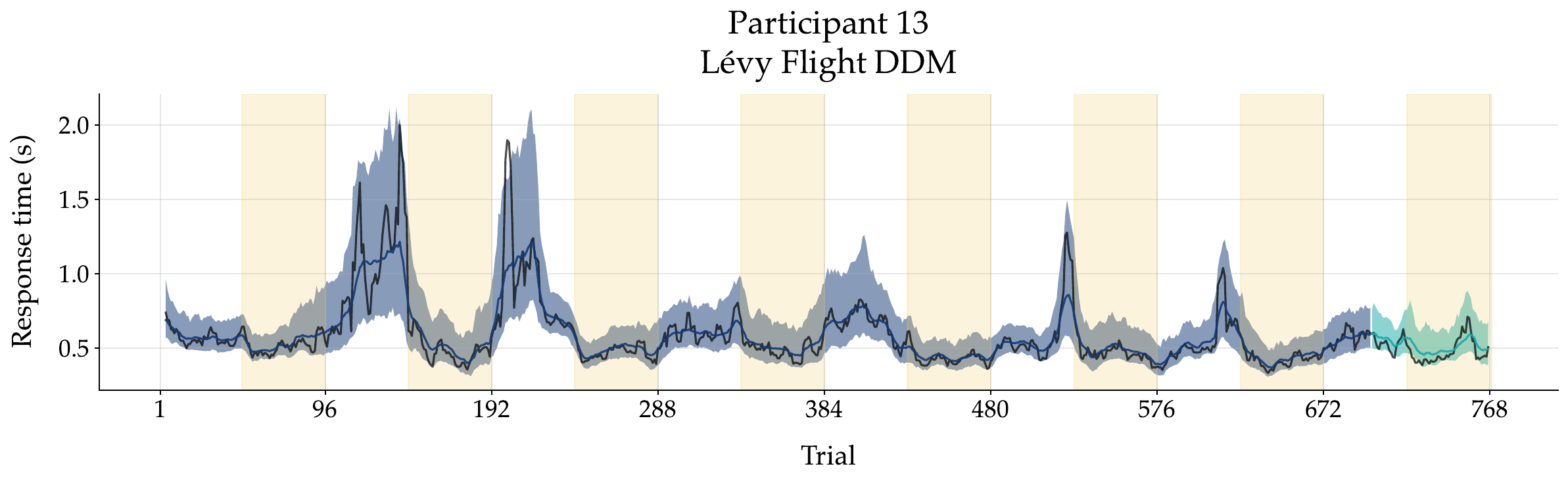}
\caption{Model fit to response time (RT) time series. The empirical RT time series of participant $13$ is shown in black. From trial $1$ to $700$, the posterior re-simulation (aka retrodictive check) using the best fitting non-stationary diffusion decision model (NSDDM) for this specific individual are shown in blue. In this instance, the results stem from a Lévy flight DDM. For the remaining trials, one-step-ahead predictions are depicted in cyan. Solid lines correspond to the median and shaded bands to $90\%$ credibility intervals (CI). The empirical, re-simulated, and predicted RT time series were smoothed via a simple moving average (SMA) with a period of $5$. The yellow shaded regions indicate trials where speed was emphasised over accuracy, while blank white areas denote instances where the opposite emphasis was applied.}
\end{figure*}

\begin{figure*}[h!]
\centering
\includegraphics[width=\textwidth]{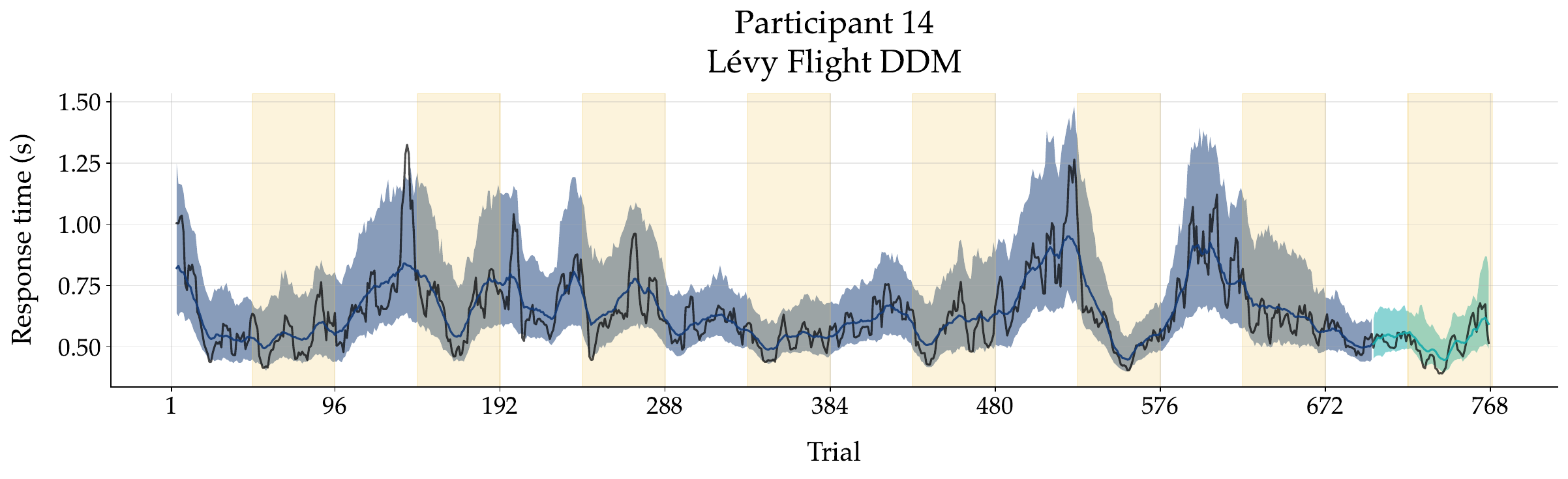}
\caption{Model fit to response time (RT) time series. The empirical RT time series of participant $14$ is shown in black. From trial $1$ to $700$, the posterior re-simulation (aka retrodictive check) using the best fitting non-stationary diffusion decision model (NSDDM) for this specific individual are shown in blue. In this instance, the results stem from a Lévy flight DDM. For the remaining trials, one-step-ahead predictions are depicted in cyan. Solid lines correspond to the median and shaded bands to $90\%$ credibility intervals (CI). The empirical, re-simulated, and predicted RT time series were smoothed via a simple moving average (SMA) with a period of $5$. The yellow shaded regions indicate trials where speed was emphasised over accuracy, while blank white areas denote instances where the opposite emphasis was applied.}
\end{figure*}

\FloatBarrier
\newpage

\section{Parameter Trajectories}

In the following, we present the inferred parameter trajectories for the remaining participants. For each visualisation the model with the highest posterior model probability for that specific individual was used.

\begin{figure*}[h!]
\centering
\includegraphics[width=0.99\textwidth]{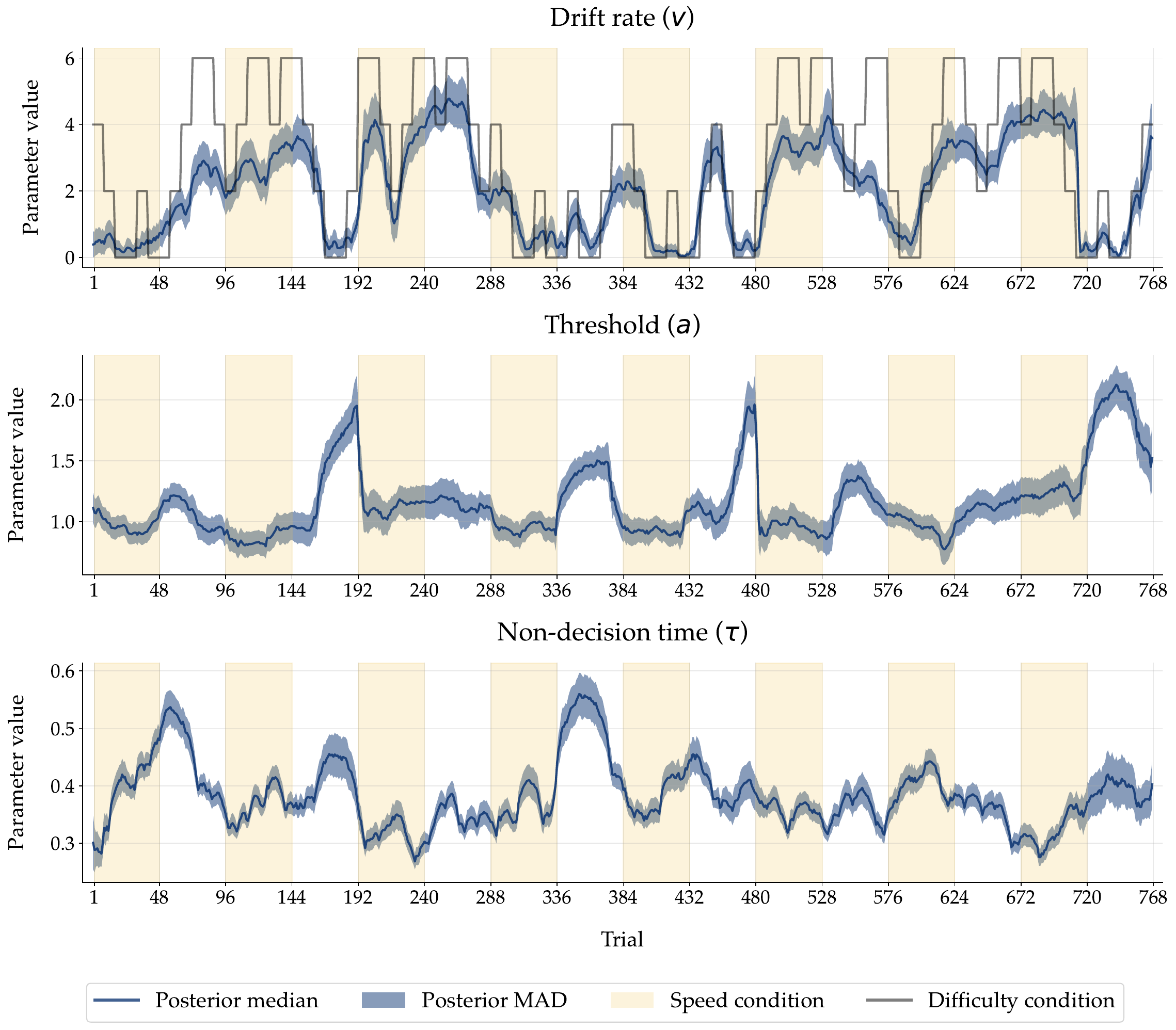}
\caption{Posterior parameter trajectory inferred with the best fitting NSDDM of participant $1$ (a Lévy flight DDM in this case) for all three DDM parameters (drift rate, threshold, and non-decision time) separately. The yellow shaded areas indicate trials where speed was emphasised over accuracy and blank white area indicated where the opposite was asked for. In the top panel, the task difficulty levels sequence is depicted in black lines.}
\end{figure*}

\FloatBarrier

\begin{figure*}[h!]
\centering
\includegraphics[width=0.99\textwidth]{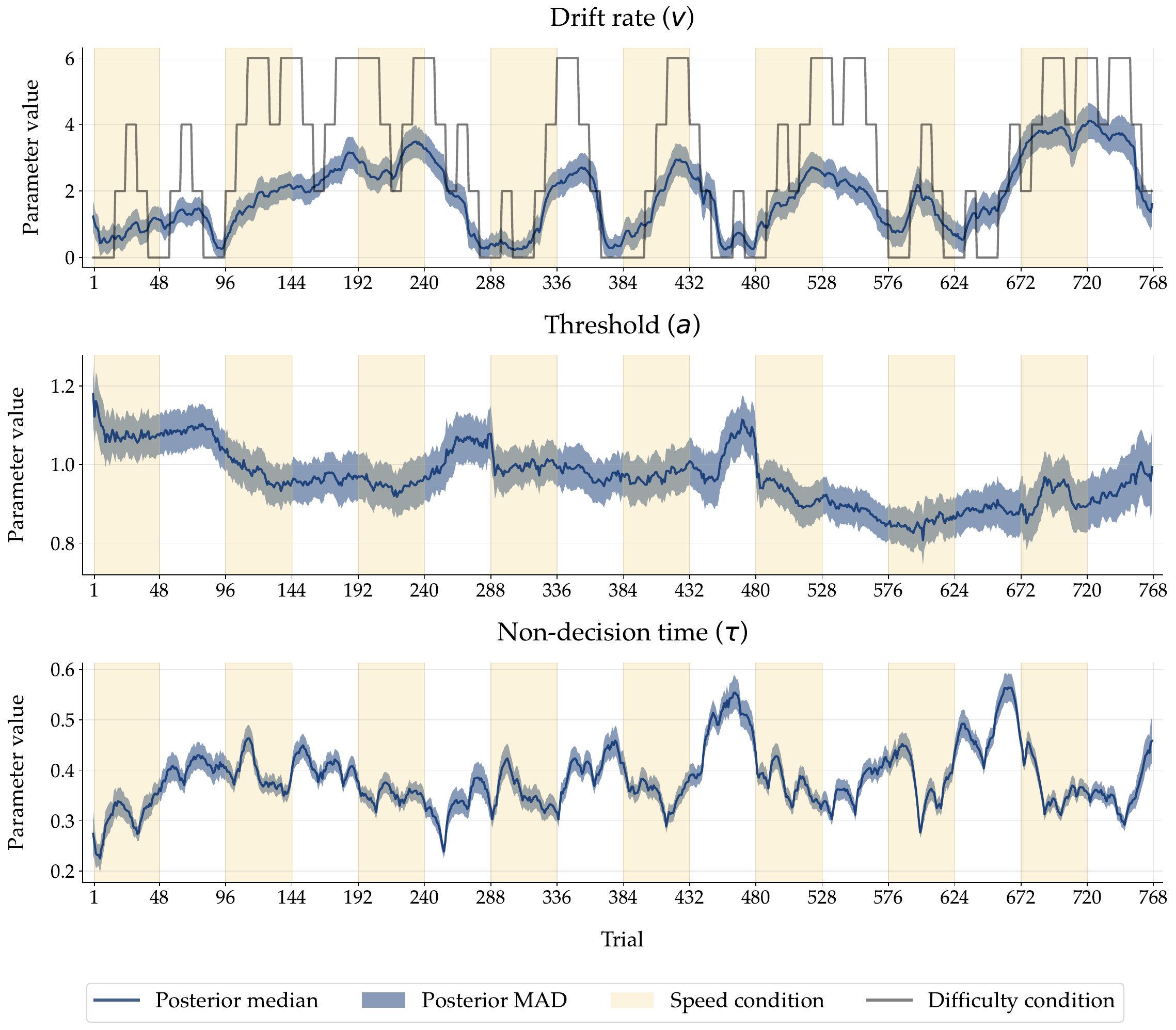}
\caption{Posterior parameter trajectory inferred with the best fitting NSDDM of participant $2$ (a Lévy flight DDM in this case) for all three DDM parameters (drift rate, threshold, and non-decision time) separately. The yellow shaded areas indicate trials where speed was emphasised over accuracy and blank white area indicated where the opposite was asked for. In the top panel, the task difficulty levels sequence is depicted in black lines.}
\end{figure*}

\FloatBarrier

\begin{figure*}[h!]
\centering
\includegraphics[width=0.99\textwidth]{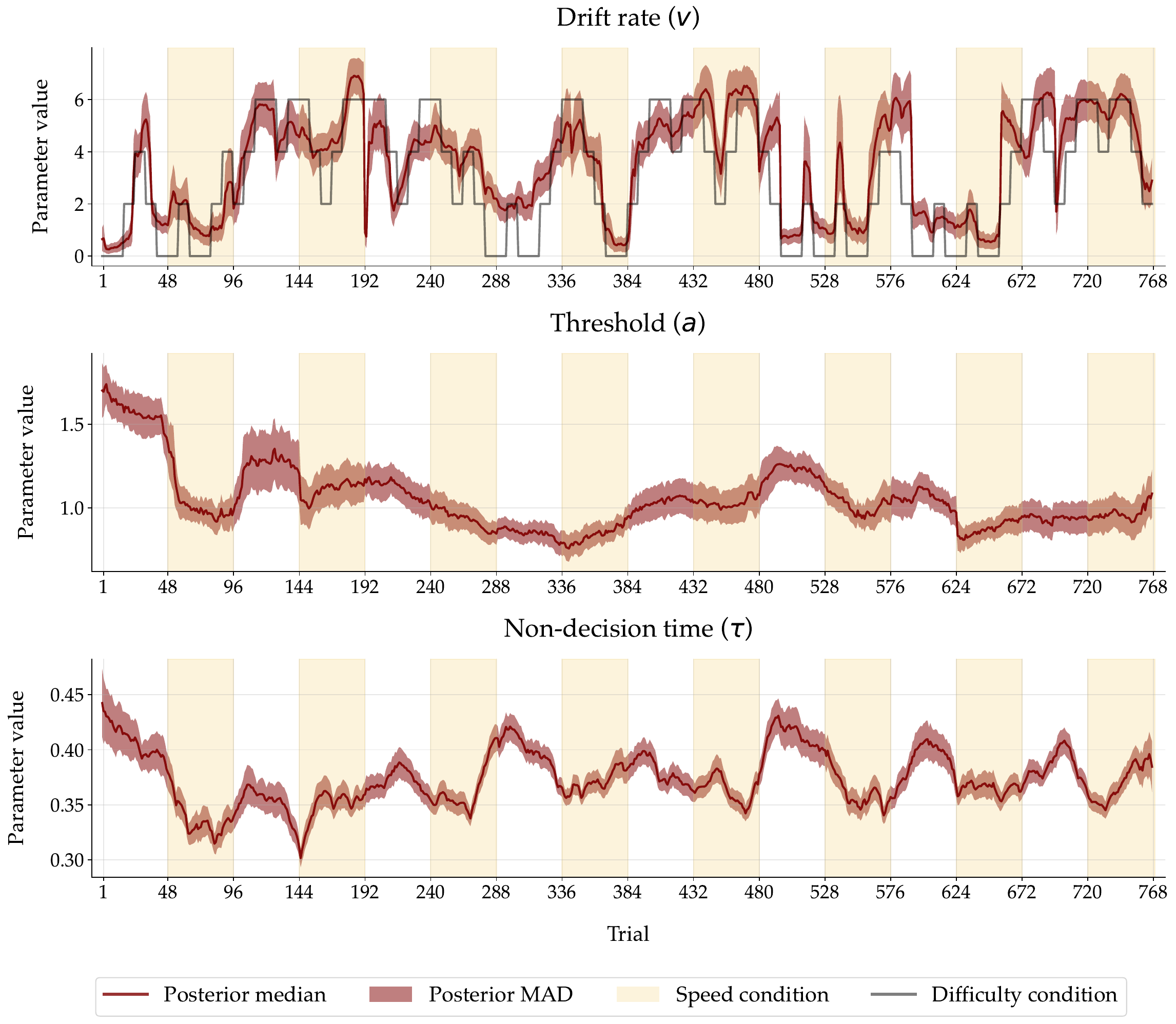}
\caption{Posterior parameter trajectory inferred with the best fitting NSDDM of participant $3$ (a mixture random walk DDM in this case) for all three DDM parameters (drift rate, threshold, and non-decision time) separately. The yellow shaded areas indicate trials where speed was emphasised over accuracy and blank white area indicated where the opposite was asked for. In the top panel, the task difficulty levels sequence is depicted in black lines.}
\end{figure*}

\FloatBarrier

\begin{figure*}[h!]
\centering
\includegraphics[width=0.99\textwidth]{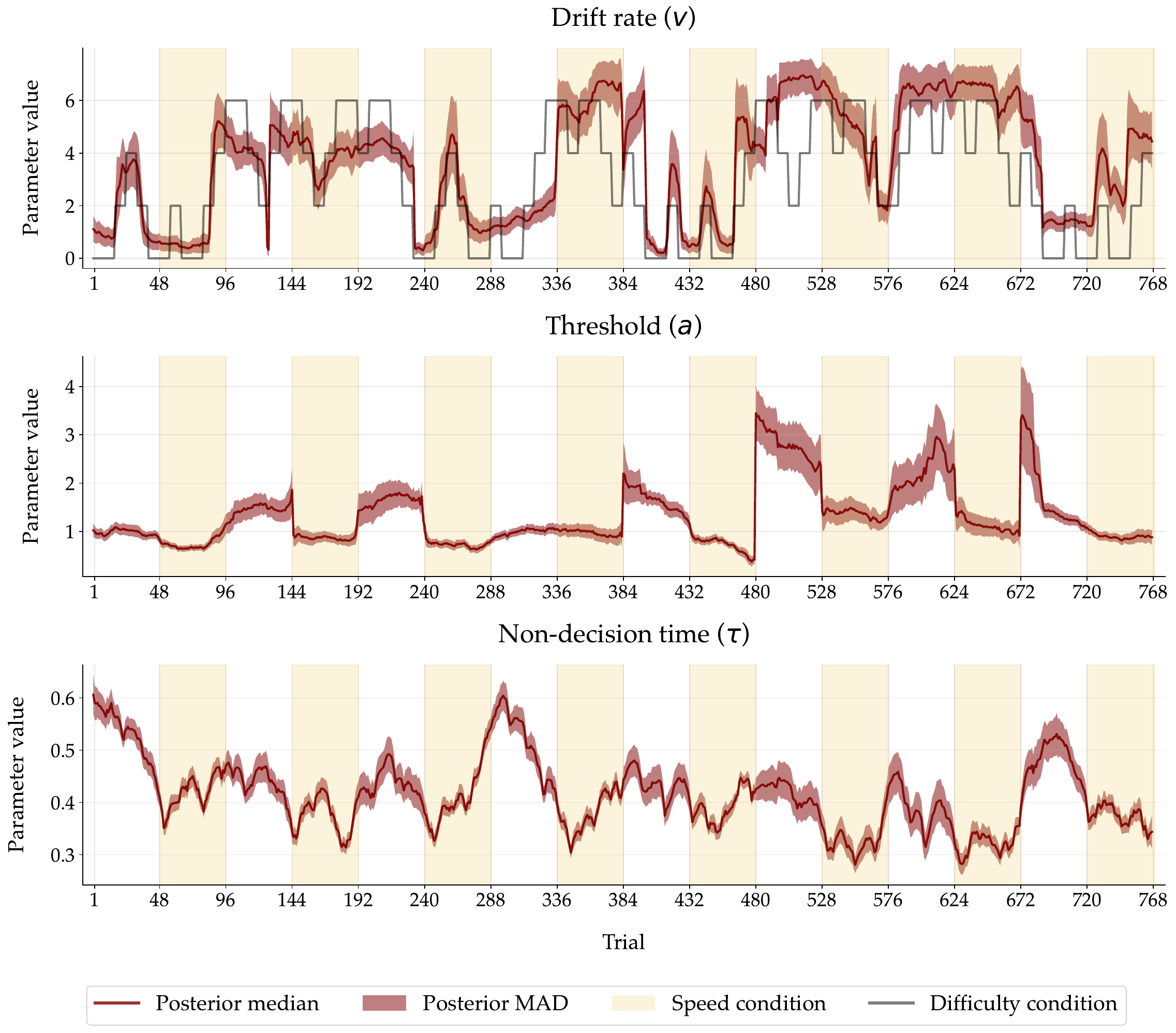}
\caption{Posterior parameter trajectory inferred with the best fitting NSDDM of participant $4$ (a mixture random walk DDM in this case) for all three DDM parameters (drift rate, threshold, and non-decision time) separately. The yellow shaded areas indicate trials where speed was emphasised over accuracy and blank white area indicated where the opposite was asked for. In the top panel, the task difficulty levels sequence is depicted in black lines.}
\end{figure*}

\FloatBarrier

\begin{figure*}[h!]
\centering
\includegraphics[width=0.99\textwidth]{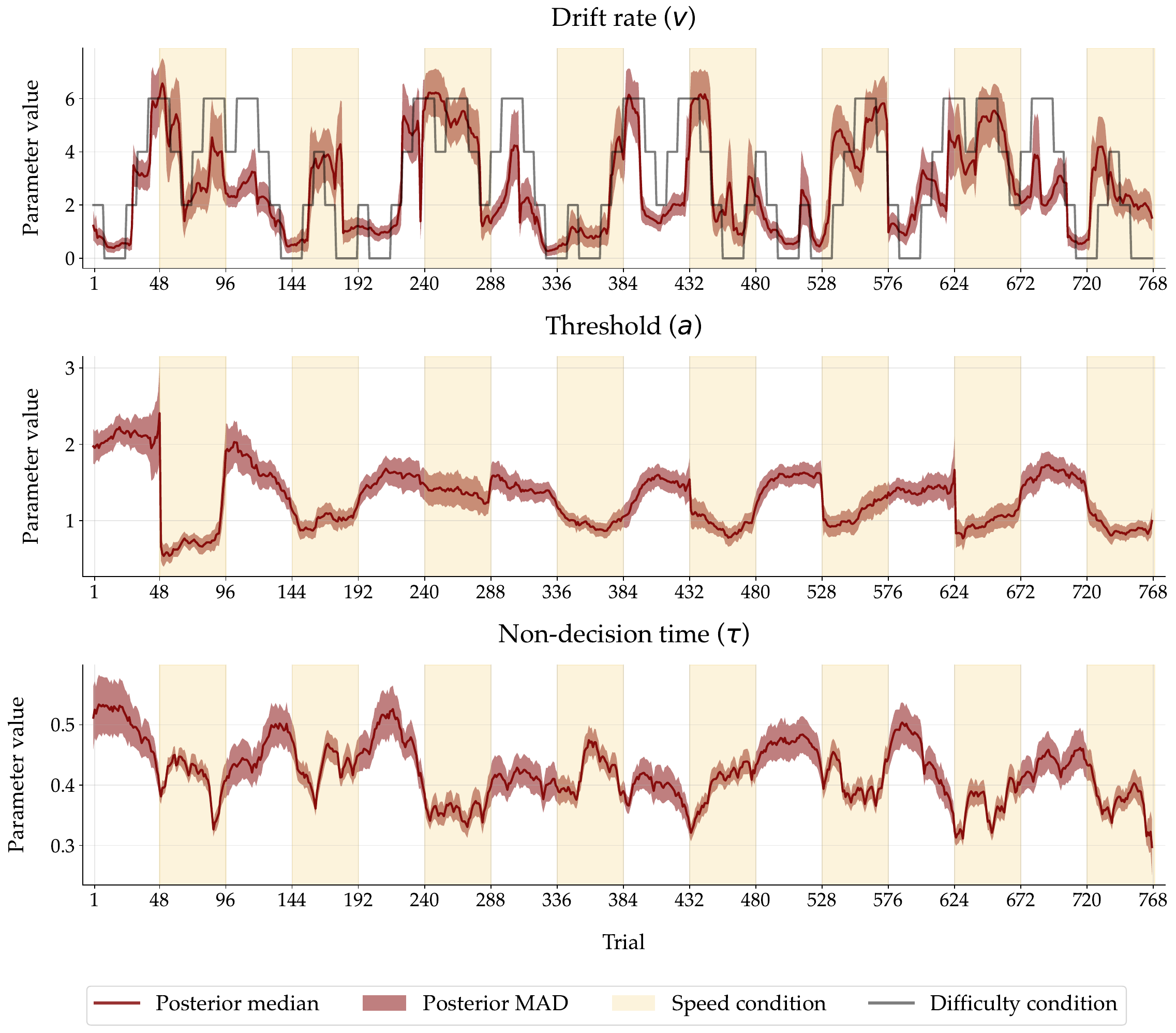}
\caption{Posterior parameter trajectory inferred with the best fitting NSDDM of participant $5$ (a mixture random walk DDM in this case) for all three DDM parameters (drift rate, threshold, and non-decision time) separately. The yellow shaded areas indicate trials where speed was emphasised over accuracy and blank white area indicated where the opposite was asked for. In the top panel, the task difficulty levels sequence is depicted in black lines.}
\end{figure*}

\FloatBarrier

\begin{figure*}[h!]
\centering
\includegraphics[width=0.99\textwidth]{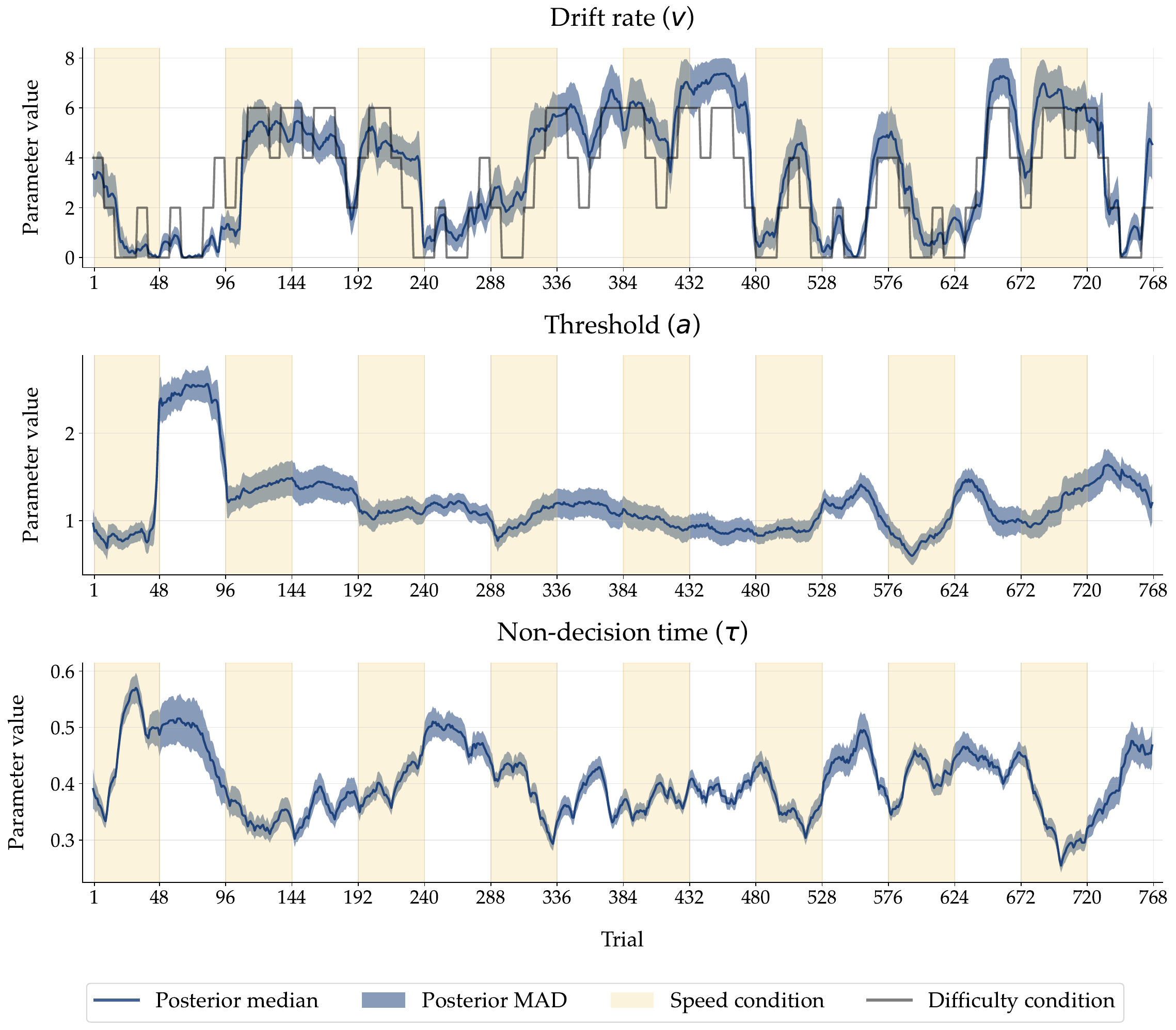}
\caption{Posterior parameter trajectory inferred with the best fitting NSDDM of participant $7$ (a Lévy flight DDM in this case) for all three DDM parameters (drift rate, threshold, and non-decision time) separately. The yellow shaded areas indicate trials where speed was emphasised over accuracy and blank white area indicated where the opposite was asked for. In the top panel, the task difficulty levels sequence is depicted in black lines.}
\end{figure*}

\FloatBarrier

\begin{figure*}[h!]
\centering
\includegraphics[width=0.99\textwidth]{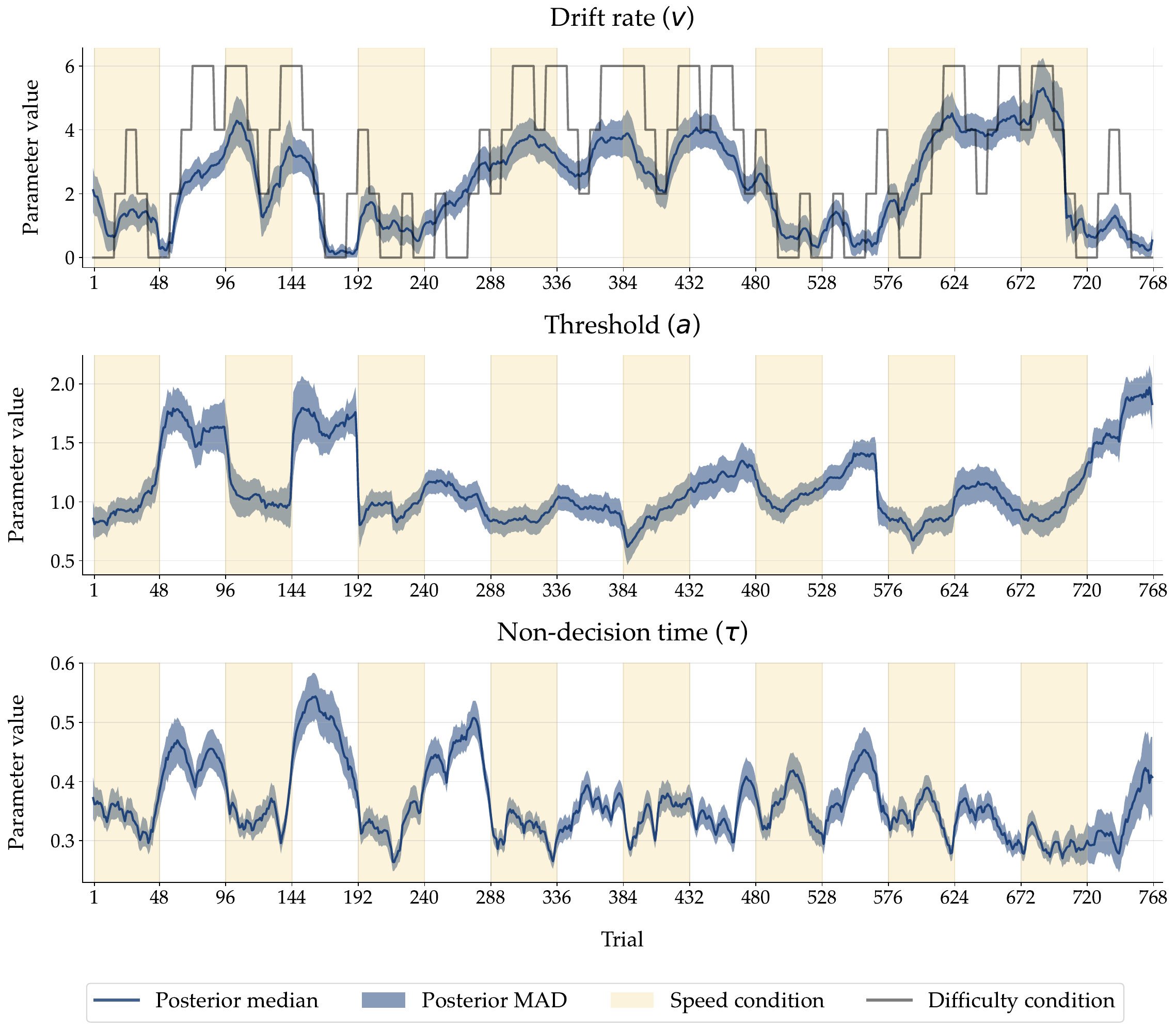}
\caption{Posterior parameter trajectory inferred with the best fitting NSDDM of participant $8$ (a Lévy flight DDM in this case) for all three DDM parameters (drift rate, threshold, and non-decision time) separately. The yellow shaded areas indicate trials where speed was emphasised over accuracy and blank white area indicated where the opposite was asked for. In the top panel, the task difficulty levels sequence is depicted in black lines.}
\end{figure*}

\FloatBarrier

\begin{figure*}[h!]
\centering
\includegraphics[width=0.99\textwidth]{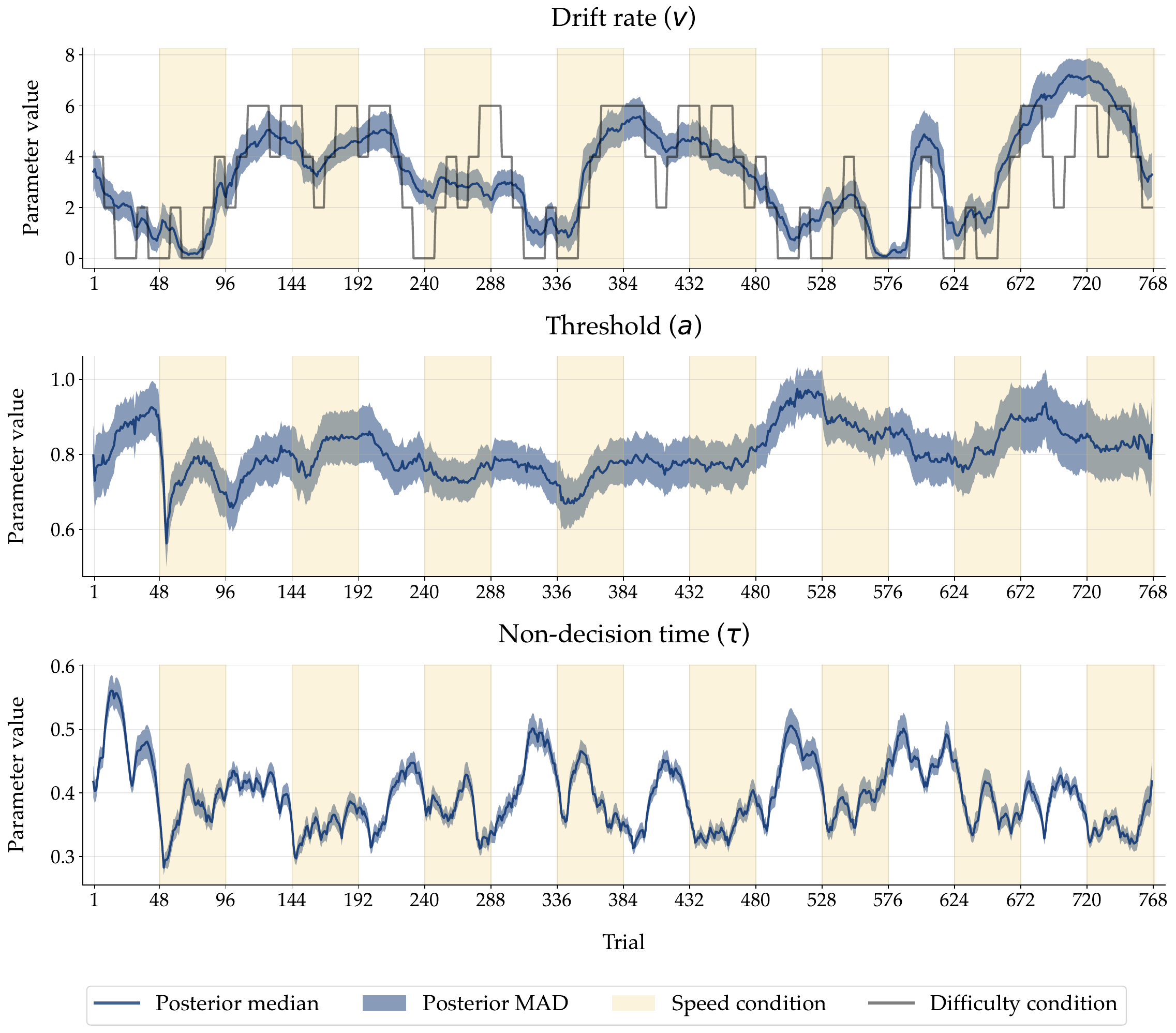}
\caption{Posterior parameter trajectory inferred with the best fitting NSDDM of participant $9$ (a Lévy flight DDM in this case) for all three DDM parameters (drift rate, threshold, and non-decision time) separately. The yellow shaded areas indicate trials where speed was emphasised over accuracy and blank white area indicated where the opposite was asked for. In the top panel, the task difficulty levels sequence is depicted in black lines.}
\end{figure*}

\FloatBarrier

\begin{figure*}[h!]
\centering
\includegraphics[width=0.99\textwidth]{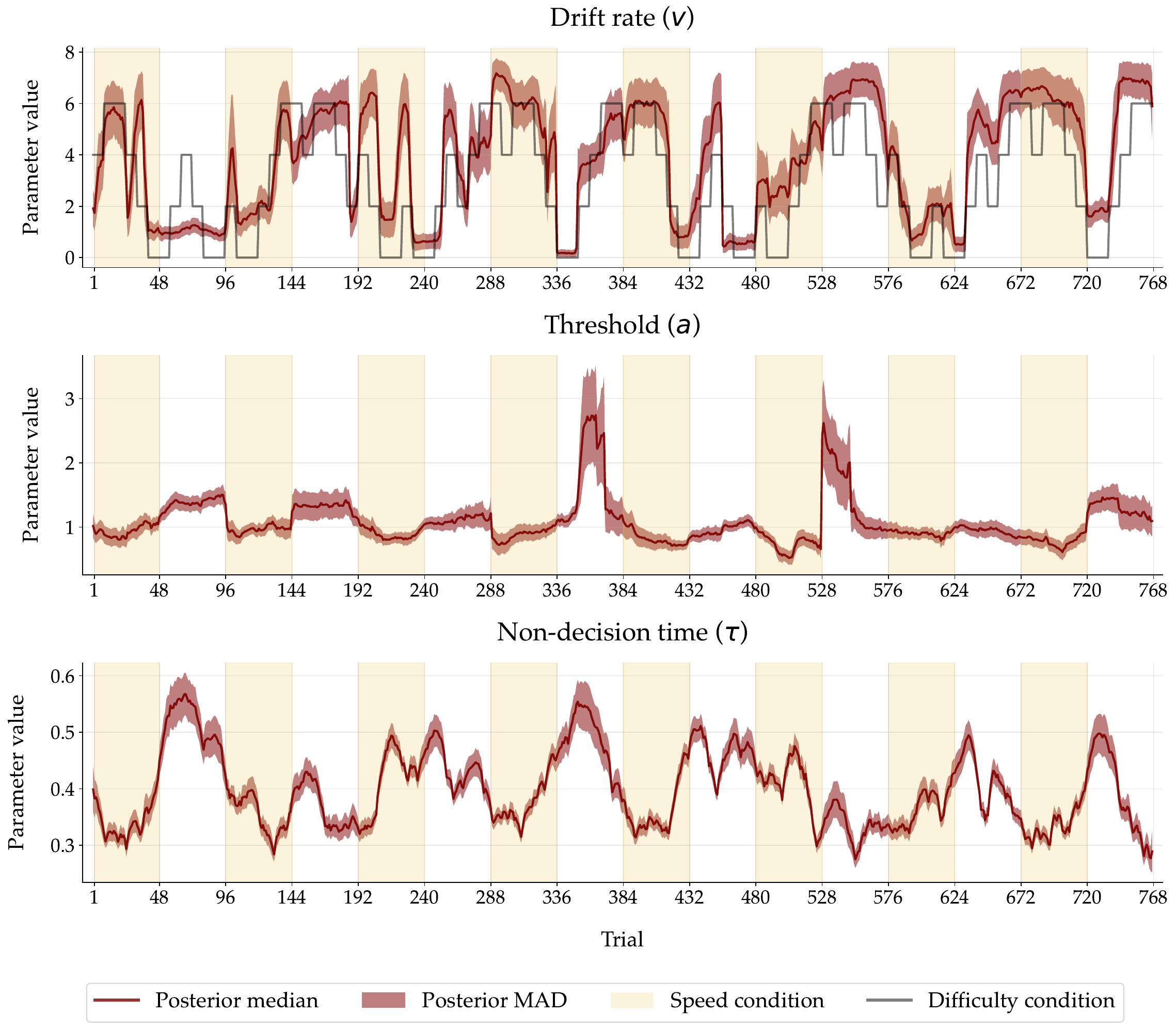}
\caption{Posterior parameter trajectory inferred with the best fitting NSDDM of participant $10$ (a mixture random walk DDM in this case) for all three DDM parameters (drift rate, threshold, and non-decision time) separately. The yellow shaded areas indicate trials where speed was emphasised over accuracy and blank white area indicated where the opposite was asked for. In the top panel, the task difficulty levels sequence is depicted in black lines.}
\end{figure*}

\FloatBarrier

\begin{figure*}[h!]
\centering
\includegraphics[width=0.99\textwidth]{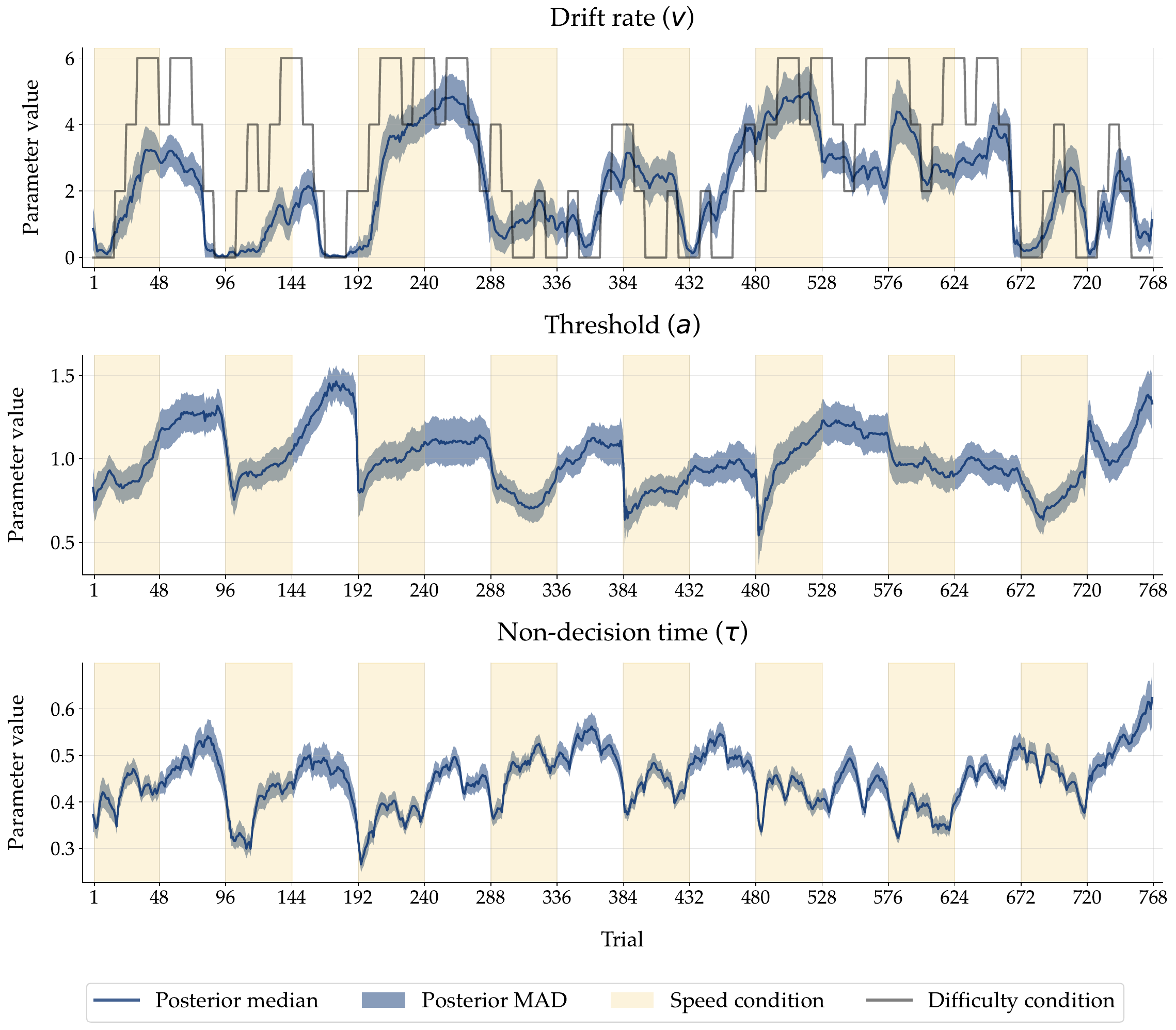}
\caption{Posterior parameter trajectory inferred with the best fitting NSDDM of participant $12$ (a Lévy flight DDM in this case) for all three DDM parameters (drift rate, threshold, and non-decision time) separately. The yellow shaded areas indicate trials where speed was emphasised over accuracy and blank white area indicated where the opposite was asked for. In the top panel, the task difficulty levels sequence is depicted in black lines.}
\end{figure*}

\FloatBarrier

\begin{figure*}[h!]
\centering
\includegraphics[width=0.99\textwidth]{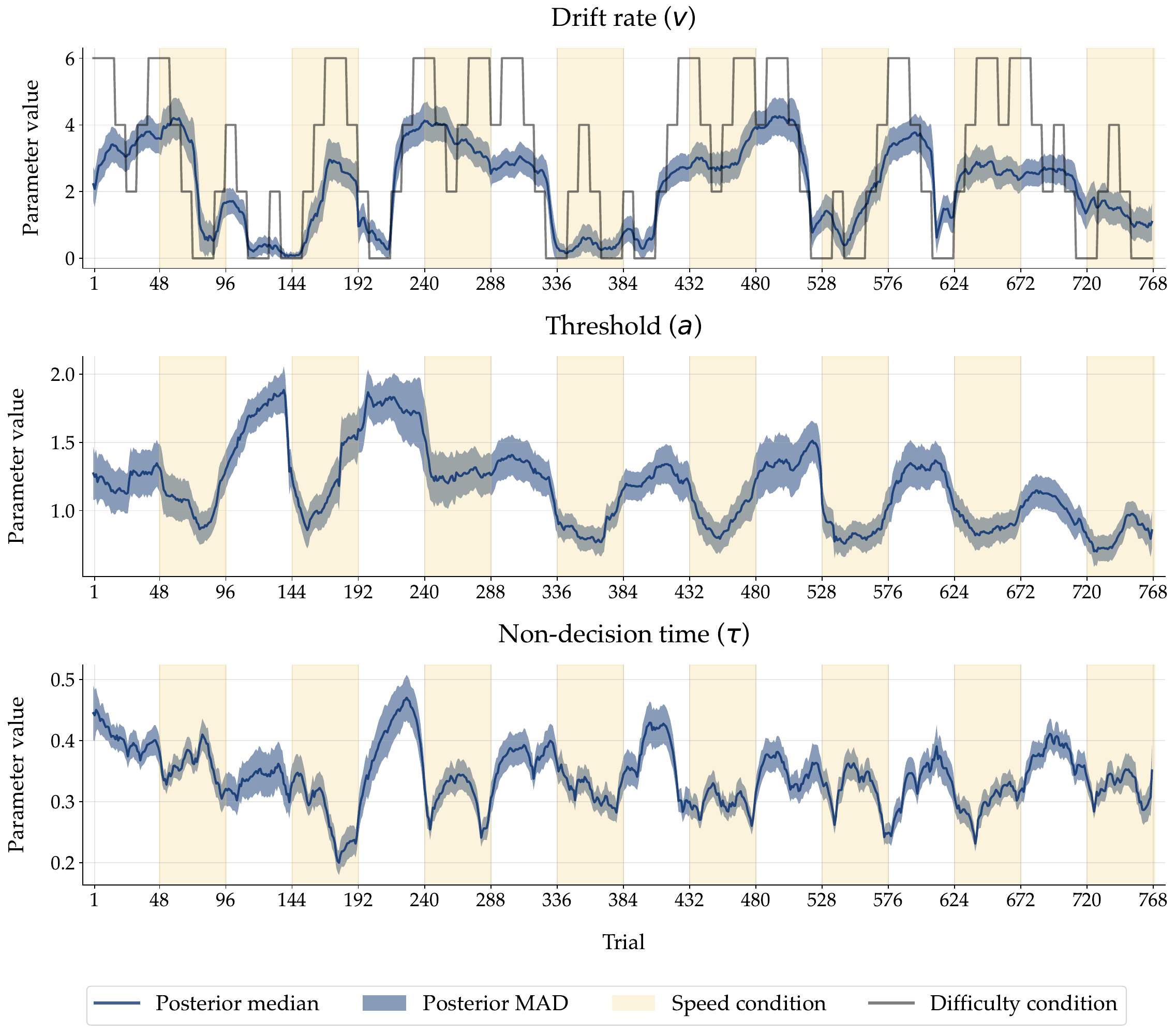}
\caption{Posterior parameter trajectory inferred with the best fitting NSDDM of participant $13$ (a Lévy flight DDM in this case) for all three DDM parameters (drift rate, threshold, and non-decision time) separately. The yellow shaded areas indicate trials where speed was emphasised over accuracy and blank white area indicated where the opposite was asked for. In the top panel, the task difficulty levels sequence is depicted in black lines.}
\end{figure*}

\FloatBarrier

\begin{figure*}[h!]
\centering
\includegraphics[width=0.99\textwidth]{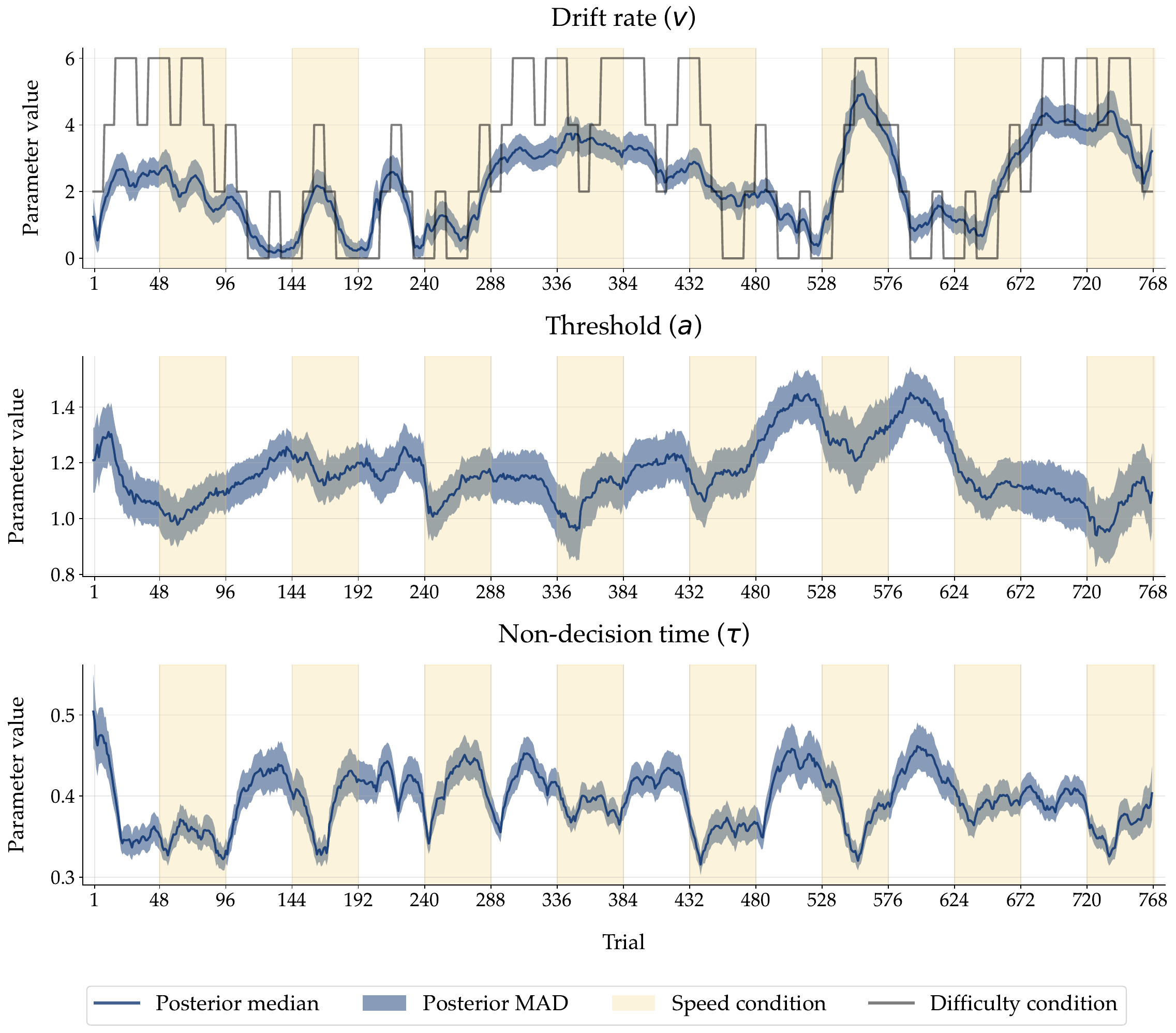}
\caption{Posterior parameter trajectory inferred with the best fitting NSDDM of participant $14$ (a Lévy flight DDM in this case) for all three DDM parameters (drift rate, threshold, and non-decision time) separately. The yellow shaded areas indicate trials where speed was emphasised over accuracy and blank white area indicated where the opposite was asked for. In the top panel, the task difficulty levels sequence is depicted in black lines.}
\end{figure*}

\FloatBarrier
\end{document}